\documentclass[aps, prd, twocolumn, superscriptaddress, preprintnumbers,nofootinbib]{revtex4}
\usepackage{graphicx}
\usepackage{dcolumn}
\usepackage{amssymb}
\usepackage{amsmath}
\usepackage{color}
\usepackage{verbatim}
\usepackage{multirow}
\usepackage{soul}

\def\al{\alpha} 
\def\be{\beta} 
\def\ga{\gamma}

\def\la{\lambda}

\def\si{\sigma}
\def\ta{\tau}

\def\La{\Lambda}

\newcommand{\ben}{\begin{equation}}
\newcommand{\een}{\end{equation}}
\newcommand{\bea}{\begin{eqnarray}}
\newcommand{\eea}{\end{eqnarray}}
\newcommand{\ba}{\begin{array}}
\newcommand{\ea}{\end{array}}
\newcommand{\bit}{\begin{itemize}}
\newcommand{\eit}{\end{itemize}}

\newcommand{\vev}[1]{\left\langle#1\right\rangle}

\newcommand{\bk}{\textbf{k}}

\newcommand{\ie}{{\textit{i.e.}}}

\newcommand{\tsim}{\tau_\mathrm{sim}}

\newcommand{\barC}{\bar{C}}
\newcommand{\tStart}{\tau_\mathrm{start}}
\newcommand{\tDiff}{\tau_\mathrm{diff}}
\newcommand{\tCG}{\tau_\mathrm{cg}}
\newcommand{\tRef}{\tau_\mathrm{ref}}
\newcommand{\tEq}{\tau_\mathrm{eq}}
\newcommand{\tLam}{\tau_\mathrm{\La}}
\newcommand{\tNow}{\tau_0}
\newcommand{\tEnd}{\tau_\text{end}}
\newcommand{\tMax}{\tau_\text{max}}

\newcommand{\tInit}{\ta_\text{i}}

\newcommand{\bC}{\bar{C}}

\newcommand{\barE}{\bar{E}}

\newcommand{\eVec}{c}
\newcommand{\eVal}{\la}

\newcommand{\Nbin}{N_\text{b}}

\newcommand{\xPeak}{x_\text{p}}

\begin{document}

\title{Energy-momentum correlations for Abelian Higgs cosmic strings}

\newcommand{\addressSussex}{Department of Physics \& Astronomy, University of Sussex, Brighton, BN1 9QH, United Kingdom}

\author{David Daverio}
\email{david.daverio@unige.ch}
\affiliation{D\'epartement de Physique Th\'eorique and Center for Astroparticle Physics, Universit\'e de Gen\`eve, 24 quai Ansermet, CH--1211 Gen\`eve 4, Switzerland}

\author{Mark Hindmarsh} 
\email{m.b.hindmarsh@sussex.ac.uk}
\affiliation{\addressSussex}
\affiliation{
Department of Physics and Helsinki Institute of Physics,
PL 64,  
 00014 University of Helsinki,
Finland
}
\author{Martin Kunz}
\email{martin.kunz@unige.ch}
\affiliation{D\'epartement de Physique Th\'eorique and Center for Astroparticle Physics, Universit\'e de Gen\`eve, 24 quai Ansermet, CH--1211 Gen\`eve 4, Switzerland}
\affiliation{African Institute for Mathematical Sciences, 6 Melrose Rd, Muizenberg, 7945, Cape Town, South Africa}

\author{Joanes Lizarraga}
\email{joanes.lizarraga@ehu.eus}
\affiliation{Department of Theoretical Physics, University of the Basque Country UPV-EHU, 48040 Bilbao, Spain}

\author{Jon Urrestilla}
\email{jon.urrestilla@ehu.eus}
\affiliation{Department of Theoretical Physics, University of the Basque Country UPV-EHU, 48040 Bilbao, Spain}

\date{\today}

\begin{abstract}
We report on the energy-momentum correlators obtained with recent numerical simulations of the Abelian Higgs model, essential for the computation of cosmic microwave background and matter perturbations of cosmic strings. Due to significant improvements both in raw computing power and in our parallel simulation framework, the dynamical range of the simulations has increased four-fold both in space and time, and for the first time we are able to simulate strings with a constant physical width in both the radiation and matter eras. The new simulations improve the accuracy of the measurements of the correlation functions at the horizon scale and confirm the shape around the peak. The normalization is slightly higher in the high wave-number tails, due to a small increase in the string density.
We study for the first time the behaviour of the correlators across cosmological transitions, and discover that the correlation functions evolve adiabatically, \ie\ the network adapts quickly to changes in the expansion rate. We propose a new method for constructing source functions for Einstein-Boltzmann integrators, comparing it with two other methods previously used.  The new method is more consistent, easier to implement, and significantly more accurate.
\end{abstract}

\keywords{cosmology: topological defects: CMB anisotropies}
\pacs{}

\maketitle


\section{Introduction}
\label{sec:Intro}

Cosmic strings \cite{VilShe94,Hindmarsh:1994re,Copeland:2011dx,Hindmarsh:2011qj}
are relics of the phase transitions occurring in the earliest stages of the universe,
predicted in many well-motivated models of high energy particle physics and cosmology 
 \cite{Sarangi:2002yt,Jones:2003da,Jeannerot:2003qv}.
Increasingly accurate observations of the cosmic microwave background (CMB) \cite{Dvorkin:2011aj,Hinshaw:2012aka,Ade:2013xla,Ade:2013zuv} 
and increasingly robust theoretical predictions 
\cite{Wyman:2005tu,Battye:2006pk,Fraisse:2007nu,Bevis:2007gh,Urrestilla:2008jv,Battye:2010xz,Urrestilla:2011gr,Lazanu:2014eya} 
have established that strings do not contribute more than a few percent to the temperature perturbations, corresponding to an upper bound on the symmetry-breaking scale of about $4\times10^{15}$ GeV. The majority of the temperature perturbations can be accounted for by an inflationary model with scalar spectral index $n_s = 0.968\pm0.006$ and tensor-to-scalar ratio $r < 0.12$ (95\% CL) \cite{Ade:2015lrj}.

Since the contribution of defects to CMB temperature fluctuations is small, accurate measurements of the CMB polarization channels acquire major importance, especially B-modes, where strings could still contribute at the same level as inflation. 
Recently,  the BICEP2 \cite{Ade:2014xna} collaboration released the first measurement of a B-mode signal at angular scales relevant for early universe physics. A careful cross-correlation with Planck data showed strong evidence that its origin was galactic dust:  no significant evidence for tensor modes was found \cite{Ade:2015tva}. In any case, the signal cannot be entirely produced by cosmic defects \cite{Lizarraga:2014eaa,Moss:2014cra,Lizarraga:2014xza}. Nevertheless, these works and others 
\cite{Seljak:2006hi,Pogosian:2007gi,Bevis:2007qz,Mukherjee:2010ve} 
highlight the sensitivity of B-mode measurements to strings and other topological defect models. Indeed, further progress in constraining or detecting strings from the CMB will come from future B-mode data and experiments at small angular scales.

The continuing improvement in data motivates us to reduce the remaining theoretical uncertainties in the cosmic string CMB calculations. In previous works we calculated energy-momentum and CMB power spectra contributions from cosmic string networks using field theory simulations of the Abelian Higgs (AH) model \cite{Bevis:2006mj,Bevis:2007qz,Bevis:2010gj}. 
The principal uncertainties in this approach are due to approximations used to handle a field theory in an expanding universe; modelling of the strings across cosmological transitions between the radiation, matter and $\Lambda$ eras; and extrapolations to large times and small angular scales.

The aim of a numerical simulation is to compute the unequal time correlators (UETCs) of the energy momentum tensor of the strings, from which CMB power spectra can be computed 
\cite{Hindmarsh:1994re, Turok:1996ud, Pen:1997ae,Durrer:1998rw,Durrer:2001cg}.   
The UETC approach has been widely used in field theory simulations, and in recent years, it has also been adapted to other cosmic string simulation schemes such as the Unconnected Segment Model (USM) \cite{Avgoustidis:2012gb} and the Nambu-Goto (NG) approximation \cite{Lazanu:2014xxa}.

In this paper we present updated energy-momentum correlations obtained from the largest 
field theory simulations performed to date. Thanks to a considerable increase in computational resources and in their programming management through the LATField2  \cite{LatField2d} framework, significant progress has been possible. These improvements enabled us to tackle the challenges outlined above.
We have been able to increase the size of the simulation box from $1024^3$ to $4096^3$ lattice sites (``4k'' simulations), so that we cover a patch of the universe 64 times bigger than in \cite{Bevis:2010gj}, and to simulate for four times longer. Therefore, some of the scales that could only be accessed by extrapolation in previous works can now be directly simulated. 

The first uncertainty mentioned above comes from the requirement that we simulate a massive field theory in an expanding universe. While the cosmic string core width is set by the mass scale of the field theory and remains unaltered by the expansion, 
the field equations are solved on lattices with comoving coordinates. As the universe expands 
the comoving string width shrinks. 
At some point in the evolution the comoving string width becomes less than the separation of adjacent lattice points, and we can no longer resolve the string core on the grid.

An effective proposal to avoid that situation has been to change the equation of motions so that the cosmic strings have an artificially growing physical core width, 
so that they can be resolved on a comoving lattice throughout the simulation \cite{Press:1989yh,Bevis:2006mj}. 
It was shown that the uncertainties thereby introduced were less than those originating from the limited volume and time of the simulations. However, the great increase in both volume and time of the simulations demand a re-examination of the core growth technique.
Our new resources have  made it possible to simulate 
string networks following the {true} equations of motion for both matter and radiation eras, 
at the cost of some dynamic range, as the system takes longer to settle into its scaling evolution. 
We find that, as argued previously, the differences in the UETCs with and without core growth are small, in the range 10-20\% near the peak of the correlators.

The new simulations extend the wave-number range of previous measurements both at low and high wave-number. We measure correlators at the horizon scale more accurately, and we are able to measure directly at values of $k$ a factor four higher than before, which we had previously reached only by extrapolation. We confirm the power-law behaviour of the scalar correlators at large wave-numbers, although the behaviour of the vector and tensor correlators is less clear.
We provide fits to the UETCs in closed form which can be used for modelling purposes.

We also address the modelling of the cosmological transitions in our simulations, not only the transition from radiation to matter but also that to a universe dominated by a cosmological constant. 
We perform the first simulations of Abelian Higgs strings across cosmological transitions, 
essential for checking and improving previous modelling. 
We find that string networks evolve in a close to adiabatic way across the radiation-matter transition; their properties are at all times close to those of a network simulated with a constant expansion rate equal to the instantaneous rate. 

We introduce a new technique for deriving the source functions for Einstein-Boltzmann integrators, which are a crucial step in the pipeline for calculating CMB and matter perturbations, and a source of significant uncertainty in the past.   We call our new method 
fixed-$k$ UETC interpolation.
We compare it to previous methods \cite{Bevis:2006mj,Fenu:2013tea}, finding that it is significantly more accurate.  

The paper is structured as follows. In Sec~\ref{sec:Model} we summarize the AH model and the UETC formalism, detailing the field theory simulations and scaling. We show how we merge the data from our simulations and correct for the effects of the finite string width in Sec~\ref{sec:UETC} and we describe in Sec~\ref{sec:Evec} the three methods for deriving transition-era source functions from the correlators, including our new one, comparing to new numerical simulations of transition-era correlators. We discuss and conclude in Sec~\ref{sec:Disc}. Two appendices contain a comparison of the new ETCs with those in Ref.\ \cite{Bevis:2010gj}, and a table of fitting functions which can be used to model the UETCs.


\section{Model and method overview}
\label{sec:Model}

We simulate local cosmic strings based on the simplest field theory model that contains them: the Abelian Higgs model, which has a $U(1)$ gauge symmetry. Following the notation used in previous works, we define the Lagrangian density as:

\begin{equation}
\mathcal{L} = -\frac{1}{4e^2} F_{\mu\nu}F^{\mu\nu} \\+ (D_{\mu}\phi)^{*}(D^{\mu}\phi) - \frac{\lambda}{4} (|\phi|^2 - \phi_0^2)^2\,,
\end{equation}
where $D_{\mu} = \partial_{\mu} + i A_{\mu}$ and $F_{\mu\nu} = \partial_{\mu}A_{\nu} - \partial_{\nu}A_{\mu}$, and where $e$ and $\lambda$ are dimensionless coupling constants. In a spatially flat Friedmann-Lema\^\i tre-Robertson-Walker (FLRW) cosmology with scale factor $a$ and choosing the temporal gauge ($A_0=0$), the equations of motion read:
\begin{equation}
\ddot{\phi} + 2 \frac{\dot{a}}{a} \dot{\phi} - D_jD_j\phi = -a^2\frac{\lambda}{2}(|\phi|^2 - \phi_0^2)\phi\,,
\end{equation}
\begin{equation}
\dot{F}_{0j} - \partial_iF_{ij} = -2a^2e^2 {\rm Im} (\phi^*D_j\phi)\,,
\end{equation}
where $F_{0i} = \dot A_i$, which are supplemented with the Gauss law constraint  
\begin{equation}
-\partial_iF_{0i} = -2a^2e^2 {\rm Im} (\phi^*\dot{\phi})\,.
\label{AHcons}
\end{equation}
In these equations, the  dot represents derivatives with respect to the conformal time; and the spatial derivatives are taken  with respect to the comoving coordinates. 

As mentioned in the introduction, in order to be able to simultaneously resolve the width of the string and the expansion of the universe in comoving coordinates, the equation of motion can be modified so that the physical width grows, and the comoving width does not shrink as fast as it should \cite{Press:1989yh,Moore:2001px,Bevis:2006mj}. 
One can also allow for non-standard damping terms to preserve the decay of proper momentum of a straight string.  The modified equations are then 
\bea
\ddot{\phi} + r_\phi \frac{\dot{a}}{a} \dot{\phi} - D_jD_j\phi &=& -a^{2s_\phi}\frac{\lambda}{2}(|\phi|^2 - \phi_0^2)\phi \, ,
\label{AHPRS1}\\
\dot{F}_{0j} + r_A\frac{\dot{a}}{a}F_{0j}- \partial_iF_{ij} &=& -2a^{2s_A}e^2 {\rm Im} (\phi^*D_j\phi) \, .
\label{AHPRS2}
\eea
with $r_\phi$, $r_A$, $s_\phi$ and $s_A$ constants.

This method will certainly violate energy conservation, and also runs the risk of violating Gauss's law. 
However, if we derive the field equations from a gauge-invariant action with time dependent coupling constants, Gauss's law will be maintained \cite{Bevis:2006mj,Bevis:2010gj}. 
Writing the time varying coupling constants as 
\begin{equation}
\lambda = \frac{\lambda_0}{a^{2(1-s)}}\,, \quad e = \frac{e_0}{a^{1-s}}\,,
\end{equation}
where we call $s$ the core growth parameter, 
leads to the equations
\bea
\ddot{\phi} + 2 \frac{\dot{a}}{a} \dot{\phi} - D_jD_j\phi &=& -a^{2s}\frac{\lambda_0}{2}(|\phi|^2 - \phi_0^2)\phi ,\;\;
\label{AHMod1}\\
\dot{F}_{0j} + 2(1-s)\frac{\dot{a}}{a}F_{0j}- \partial_iF_{ij} &=& -2a^{2s}e_0^2 {\rm Im} (\phi^*D_j\phi) .\;\;
\label{AHMod2}
\eea
These equations preserve Gauss's law and reduce to the true field equations when $s = 1$. 
We can write the Gauss law preserving parameters in Eqs. (\ref{AHPRS1},\ref{AHPRS2}) as $r_\phi = 2$, $r_A = 2(1-s)$, $s_\phi = s_A = s$. 
We note that the Abelian Higgs simulations of Ref.\ \cite{Moore:2001px} took $r_\phi = 2$, $r_A=0$ and $s_\phi = s_A = 0$, which violate Gauss's law.

With the help of the core growth parameter we can write the comoving string width as:
\begin{equation}
w = \frac{w_0}{a^s} \, .
\label{e:StrWid}
\end{equation}
If $s\leq1$, the strings have growing physical width.
However, the string mass per unit length and tension is preserved, and therefore the string dynamics are unaffected for configurations where the width can be neglected.
The extreme case is given by $s=0$ in which the width of the string is constant in comoving coordinates. 
Extensive testing showed that taking  $s=0$ was an acceptable approximation, with errors which were subdominant to those introduced by the finite size and finite duration of the simulations \cite{Bevis:2006mj}.

The evolution of the string network perturbs the background space-time; those perturbations evolve and affect the contents of the universe, eventually creating CMB anisotropies. In contrast to the inflationary perturbations, which were seeded primordially and then evolve ``passively'', defects induce perturbations actively during their whole existence. Those are estimated to be roughly of the order of the magnitude of $G\mu$, where $G$ is Newton's constant and $\mu$ the string tension. Current bounds on $G\mu$ constrain its value to be below $10^{-6}$ \cite{Urrestilla:2011gr,Ade:2013xla,Lizarraga:2014xza,Lazanu:2014xxa}. 
The gravitational back-reaction experienced by the network is not taken into account, since its magnitude is of order $(G\mu)^2$.

Energy-momentum correlations are an effective statistical tool used to describe defect induced perturbations. Indeed, the unequal time correlators of the energy-momentum tensor are the only object needed to derive the power spectrum of CMB anisotropies. UETCs are defined as follows:
\begin{equation}
{U}_{\lambda\kappa\mu\nu} (\textbf{k},\tau,\tau') = \langle{\mathcal{T}}_{\lambda\kappa}(\textbf{k},\tau){\mathcal{T}}^*_{\mu\nu}(\textbf{k},\tau')\rangle\,,
\label{emtens}
\end{equation}
where ${\mathcal{T}}_{\alpha\beta}(\textbf{k},\tau)$ is the AH energy-momentum tensor.

In principle considering all possible degrees of freedom of the energy-momentum tensor (\ref{emtens}), there seem to be $\frac{1}{2}10(10+1) = 55$ such correlators that would be functions of 5 variables (3 components of $\bf{k}$ plus two times). Fortunately, rotational symmetry simplifies the problem considerably and reduce the UETC group to 5 independent correlators that depend on 3 variables: $k$ (the magnitude of $\bk$), $\tau$ and $\tau'$. 

The correlations are calculated between projected components of the energy-momentum tensor
\ben
S_a(\bk,t) = P_a^{\mu\nu}(\bk) {\mathcal{T}}_{\mu\nu}(\textbf{k},\tau'),
\een
where $P_a^{\mu\nu}(\bk)$ project onto scalar, vector and tensor parts.  In principle there are two of each, but the two vector and the two tensor components are related by parity for a symmetric source like Abelian Higgs strings. 
Hence, we may consider that the indices $a$, $b$ take four values corresponding to the independent components of the energy momentum tensor: two scalar, one vector and one tensor. 
We will denote the scalar indices $1$ and $2$ (corresponding to the 
longitudinal gauge potentials $\phi$ and $\psi$), the vector component with `v' and the tensor component with `t'.

Thus we can write 
\begin{equation}
{U}_{ab}(\textbf{k} ,\tau,\tau') = \frac{\phi_0^4}{\sqrt{\tau\tau'}}\frac{1}{V}{C}_{ab}(k,\tau,\tau'),
\label{UETCdecom}
\end{equation}
where $\phi_0$ is the symmetry breaking scale, $V$ a formal comoving volume factor, and the functions $C_{ab}(k,\tau,\tau')$ defined by this equation are dimensionless.
Note that the scalar, vector and tensor contributions are decoupled for linearized cosmological perturbations, and therefore cross correlators between them vanish, except in the scalar sector: hence the 5 independent correlators. Note also that the definition of $C_\text{vv}$ is different by a factor $(k\tau)^2$ from that in Ref.\ \cite{Bevis:2010gj}.

The UETCs give the power spectra of cosmological perturbations when convolved with the appropriate Green's functions. In practice, they are decomposed into a set of functions derived from the 
eigenvectors of the UETCs, 
which are used as sources for an Einstein-Boltzmann integrator. 
The power spectrum of interest is reconstructed as the sum of power spectra from each of the  source functions . 

A further simplification occurs when the times  $\tau$ and $\tau'$ are both in epochs 
during which the scale factor grows with the same constant power of conformal time. In this case 
the correlation functions do not depend on $k$, $\tau$ and $\tau'$ separately, but only on $k\tau$ and $k\tau'$. 
This behaviour is called scaling, and scaling correlators can be written 
\begin{equation}
{U}_{ab}(\textbf{k} ,\tau,\tau') = \frac{\phi_0^4}{\sqrt{\tau\tau'}}\frac{1}{V}{\barC}_{ab}(k\sqrt{\tau\tau'},\tau'/\tau)\, .
\label{UETCscaling}
\end{equation}
Here, the overbar represents the scaling form of the UETC in a FLRW background.
We will sometimes write $z = k\sqrt{\tau\tau'}$, $r = \tau'/\tau$.  An alternative pair of scaling variables is 
$x,x' = k\tau,k\tau'$. A scaling UETC will have eigenvectors which depend on $k$ and $\tau$ only through the combination $x$.

Scaling is an immensely valuable property, as it allows to extrapolate numerical simulations to the required cosmological scales.  However, perfect scaling is not a feature of the true UETCs, as the universe undergoes a transition from radiation-dominated to matter-dominated expansion during times of interest, and more recently to accelerated expansion.  Hence the UETCs also depend explicitly on $\tEq$ and $\tLam$, the times of equal radiation and matter density, and equal matter and dark energy density. Exploring UETCs with broken scaling, and improving our previous method of accounting for cosmological transitions, are an important part of this paper.


\section{UETCs from the simulations}
\label{sec:UETC}

In this section we present the details of the numerical simulations from which the UETC data was collected, and how the data was merged into a set of 10 scaling UETCs, 5 each in the matter and radiation eras.  
These merged scaling UETCs are the inputs for the next section, in which the eigenvector decomposition methods are discussed.

\subsection{Simulation details}

The data was obtained from two years of production on the supercomputers Monte Rosa and Piz Daint, the two largest systems of the Swiss National Supercomputer Center (CSCS). On both of those systems we have used 34816 cores/MPI processes, 32768 for computation and 2048 for efficient output operations.

The field equations were evolved on $4096^3$ lattices with comoving spatial separation of $dx=0.5$ and time steps of $dt=0.1$, in units where $\phi_0=1$. The simulation volume therefore has comoving size $L = 2048$.
The couplings were $\la_0 = 2$ and $e_0 = 1$, chosen so that the mass of the gauge and scalar fields, $\la\phi_0/\sqrt{2}$ and $\sqrt{2}e\phi_0$, are the same, and equal to $\sqrt{2}\phi_0$ at the end of the simulation.  The inverse mass of the fields sets the length scale of the string width.  
With these couplings, the mass per unit length of the string in the continuum is $\mu = 2\pi\phi_0^2$.

We performed 7 individual runs in pure radiation and in pure matter domination eras to determine the scaling form of the UETCs, for two values of the string core growth parameter, $s=0$ and $s=1$ (see section \ref{sec:Model} for the definition of $s$). 
We also performed runs across the radiation-matter cosmological transitions on   $1024^3$ lattices, with  $s=0$. 
In total, we used UETCs from 28 4k and  35  1k production runs. Each 4k run took approximately 400k core-hours.

In the initial field configuration, only the scalar field is non-zero, and set to be a stationary Gaussian random field 
with a power spectrum
\ben
P_\phi(\bk) = \frac{A}{1 + (k L_\phi)^2},
\een
with $A$ chosen so that $\vev{|\phi^2|} = \phi_0^2$, and $L_\phi = 5\phi_0^{-1}$.

The UETCs cannot be calculated until cosmic strings are formed and reach their scaling configuration. These early phases contain a huge amount of excess energy induced by the random initial conditions, therefore we smooth the field distribution by applying a  period of diffusive evolution 
\bea
\dot{\phi} &=& D_jD_j\phi - \frac{\lambda}{2}(|\phi|^2 - \phi_0^2)\phi \, ,
\label{AHdiff1}\\
F_{0j} &=& \partial_iF_{ij} - e^2 {\rm Im} (\phi^*D_j\phi) \, ,
\label{AHdiff2}
\eea
between the start time of the simulation $\tStart = 50$ and a time $\tDiff = 70$. The timestep was $1/30$, in units where $\phi_0=1$.   

 We follow the same technique as in Ref.\ \cite{Bevis:2006mj} to accelerate the formation of the strings in the $s=1$ case, by setting $s$ negative, so that the cores of the strings grow with the comoving horizon until a time $\tCG$, staying at most 1/10 of the horizon size at all times. 
The cooling and the core growth optimize the speed of approach to a scaling field configuration. 
The run is stopped soon after half a light-crossing time of the simulation volume, to ensure there are no artefacts from the periodic boundary conditions.

With our current computing power we are able to get scaling string configurations following the real equations of motion, \ie, equations with $s=1$,  even in the matter era where the string width shrinks as the square of conformal time in comoving coordinates. However, in general, the closer the evolution to the true dynamics, the larger the initial relaxation period where UETCs cannot be collected, and the shorter the period during which they exhibit scaling. 
Conversely, $s=0$ simulations reach  the scaling regime much more quickly: in our current simulation box $s=0$ simulations scale for a period 4 times longer than $s=1$ networks. 

We measure the UETC by recording the mean value of $C_{ab}(k,\tRef,\tau)$ for wavevectors binned in the range $2\pi (n-1)/L < |\bk| \le  2 n/L$ ($1 \le n <  \Nbin $), with $\Nbin = 3458$, and  150 logarithmically-spaced times between $\tRef$ and $\tEnd = 1100$.  
The wavenumber of the $n$th bin $k_n$ is set to the mean value of $|\bk|$ in that bin. Table~\ref{table_tref} shows the values of $\tRef$ taken.

We also record the equal time correlators (ETCs) at each time the UETC is evaluated, with which we can monitor the quality of the scaling.  Perfect scaling would mean that the ETCs collapse to a single line when plotted against $x = k\tau$.  As mentioned above, the network takes some time to relax to scaling, and in the $s=1$ case we see some evidence that the vector ETCs depart from scaling towards the end of the simulation, which we believe is a lattice resolution effect.  We therefore conservatively take UETC data up to a time $\tMax$.  Table~\ref{table_tref} also shows $\tMax$, and derived parameters which describe the dynamic range of the  simulation.

\begin{table}[h!]
\renewcommand{\arraystretch}{1.2}
\begin{tabular}{|c||c|c||c|c|}
\hline
 Model & \multicolumn{2}{c||}{$s=1$} & \multicolumn{2}{c|}{$s=0$}  \\\hline
 Cosmology & Radiation & Matter & Radiation & Matter \\ \hline
 $\tau_\text{cg}$ & 204 & 366 & -- & -- \\
 $s_\text{cg}$ & -1 & -0.5 & -- & -- \\
 $\tau_{\mathrm{ref}}$ & 450 & 600 & 200 & 200 \\
 $\tau_{\mathrm{max}}$ & 600 & 800 & 1100 & 1100 \\
 $r^{\mathrm{max}}$ & 1.33 & 1.33 & 5.5 & 5.5\\
 $x^{\mathrm{min}}$ & 1.38 & 1.84 & 0.61 & 0.61 \\
 $x^{\mathrm{max}}/10^{3}$ & 4.90 & 6.53 & 2.18 & 2.18 \\
 \hline
\end{tabular}
 \caption{\label{table_tref} 
Core growth time $\tCG$, and the value of the core growth parameter $s$ during the core growth phase of the simulation.
 Also given are 
  UETC reference times $\tRef$, the maximum time at which data is taken for the UETC $\tMax$, the ratio between the two $r^{\mathrm{max}}$, and the minimum and maximum values of $x = k\tRef$,  $x^{\mathrm{min}}$ and  $x^{\mathrm{max}}$,  for each of the sets of 4k simulations in the radiation and matter eras, without ($s=1$) and with ($s=0$) the string core growth approximation.
Times are given in units where  $\phi_0 = 1$. }
\end{table}

Despite the modest scaling range of the $s=1$ simulation, it is enough to characterize the region around the peak of the UETCs. This region contains the ETC obtained at the reference time and its surrounding area, where the maximum correlation within the network is set.  Although it does not supply all the information required for a CMB calculation, it gives the major contribution to the power spectra. In contrast, because $s=0$ simulations scale earlier and for a longer period of time, 
they probe higher time-ratios and larger length scales.

We will see that the $s\!=\!0$ and $s\!=\!1$ ETCs are very similar, when networks with the same string separation are compared. 
This similarity motivates a new merged structure for the UETCs,  
incorporating contributions from simulations with maximum fidelity and with maximum dynamic range: the $s=1$ measurements establish the central part of the UETCs, while the $s=0$ are used at large time ratios and large length scales (low $k\tau$).

\subsection{Scaling}

In the merging process special care must be taken 
concerning the role of the simulation time parameter $\tsim$. Simulations for different values of the core growth parameter follow different equations of motion (see Eqs.\ (\ref{AHMod1}) and (\ref{AHMod2}) which depend explicitly on $s$). In addition, each simulation starts from different initial conditions and  applies different amounts of core growth, depending on the expansion rate and the value of $s$ (see Table \ref{table_tref}). 
For these reasons, one cannot directly compare simulations with different $s$ at the same simulation time $\tsim$.

Hence it is better to define the `physical' time based on the state of the string network itself, and in particular to use a length scale in the network. Specifically, the comoving string separation $\xi$ has been identified as a useful quantity to determine compatible simulation stages. The string separation is defined in terms of the mean string length $L_\text{s}$ in a horizon volume $V$ as
\begin{equation}
\xi = \sqrt{\frac{V}{L_\text{s}}}\,.
\end{equation}

\begin{figure}[h!]
\centering
\includegraphics[width=0.5\textwidth]{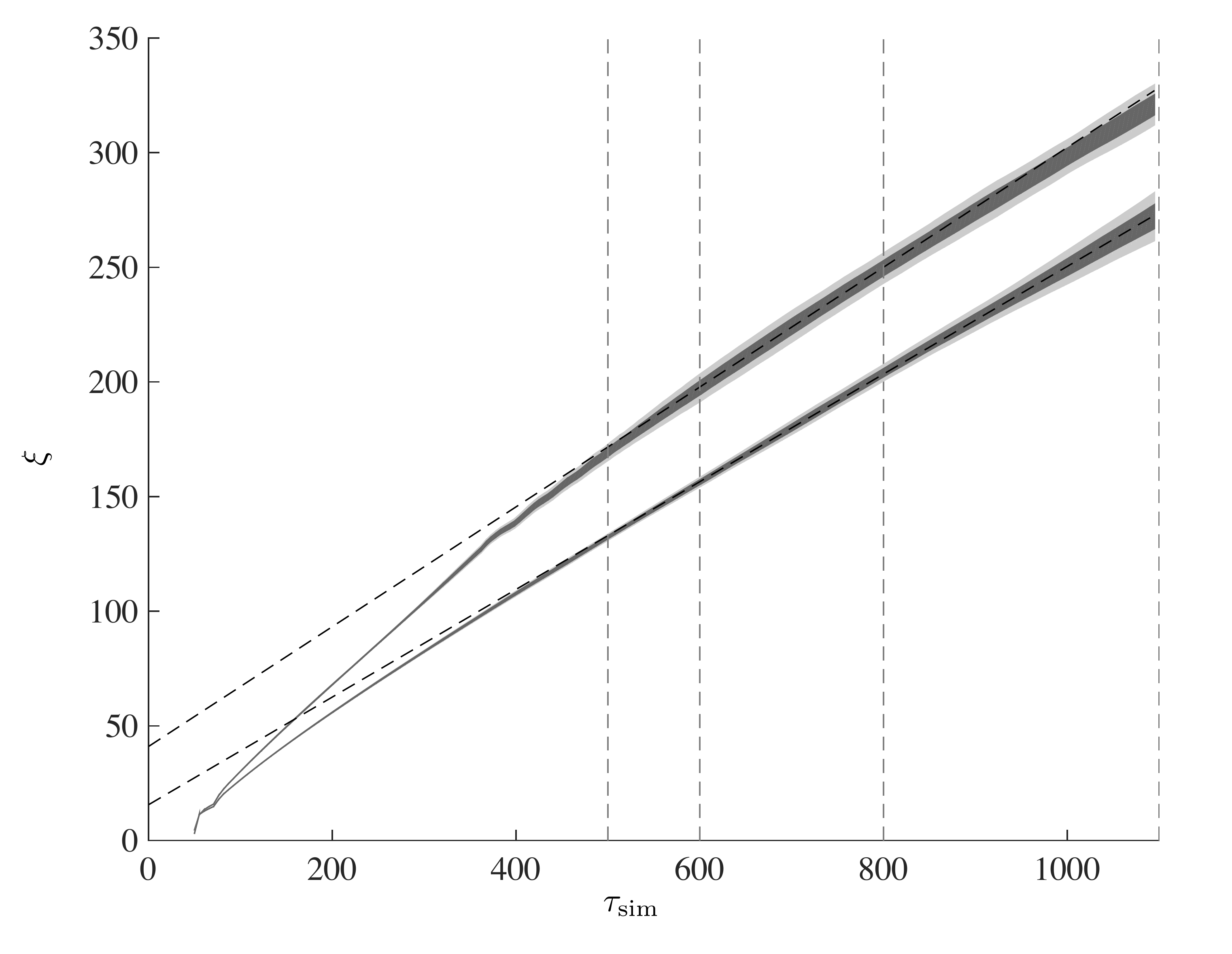}
\caption{Average string separation $\xi$ from $s=1$ (top line) and $s=0$ (bottom line) simulations in the matter era, with $\xi$ obtained from the Lagrangian length measure (\ref{e:LagLen}). Shaded regions correspond to the $1\sigma$ and $2\sigma$ deviations from the mean value obtained by averaging over all seven realizations. We also included the linear fit of the function (dashed black line) where the time intervals fitted have been $\tsim \in [600,\ 800]$ ($s=1$)  and $\tsim \in [500,\ 1100]$ ($s=0$).}
\label{xi_s0ands1}
\end{figure}

The mean string length is usually derived by directly measuring the comoving length of each string (see details in \cite{Kajantie:1998bg,Scherrer:1997sq,Bevis:2006mj,Bevis:2010gj,TestPaper}). One way of obtaining the length of string is by summing the number of plaquettes pierced by strings. 
Such plaquettes are identified from the winding of a gauge-invariant phase around them \cite{Kajantie:1998bg}. 
We correct the ``Manhattan'' length so obtained by a factor $\pi/6$ \cite{Scherrer:1997sq}.
An alternative way, and the one we use in this work,  is to use local field theory estimators to get the above ratio. In our case we employ the mean Lagrangian density $\bar{\mathcal{L}}$, with 
\ben
{L_\text{s}} = -\bar{\mathcal{L}}V/\mu.
\label{e:LagLen}
\een 
We show the measured values of $\xi$ inferred from the mean Lagrangian density in the matter era in Fig.~\ref{xi_s0ands1}.

As we found in previous works, the asymptotic behaviour of the string separation is  very close to linear, 
\begin{equation}
\xi \to \beta (\tsim - \tau_{\rm offset})\,,
\label{XiEq}
\end{equation}
where $\tau_{\rm offset}$ is the  time  offset of the $\xi$ curve (see Fig.\ \ref{xi_s0ands1}). The time offset as well as the slope of $\xi$ in the linear regime are different for each realization due to the random initial conditions. We define the mean slope $\beta$ as the average of all different slopes from different realizations. 
Numerical values of the slopes can be found in Table~\ref{table_beta}.
 
\begin{table}[h!]
\renewcommand{\arraystretch}{1.2}
\scalebox{0.9}{
\begin{tabular}{|c||c|c||c|c|}
\hline
 & \multicolumn{2}{c||}{$s=1$} & \multicolumn{2}{c|}{$s=0$}  \\\hline
& Radiation & Matter & Radiation & Matter \\ \hline
 $\beta_{\rm W}$ & $0.265\pm0.005$ & $0.277\pm0.009$ & $0.244\pm0.005$ & $0.247\pm0.008$ \\
 $\beta_{\mathcal{L}}$ & $0.254\pm0.005$ & $0.261\pm0.008$ & $0.234\pm0.006$ & $0.235\pm0.008$ \\
 \hline
\end{tabular}}
 \caption{\label{table_beta} Values of the slope of the average string separation (see Eq.~\ref{XiEq}). In the $s=1$ case, the time intervals used to fit the function are  $\tsim \in[600\ 800]$  for matter  and  $\tsim \in[450\ 600]$  for radiation.  In the $s=0$ case, the time intervals are  $\tsim \in[500\ 1100]$, both for matter and radiation. The quantity $\beta_{\rm W}$ is obtained   by measuring the string length using the number of plaquettes pierced by strings (see text), whereas $\beta_{\mathcal{L}}$  is obtained by  using the Lagrangian density  (\ref{e:LagLen}). }
\end{table}

Simulations at same $\xi$ can be considered to be at the same stage of the evolution. 
Hence, in order to merge the UETCs from different runs,  
they should be converted to functions of $\xi$ and $\xi'$ rather than $\tsim$ and $\tsim'$, according to 
\begin{equation}
C_{ab}^{(\xi)}(k,\xi,\xi') = C_{ab}^\mathrm{(sim)}(k,\tsim,\tsim')\sqrt{\frac{\tsim\tsim'}{\xi\xi'}}.
\label{merge}
\end{equation}
Besides allowing comparison of the $s=0$ and $s=1$ networks, 
it is found that the variance between simulations of the ETCs $C_{ab}^{(\xi)}(k,\xi,\xi)$ plotted against $k\xi$ is thereby reduced.

\subsection{UETC merging}

\begin{table}[h!]
\renewcommand{\arraystretch}{1.2}
\begin{tabular}{|c||c|c||c|c|}
\hline
 Model & \multicolumn{2}{c||}{$s=1$} & \multicolumn{2}{c|}{$s=0$}  \\\hline
 Cosmology & Radiation & Matter & Radiation & Matter \\ \hline
 $\xi(\tRef)$ & 146.7 & 198.0 & 55.8 & 55.9 \\
 $\xi(\tMax)$ & 183.3 & 248.4 & 269.5 & 270.1 \\
 $r_{\xi}^{\mathrm{max}}$  & 1.26 & 1.26 & 4.83 & 4.83 \\
 $x_\xi^{\mathrm{min}}$ & 0.45 & 0.60 & 0.17 & 0.17 \\
 $x_\xi^{\mathrm{max}}/10^{3}$ & 1.60 & 2.16 & 0.61 & 0.61 \\
 \hline
\end{tabular}
 \caption{\label{table_xiref} Mean string separations $\xi$ at $\tRef$ and $\tMax$, the ratio between the two, and the minimum and maximum values of $x_\xi = k\xi(\tRef)$, $x^{\mathrm{min}}_{\xi}$ and  $x^{\mathrm{max}}_{\xi}$, for simulations in the radiation and matter eras, without ($s=1$) and with ($s=0$) the string core growth approximation.
Lengths are given in units where  $\phi_0 = 1$. }
\end{table}

\begin{figure}[h!]
\centering
\includegraphics[width=0.5\textwidth]{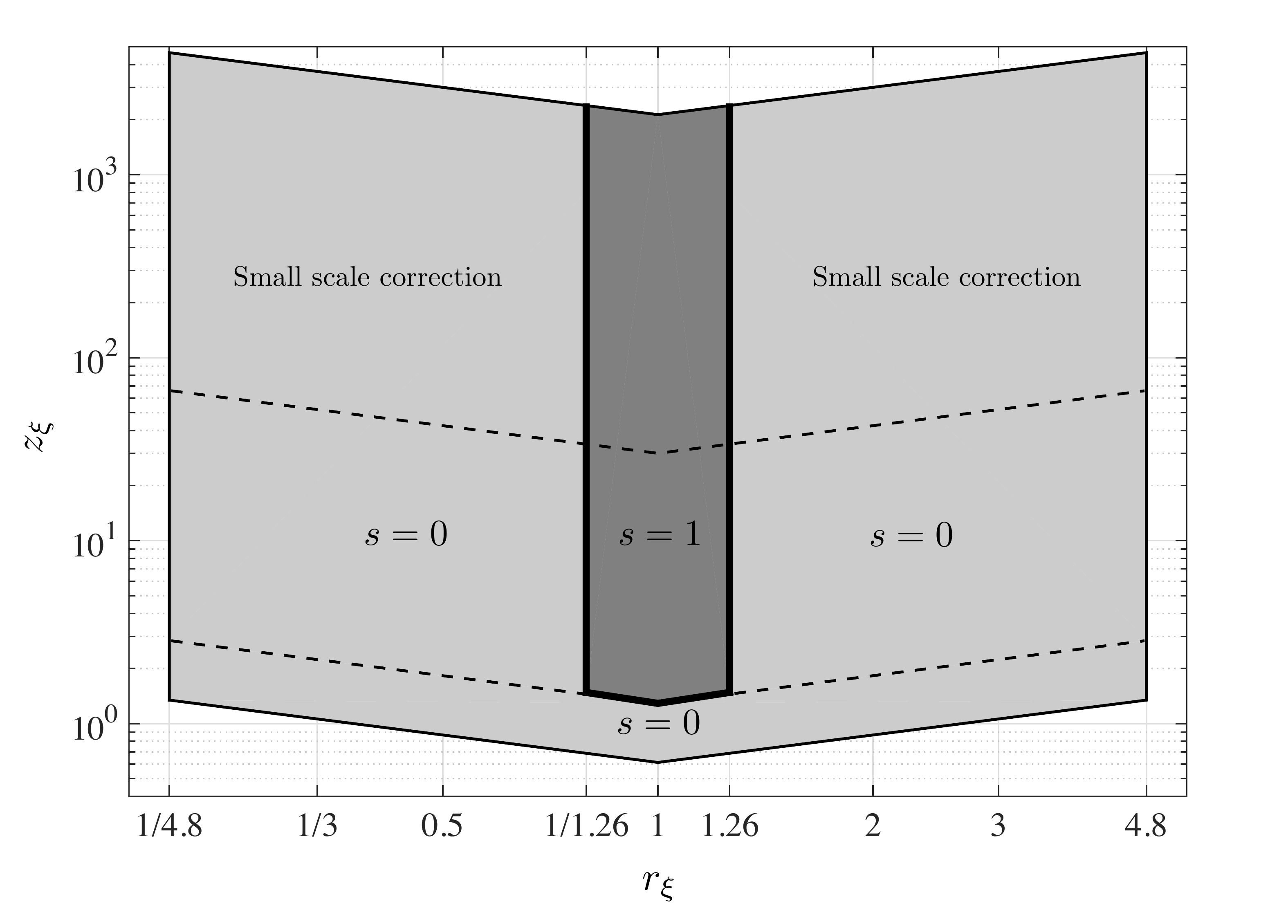}
\caption{Top view of the scheme of the merged UETC. ETCs and its surroundings are covered by $s=1$ simulations, while $s=0$ are used to extrapolate data both to large time ratios and very large scales. The variables are defined in Eq. (\ref{rxi}).}
\label{Frankie}
\end{figure}

\begin{figure}[h!]
\centering
\includegraphics[width=0.5\textwidth]{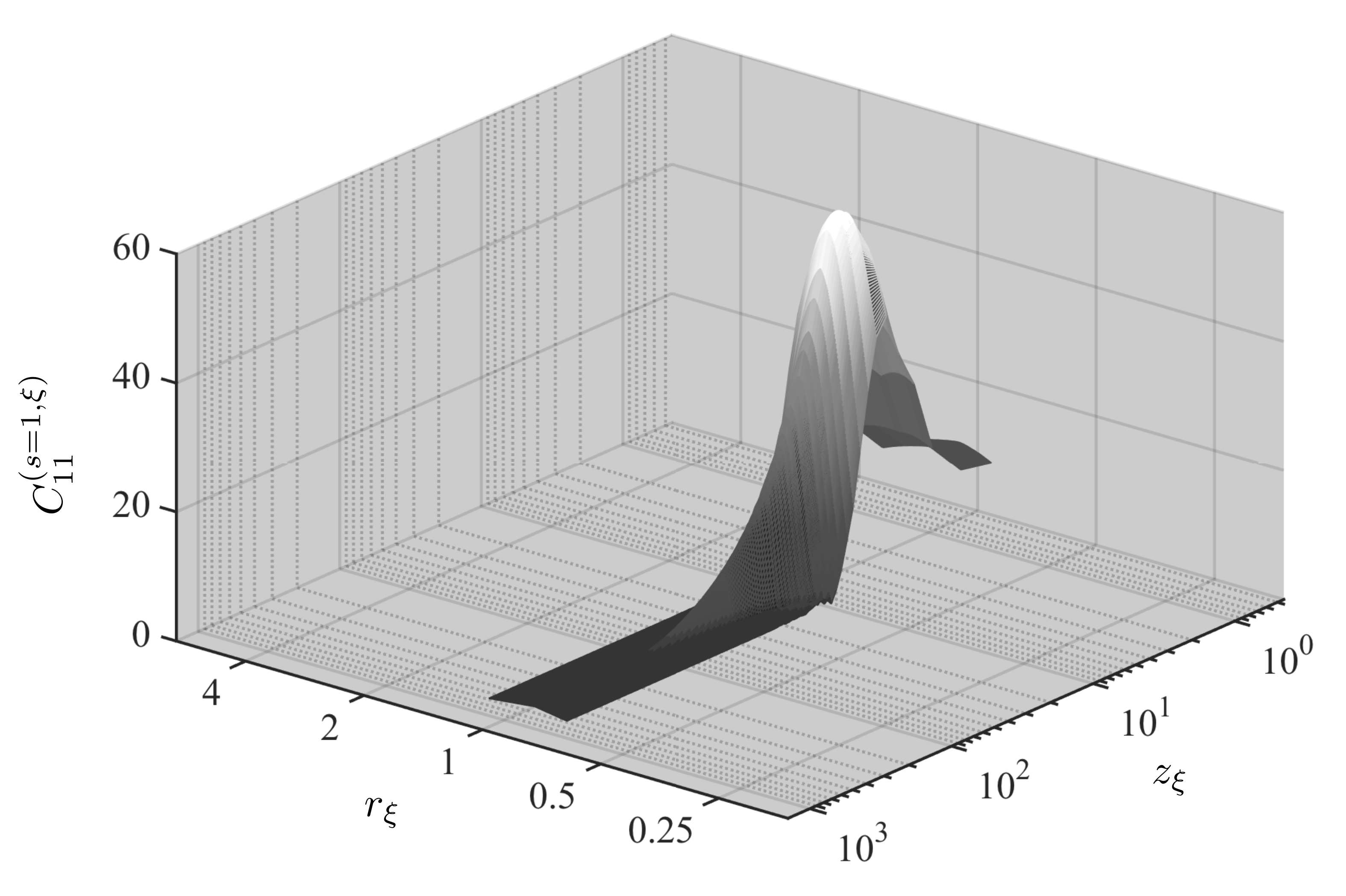}
\caption{The $C_{11}$  correlator in the matter era, simulated with $s=1$, in the $(z_\xi,r_\xi)$ region indicated in Fig.~\ref{Frankie} and Table~\ref{table_xiref}.}
\label{UETCs1}
\end{figure}

\begin{figure}[h!]
\includegraphics[width=0.5\textwidth]{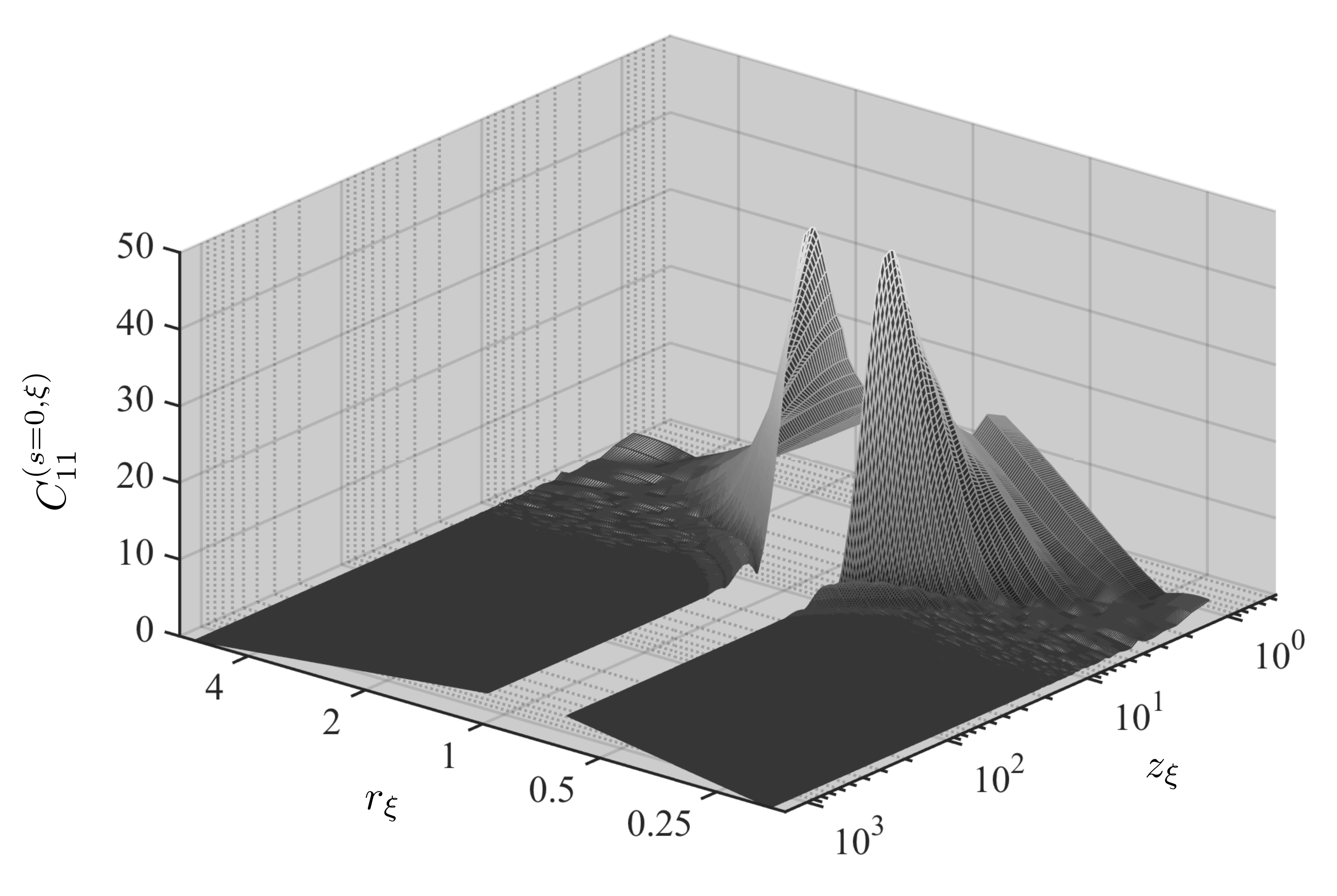}
\caption{The  $C_{11}$  correlator in the matter era, simulated with $s=0$, in the $(z_\xi,r_\xi)$ region indicated in Fig.~\ref{Frankie} and Table~\ref{table_xiref}.}
\label{UETCs0}
\end{figure}

The schematic representation of which UETC contributes to the merged UETC  can be found in Fig.~\ref{Frankie}, 
displayed in the variables 
\begin{equation}
z_\xi = k\sqrt{\xi\xi'} \quad {\rm{and}}\quad  r_{\xi}=\xi'/\xi\,.\label{rxi}
\end{equation}
 Fig.~\ref{UETCs1} and \ref{UETCs0} show how each simulation contributes to the merged UETC: the  $s=1$ simulations provide the central part of the correlation and the $s=0$ simulations  extend it to higher time-ratios. The numerical values of the limits of the different regions in the merged UETC  can be found in Table~\ref{table_xiref}.

\begin{figure}[h!]
\centering
\includegraphics[width=0.5\textwidth]{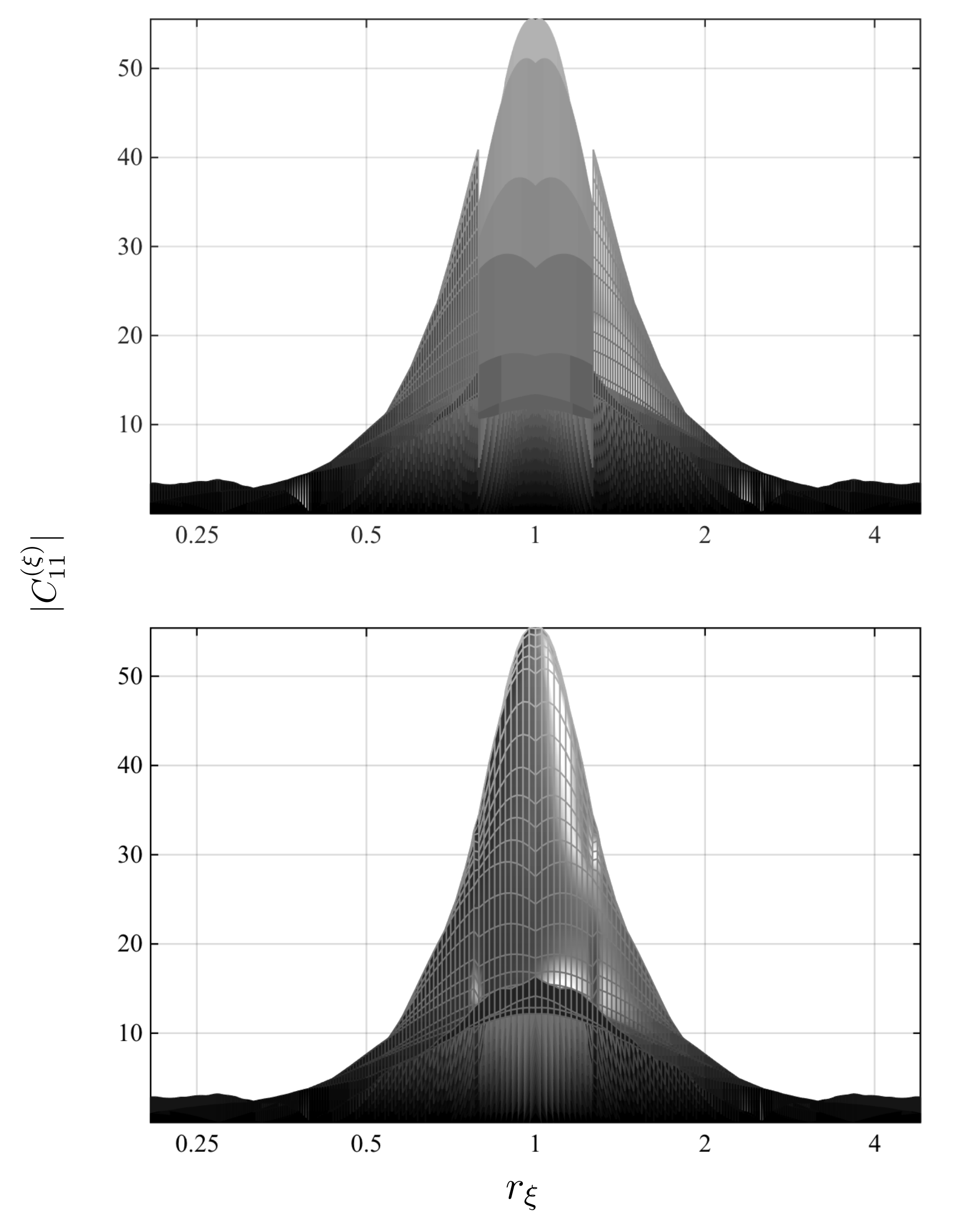}
\caption{Merged scalar $C_{11}^{(\xi)}$ matter correlator. The upper pane shows the $C_{11}$ before {ETC normalization}, whereas the lower pane represents the final merged and normalized case.\label{ETCNormMergedPro}}
\end{figure}

The resulting UETC is shown 
in the upper pane of Fig.~\ref{ETCNormMergedPro}, viewed along the $z_\xi$ axis. 
It can be seen that the $s=0$ and $s=1$ UETCs differ by less than 20\% at the junction at $r_{\xi}^{\rm lim}$, 
demonstrating that the $s=0$ simulations capture the near-equal time energy-momentum correlations rather well.

\begin{figure}[h!]
\centering
\includegraphics[width=0.5\textwidth]{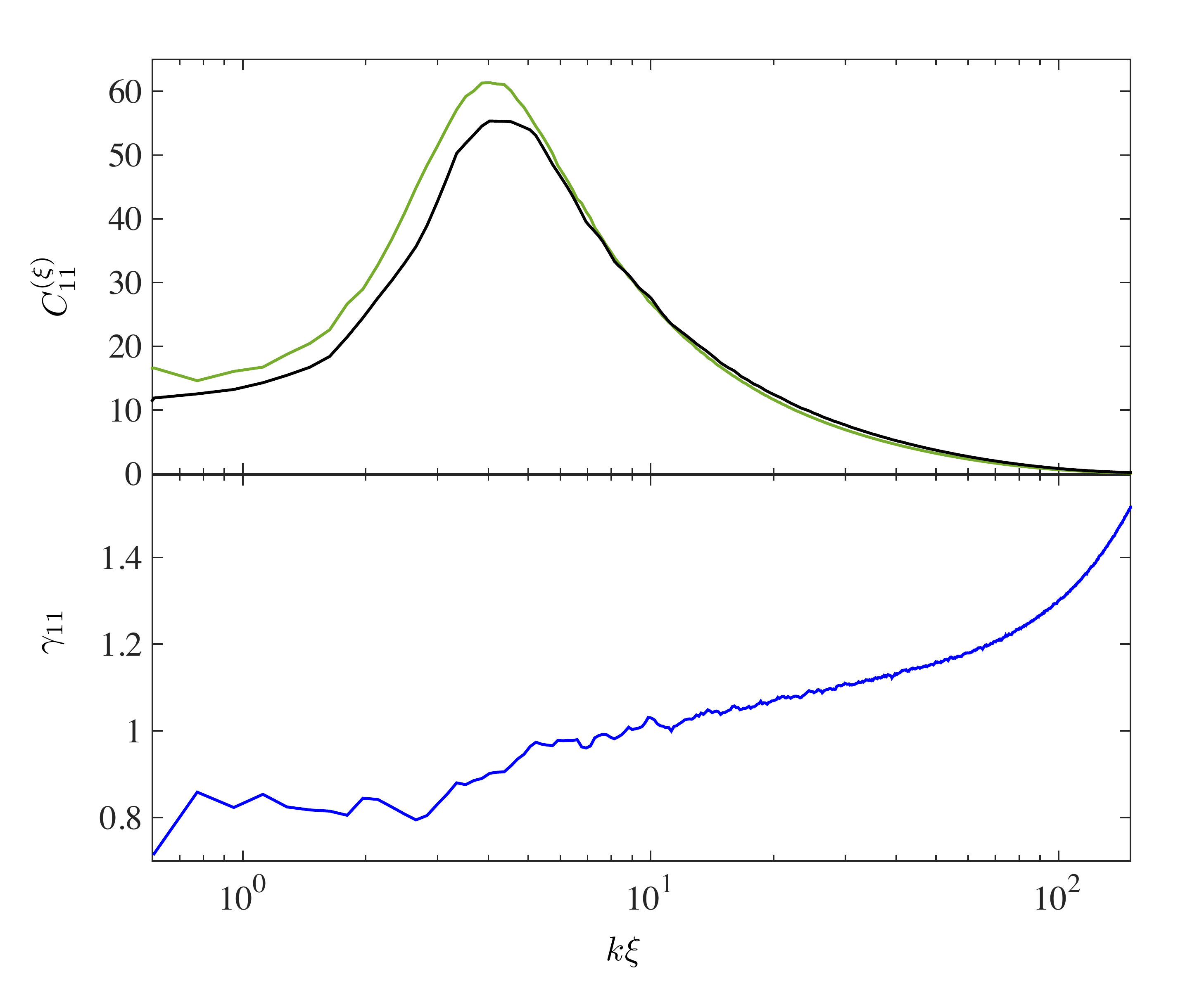}
\caption{Upper pane: ETCs corresponding to the $C_{11}^{(\xi)}$ for $s=0$ (green line) and $s=1$ (black line), in the matter era. The lower pane represents the correction factor  $\gamma_{11}(k\xi)$
 between them.}
\label{ETCNorm1}
\end{figure}

The equal-time correlations are also close.  A comparison of the ETCs and the relative factor between them, 
\begin{equation}
\gamma_{ab}(z_\xi) = \frac{C_{ab}^{ (s=1,\xi)}(z_\xi,1)}{C_{ab}^{ (s=0,\xi)}(z_\xi,1)}\,,
\label{gammaETC}
\end{equation}
can be found in Fig.~\ref{ETCNorm1}.
One can see that the $s=0$ ETC is approximately 20\% higher near the peak, although it dips below the $s=1$ ETC at higher $k\xi$ where the correlators are small.

The normalization factor $\gamma_{ab}(z_\xi)$ is applied to the UETC of the $s=0$ case, 
we call this procedure {ETC normalization}.
Therefore, the final representation of the merged UETCs for $z_\xi > 1.29\exp(|\ln(r_\xi)|)$ reads as:
\begin{multline}
C_{ab}^{\rm (tot,\xi)}(z_\xi,r_{\xi}) = \\
C_{ab}^{\rm (s=1,\xi)}(z_\xi,r_{\xi}) \theta( \ln(r_{\xi}^{\rm lim}) - |\ln(r_{\xi})|) \\ 
+ \gamma_{ab}(z_\xi)C_{ab}^{\rm (s=0,\xi)}(z_\xi,r_{\xi}) \theta(|\ln(r_{\xi})| - \ln(r_{\xi}^{\rm lim}))\, .
\label{UETCmergingGamma}
\end{multline}
Here the values of $z_\xi$ are defined from the $s=0$ simulations, with the values of the $s=1$ UETCs obtained by interpolation. 
The normalised and merged $C_{11}$ is plotted in the lower pane of  Fig.~\ref{ETCNormMergedPro}.

It is remarkable that the normalization of the ETC produces a UETC which is close to continuous at the merging boundary $r_{\xi}^{\rm lim}$.  This means that the width of the UETCs, which depends on the speed with which the strings move, is very similar.

Finally, we use $s=0$ UETC data in the range  $ 0.17 < z_\xi \exp(-|\ln(r_\xi)|) < 1.29$, normalised by the average of $\ga_{ab}(z_\xi)$ in the first six bins, weighted by the number of $\bk$ values contributing to each bin, or\footnote{See Erratum at end of paper.}
 \begin{equation}
C_{ab}^{\rm (tot,\xi)}(z_\xi,r_{\xi}) =  \bar\gamma_{ab} C_{ab}^{\rm (s=0,\xi)}(z_\xi,r_{\xi}) \, .
\label{UETCmergingGammab}
\end{equation}
The values of $\bar\gamma_{ab}$ are given in Table~\ref{table_normlowkt}.

\begin{table}[h!]
\renewcommand{\arraystretch}{1.2}
\begin{tabular}{|c||c|c|c|c|c|}
\hline
 & $C_{11}$ & $C_{12}$ & $C_{22}$ & $C_{\mathrm{vv}}$ & $C_{\mathrm{tt}}$  \\\hline
Matter & 0.76 & 0.42 & 0.77 & 0.94 & 0.91 \\ \hline
Radiation & 0.71 & 0.60 & 0.73 & 1.23 & 0.89\\
 \hline
\end{tabular}
 \caption{\label{table_normlowkt} Normalization factor $\bar\ga_{ab}$ for low $z_{\xi}$ $s=0$ data, obtained from weighted average  of the first 6 bins of $\ga_{ab}(z_\xi)$.}
\end{table}

\subsection{UETC fitting and small-scale correction}

We extend the small scale correction performed in \cite{Bevis:2010gj} (see section III-D). There it was argued that equal-time correlators decay on small scales (deep inside the horizon, $k\tau \gg 1$, but above the scales at which the string width becomes relevant) approximately as $1/k\ta$  ($1/k\xi$  in terms of the string separation scale).

\begin{figure}
\resizebox{0.49\textwidth}{!}{\includegraphics{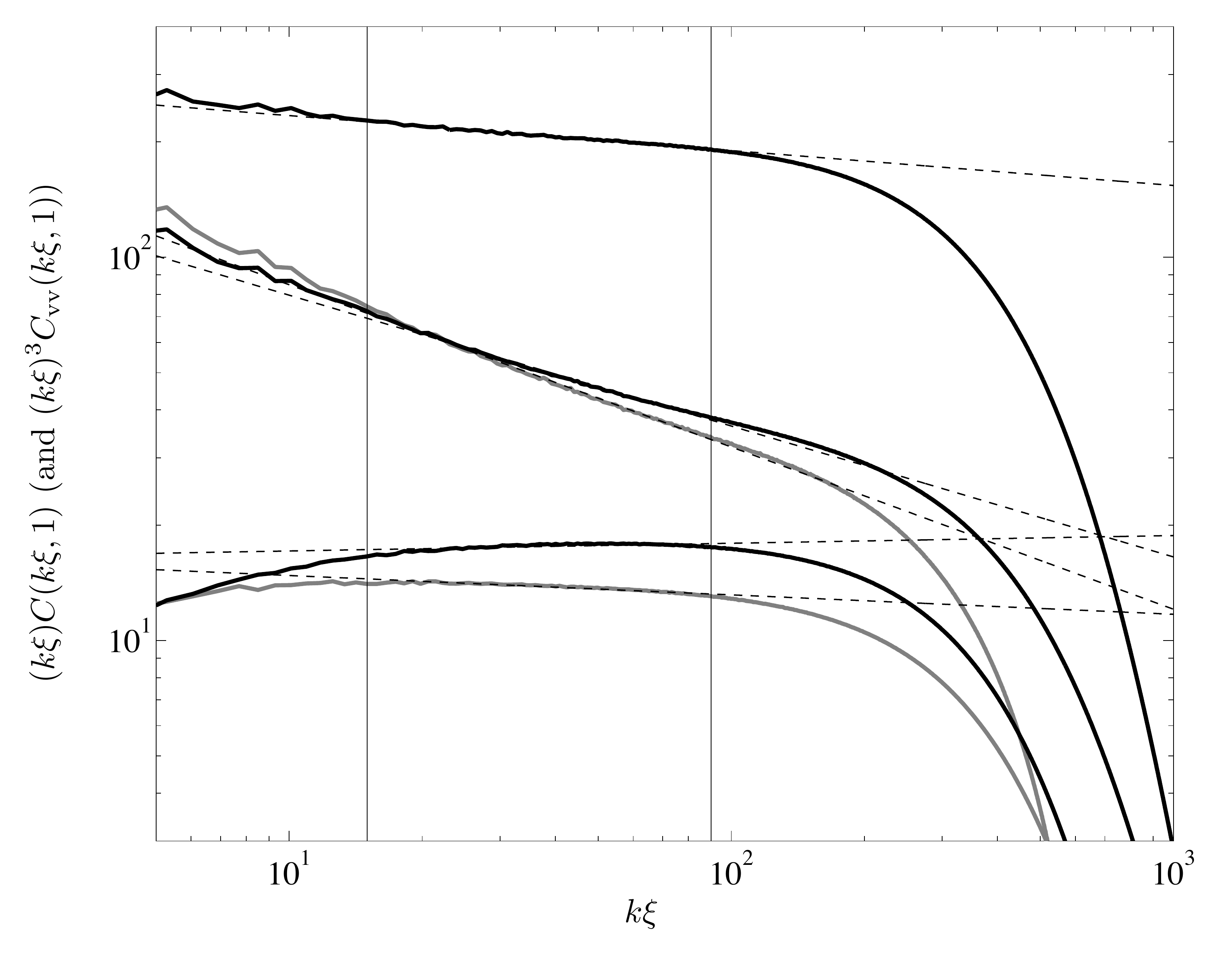}}
\resizebox{0.49\textwidth}{!}{\includegraphics{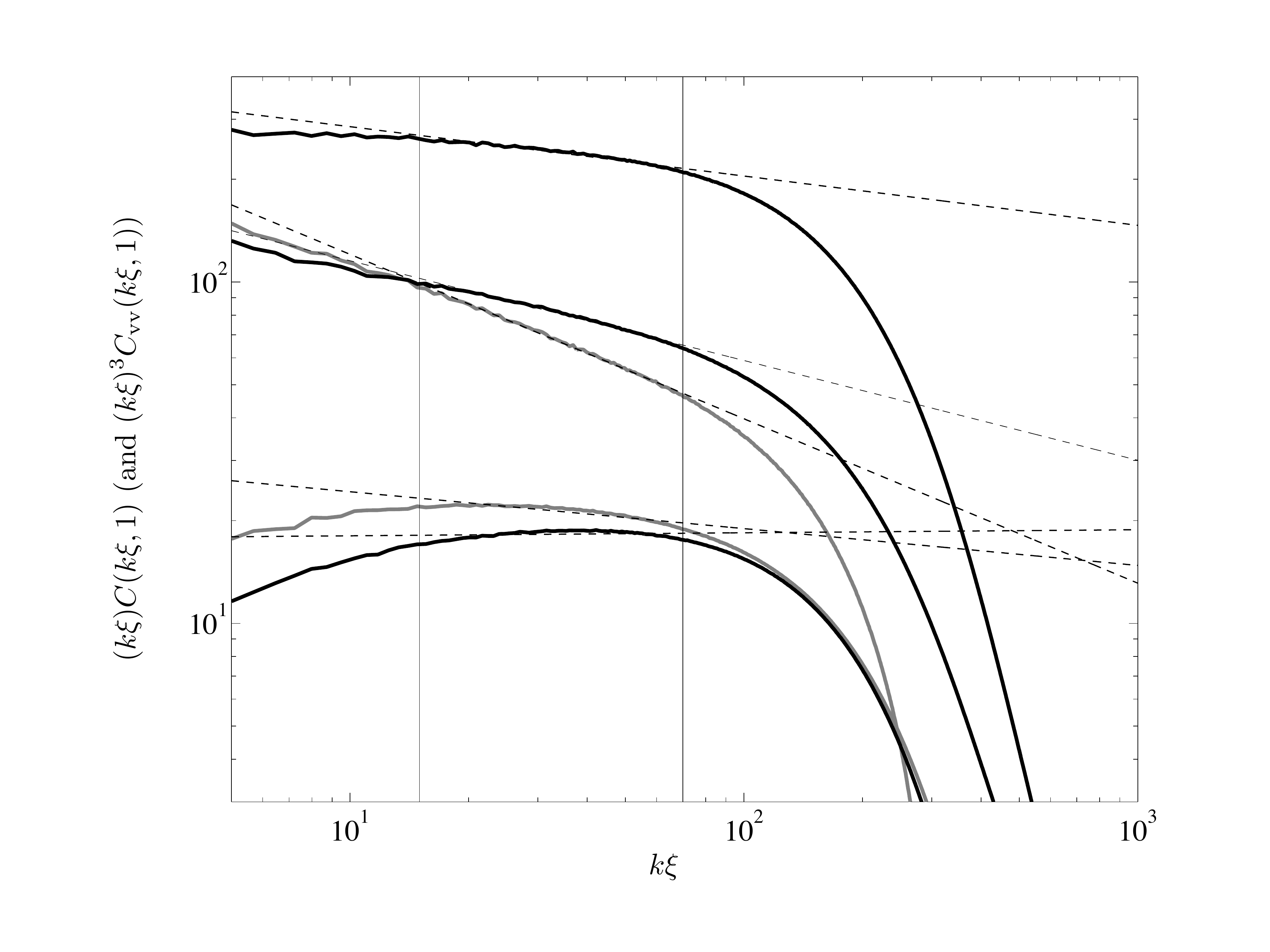}}\\
\caption{\label{ETC_PL}Power law fit for all five correlators for $s=0$ at the end of the simulation $\tau_{\mathrm{sim}}\approx1100$ (upper pane) and s=1 at the end of the scaling regime $\tau_{\mathrm{sim}}\approx800$ (lower pane). Both cases correspond to the matter era. Fitting ranges lie between the vertical lines: $k\xi\in[15\,, 90]$ for $s=0$ and $k\xi\in[15\,,70]$ for $s=1$. In both pictures the color election is the same, the uppermost line is the ETC of $C_{11}$, the middle pair of lines correspond to $C_{22}$ (black line) and $|C_{12}|$ (grey line) and finally the lower pair of lines are $C_{\mathrm{vv}}$ (grey line) and $C_{\mathrm{tt}}$ (black line).}
\end{figure}

In Fig.\ \ref{ETC_PL} we show power-law fits of $k\xi{C^{(\xi)}}$ over the range $k\xi \in [15\,,  90]$ ($s=0$) and $k\xi \in [15\,, 70]$ ($s=1$), denoted by vertical lines, giving the numerical values of the power law in Table~\ref{table_powerlaw}. We have been able to confirm that the power-law is a reasonable fit to our current ETCs, for both $s=0$ and $s=1$ simulations, 
 although the power law is less clear for the vector and tensor cases.

The power-law behaviour applies between the string separation scale $\xi$ and the string width $w$.  In our simulations, the ratio $\xi/w$ reaches a maximum of approximately 300. 
In the true Universe, the power-law behaviour would hold for much longer as the string width at late times is over 50 orders of magnitude smaller than the string separation. Thus using the extrapolation to very high $k\xi$ could be used to improve our estimates of the scaling functions at high values of $k \tau$.

We use the information of the decay trend to correct the behaviour of the UETCs at high values of the binned wave numbers $k_n$, covering scales between the string width and the lattice spacing.  We conservatively do not extrapolate the UETCs beyond the wave vectors contained in the simulations.  The UETCs are in any case very small at high $k\tau$.

To do so, we follow the procedure presented in \cite{Bevis:2010gj} and define the attenuation level:
\begin{equation}
R(k,\xi) = \frac{Q(k\xi)^{p}}{C^{(\xi)}(k\xi,1)}
\label{pQ}
\end{equation}
where $Q$ and $p$ represent constants of the power-law fit.

\begin{table}[h!]
\renewcommand{\arraystretch}{1.2}
\scalebox{0.8}{
\begin{tabular}{|c||c|c||c|c|}
\hline
 
\multirow{2}{*}{} & \multicolumn{2}{c||}{$s=1$} & \multicolumn{2}{c|}{$s=0$}  \\ \hline
 & Radiation & Matter & Radiation & Matter \\ \hline

\multirow{2}{*}{$C_{11}$}

&
$-0.22\pm0.01$  & $-0.14\pm0.01$ & $-0.14\pm0.01$& $-0.10\pm0.01$ \\
    & $2.78\pm0.01$  & $2.60\pm0.02$ & $2.60\pm0.02$& $2.47\pm0.01$ \\ \hline

\multirow{2}{*}{$C_{12}$} 

&
$-0.60\pm0.02$ & $-0.48\pm0.03$ &$-0.51\pm0.03$ & $-0.43\pm0.03$ \\
 & $2.89\pm0.03$ & $2.56\pm0.06$ & $2.66\pm0.06$ & $2.37\pm0.06$ \\ \hline
 
\multirow{2}{*}{$C_{22}$}

&
$-0.37\pm0.02$ & $-0.29\pm0.03$ & $-0.40\pm0.03$ & $-0.34\pm0.03$ \\
 & $2.63\pm0.03$ & $2.35\pm0.07$ & $2.51\pm0.07$ & $2.26\pm0.07$ \\ \hline
 
\multirow{2}{*}{$C_{\mathrm{vv}}$ }

&
$-0.134\pm0.005$  & $-0.11\pm0.01$ & $-0.059\pm0.006$ & $-0.059\pm0.003$ \\
 & $1.662\pm0.008$  & $1.49\pm0.03$ & $1.35\pm0.01$ & $1.23\pm0.02$ \\ \hline
 
\multirow{2}{*}{$C_{\mathrm{tt}}$ }

&
 $0.076\pm0.004$ & $0.009\pm0.003$ & $0.045\pm0.003$ & $0.021\pm0.003$ \\
 & $1.153\pm0.007$ & $1.25\pm0.02$ & $1.183\pm0.008$ & $1.21\pm0.01$ \\ \hline
\end{tabular}}
 \caption{\label{table_powerlaw}  
 Values of the parameters $p$ and $Q$ as defined in Eq.~(\ref{pQ}) for each correlator in matter and radiation, and for $s=0$ and $s=1$ at  $\tau=\tau_{\mathrm{end}}$. The top line gives the value of  $p$ and the second of $\log_{10}Q$. The fitting range for the values  are the same as in Fig.~\ref{ETC_PL}, except for the  $s=1$ radiation case, where the range is $k\xi\in[10\,, 60]$.}
\end{table}

This is a measure of how far  the equal time correlators are  from their corresponding power-law form. 
As mentioned above, the power-law form is not clear for the tensor and vector ETCs in either $s=0$ or $s=1$ simulations, and it is conceivable that there are contributions from the massive radiation which prevent it from ever appearing. Still larger simulations are required to establish the asymptotic form of the vector and tensor ETCs at high $k\ta$. The power laws are at least clear for the scalar correlators.

As the UETCs are quadratic functions relating two separate times, we apply the correction in the following manner:
\begin{equation}
C^{(\xi)}_{c}(k\sqrt{\xi\xi'},\xi'/\xi) = \sqrt{R(k,\xi)R(k,\xi')} \, C^{(\xi)} (k\sqrt{\xi\xi'},\xi'/\xi)\,,
\end{equation}
whenever $k\xi > 30$, for every $r_\xi$ (see the upper dashed line in Fig.~\ref{Frankie}).

\begin{figure*}
\resizebox{0.49\textwidth}{!}{\includegraphics{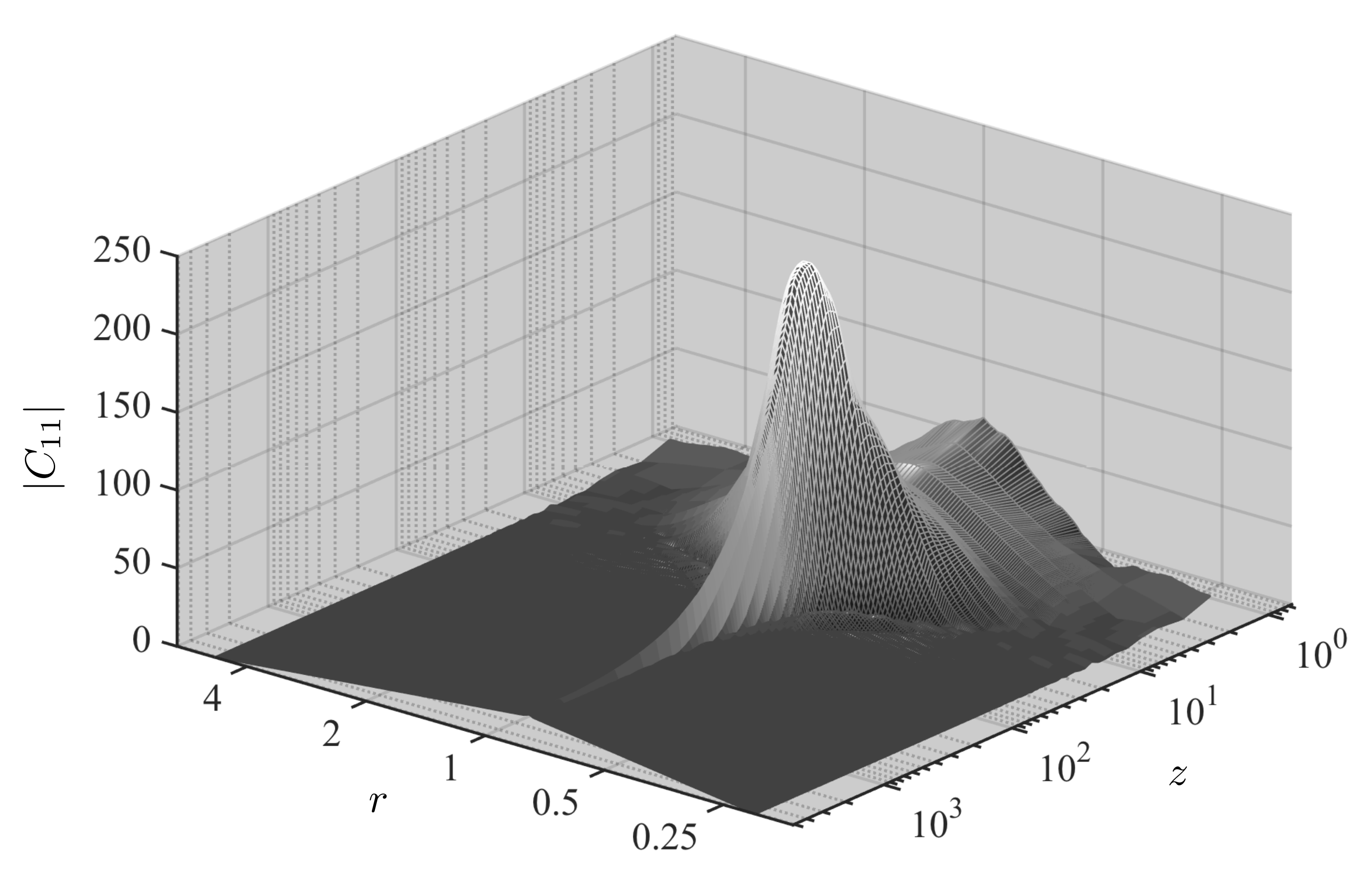}}
\resizebox{0.49\textwidth}{!}{\includegraphics{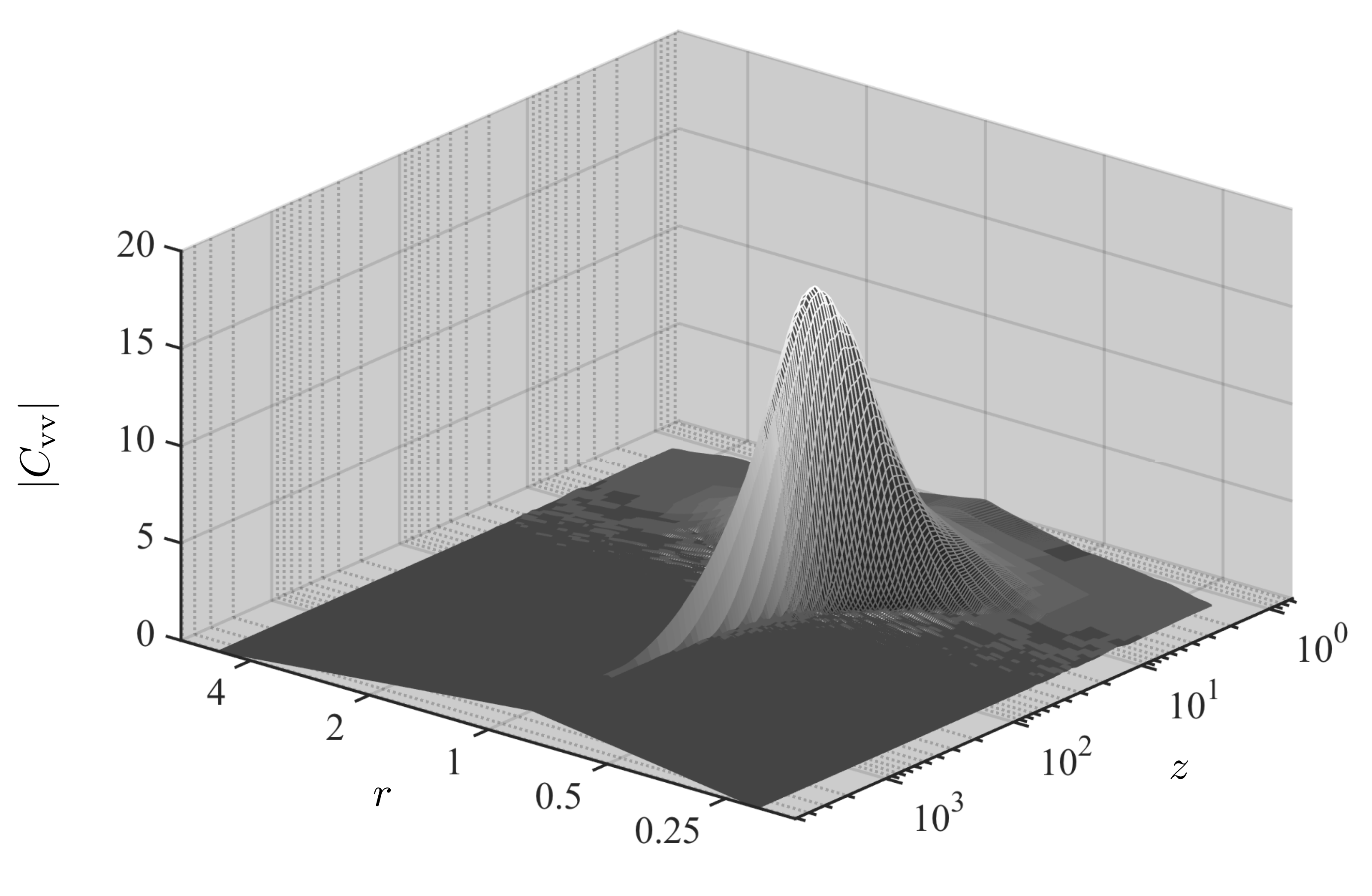}}\\
\resizebox{0.49\textwidth}{!}{\includegraphics{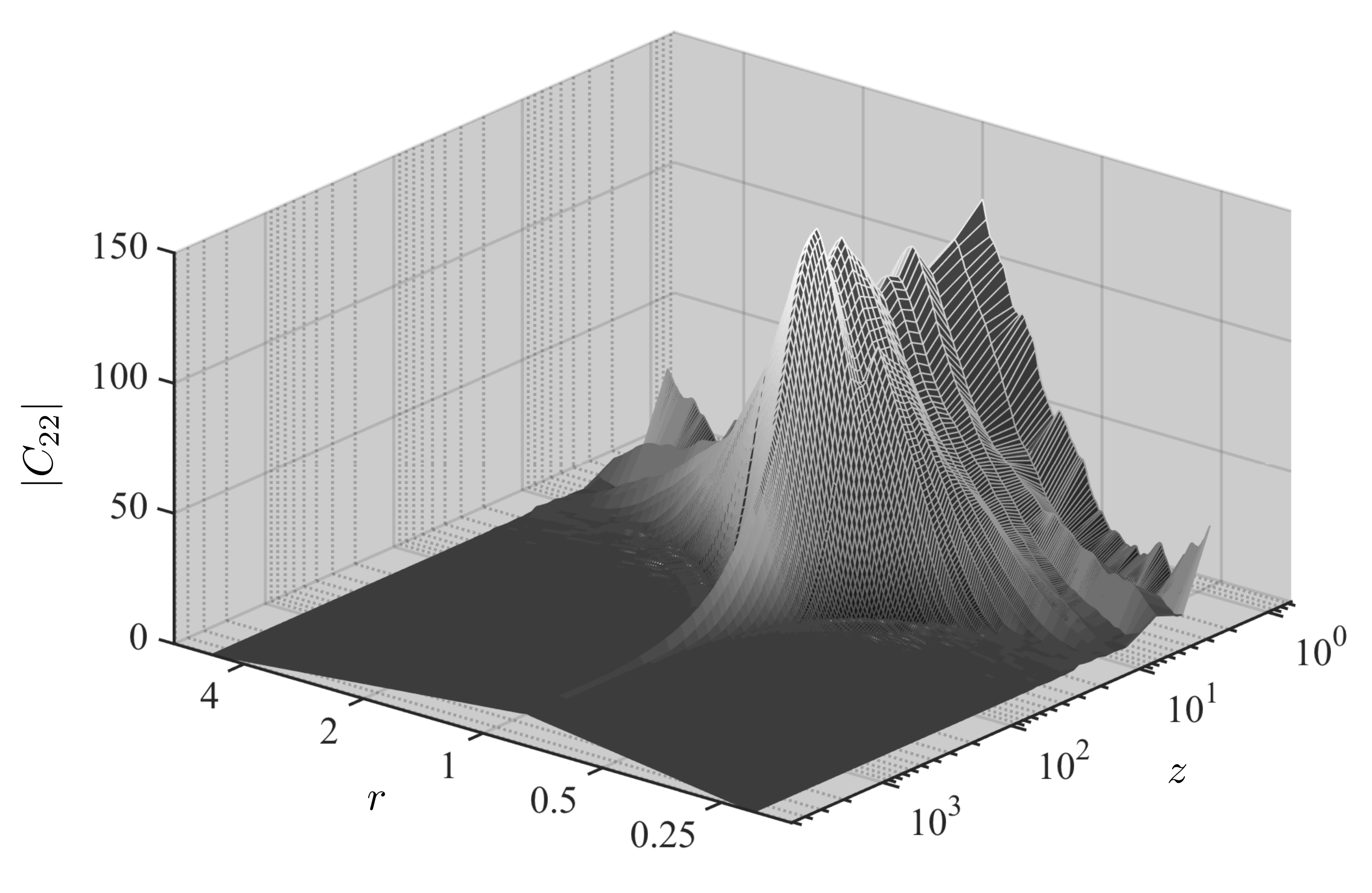}}
\resizebox{0.49\textwidth}{!}{\includegraphics{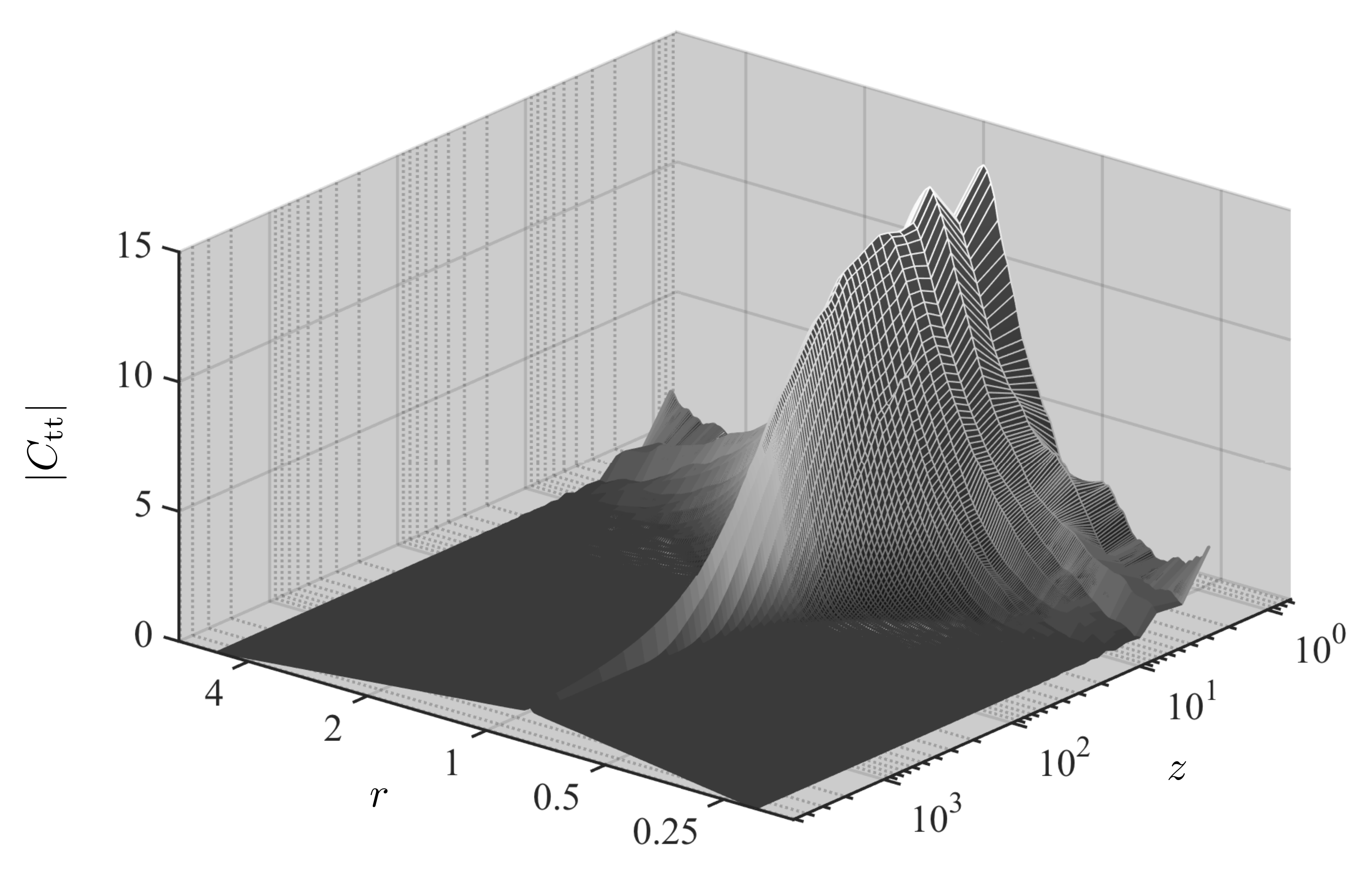}}\\
\begin{flushleft}
\resizebox{0.49\textwidth}{!}{\includegraphics{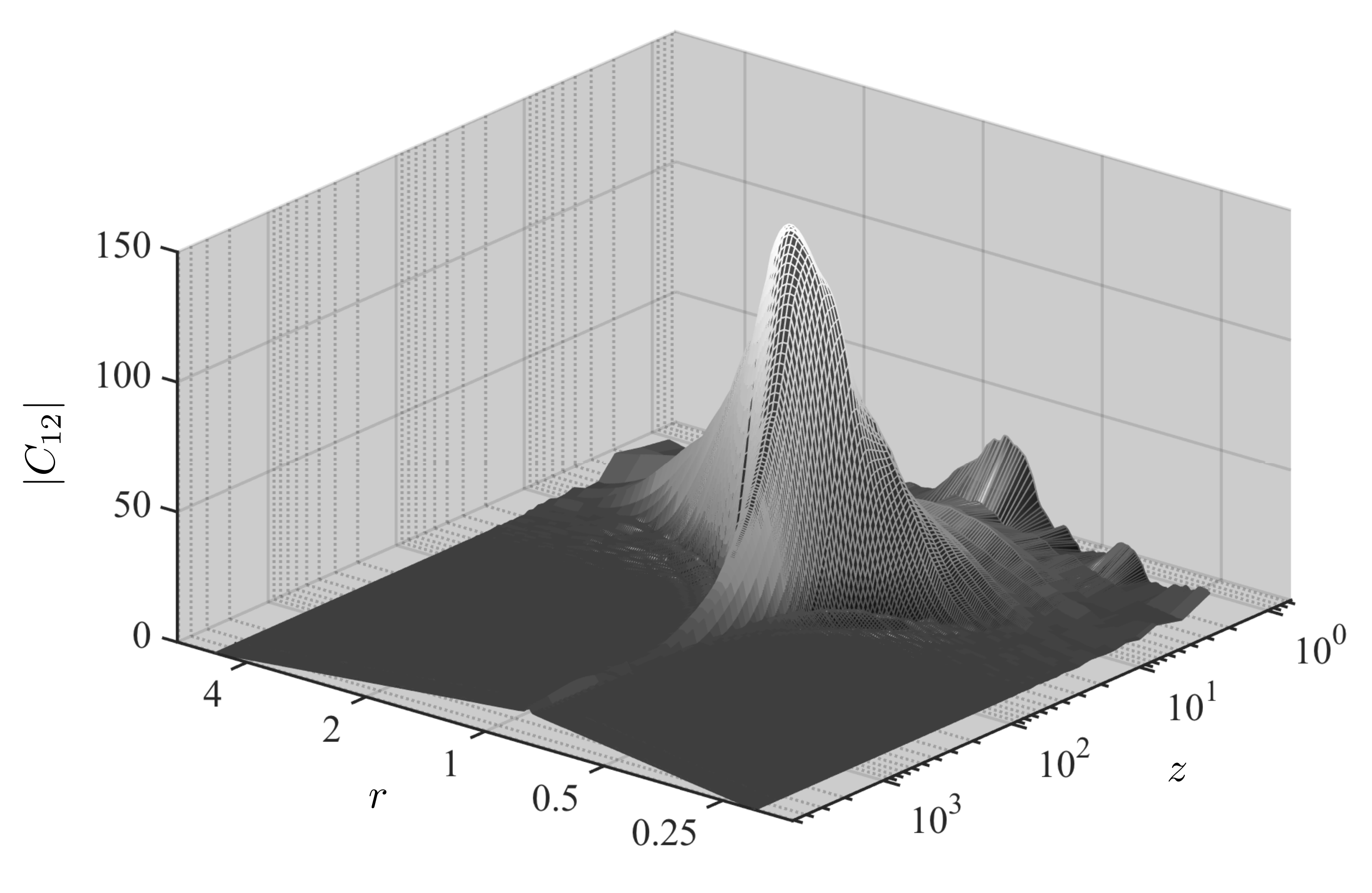}}
\end{flushleft}
\caption{\label{fig:UETCs} Full set of merged scaling UETCs the matter era, calculated from the average  of 7 $s=0$ and 7 $s=1$ runs.}
\end{figure*}

We show the set of final UETCs in the matter era in Fig~\ref{fig:UETCs}. 
Note that there is no extrapolation in $r_\xi$: the UETCs are set to zero for $|\ln(r_\xi)| > \ln(r_\xi^\text{max})$.

Finally, we transform back 
from $\xi$ to a scaling time variable $\tau$, by noting that $\xi$ and $\tau$ are proportional at large times. 
Hence we use Eq.\ (\ref{XiEq}) to write $\tau = \xi/\beta$. 
Our estimates of the scaling UETCs are therefore 
\begin{equation}
\barC_{ab}(k\sqrt{\tau\tau'},\tau/\tau') = \be^{-1}C^{(\xi)}_{ab} (k \sqrt{\xi\xi'}/\beta,\xi/\xi') \, ,
\label{eq:finalC}
\end{equation}
where $\beta$ is the mean slope obtained from the $s=1$ simulations, as given in the $\beta_{\mathcal{L}}$ row of Table~\ref{table_beta}. 
In Appendix \ref{app:fit} we provide an approximate fit to the global shape of all UETCs.


\section{Eigenvector decomposition at cosmological transitions}
\label{sec:Evec}

All the information needed to obtain the power spectra of CMB and matter perturbations is encoded in the UETCs  \cite{Turok:1996ud,Pen:1997ae,Durrer:1998rw}. 
In general, a UETC is a function of three variables 

\ben
C(k,\ta,\ta'),
\een
and the correlator is non-vanishing in the region 
\ben
\tInit \le (\tau',\tau) \le \tNow,
\een
where $\tNow$ is the current (conformal) time, and $\tInit$ is the time at which the defect-forming phase transition takes place.

The UETCs, which are real and symmetric, can be decomposed into their eigenfunctions $\eVec^n(k,\ta)$ defined through 
\ben
\int_{\tInit}^{\tNow} d\ta' C_{ab}(k,\ta,\ta') \eVec_b^n(k,\ta') = \eVal_n(k)\eVec_a^n(k,\ta).
\een
For vector and tensor modes this procedure is straight forward to implement. For the scalar case, however,  the correlation between the two potentials   is non-zero and needs to be taken into account. We do this by combining the $C_{11}$, $C_{12}$ and $C_{22}$ into a double-sized matrix as
\ben
\begin{pmatrix}
C_{11}&C_{12}\\C_{12}^T&C_{22}
\end{pmatrix}\,,
\een
and the first half of the resulting eigenvectors correspond to the $\phi$ perturbations, and the second half to the $\psi$ perturbations  (see {\it e.g.}  \cite{Bevis:2006mj} for details). 

Note that the eigenvalues $\la_n$, which are real and positive, are functions of the wavevector $k$.
As the domain is finite, there are a countable infinity of eigenvalues for each wavevector. 
The UETC is recovered through the sum
\ben
C_{ab}(k,\ta,\ta')  = \sum_n \lambda_n \eVec_a^n(k,\ta) \eVec_b^{n*}(k,\ta')\,.
\een
Formally, the power spectra and cross-correlations of a perturbation in a cosmological variable $X_a$ can be written \begin{equation}
\langle{X_a}(\textbf{k},\tau){X_b}^*(\textbf{k},\tau)\rangle = \frac{\phi_0^4}{V}\sum_n \la_n I_a^n(k,\tau) I_b^{n*}(k,\tau)\,,
\end{equation}
where the contribution of each linear term, $I_a^n(k,\ta)$, is
\begin{equation}
I_a^n(k,\ta) = \int_{{\tInit}}^{\ta} d\ta' \mathcal{G}^X_{ab}(k,\ta,\ta')\frac{\eVec_b^n(k,\ta')}{\sqrt{\ta'}}\,,
\label{e:EBInt}
\end{equation}
and  $\mathcal{G}^X$ is the Green's function for the quantity $X$. 
The integration is performed numerically, 
using a modified version of one of the standard Einstein-Boltzmann (EB) integrators 
CMBEASY \cite{Doran:2003sy}, CLASS \cite{Blas:2011rf,Lesgourgues:2011re}, or CAMB \cite{Lewis:2002ah}.
Hence, if UETCs are decomposed into their eigenfunctions, they can be used as sources for an EB solver, and the power spectra reconstructed by taking the sum of the power spectra obtained for each eigenfunction, weighted by the eigenvalue.
In practice, the square root of the eigenvalue (which should be positive) and the eigenfunction are combined together into an object we call the source function.

The EB time integration range is generally much larger than the range of any conceivable defect simulation.   However, the scaling property of the UETCs allows us to reconstruct the eigenvectors in an economical way.
Scaling means that
\ben
C_{ab}(k/\si,\si\ta,\si\ta') = C_{ab}(k,\ta,\ta'), 
\een
and therefore scaling UETCs can be written as a function of two variables, which when diagonalising are most conveniently chosen to be $x = k\ta$, $x' = k\ta'$,
\ben
C_{ab}(k,\ta,\ta') = \bC_{ab}(x,x'). 
\een
As before, the overbar represents the scaling form of the UETC in a FLRW background.

Strictly speaking, scaling is broken by $\tEq$ and $\tLam$, 
but given that UETCs decay quickly for $(x,x') \gg \xPeak$, where the peak of the UETC is $\xPeak = \text{O}(10)$, they will closely approximate radiation era scaling UETCs for $k \gg \xPeak\tEq^{-1}$, matter era scaling UETCs for $\xPeak\tLam^{-1} \gg k \gg \xPeak\tEq^{-1}$, and $\La$-era UETCs for $\xPeak\tLam^{-1} \gg k$.
The eigenfunctions of the scaling UETCs will approximate the true eigenfunctions under the same conditions. 
Scaling eigenfunctions are functions of $k\ta$, and so we infer that the eigenfunctions of the true UETCs will be well approximated by radiation era scaling eigenfunctions for $\ta \ll \tEq$, by matter era scaling eigenfunctions for $\tEq \ll \ta \ll \tLam$, and $\La$-era scaling eigenfunctions for  $\tLam \ll \ta$.

These observations underlie our discussion of methods to construct the true eigenfunctions from the scaling UETCs, which are derived from numerical simulations as discussed above.  We will discuss two existing methods, based on interpolating between sets of eigenfunctions in time, and introduce a third,  which interpolates between UETCs in $k$-space. The new method 
is superior: it reproduces better the actual (non-scaling) UETC during the radiation-matter transition, which we have measured for the first time, and also maintains the orthogonality of the approximate eigenfunctions.

\subsection{Simple eigenvector interpolation}

In the simple eigenvector interpolation method \cite{Durrer:1998rw,Bevis:2006mj,Bevis:2010gj}, 
scaling UETCs are extracted from radiation and matter cosmologies separately. Each correlator is diagonalised to obtain two sets of eigenfunction. 

The diagonalisation proceeds by sampling the numerically measured UETCs 
$\barC_{ab}(x,x')$ at a number of values of $x$ in the range available from the numerical simulations. 
For the results from our 4k simulations we took $N_i = 2048$ linearly spaced values in the interval $0.6 \leq (x,x') \leq 2300$ .

The two sets of eigenvectors, one from the radiation era and one from the matter era, are ordered by the magnitude of their eigenvalues, so that the first ones correspond to the most important contributions. We assume that the eigenvectors ordered by eigenvalue size form matching pairs, and choose the relative sign by requiring that the scalar product of the two eigenvectors is positive. Through this pairing we then define the   source function for the EB integrator as
\ben
\sqrt{\la_n}c_n(k,\tau) = e(\tau) \sqrt{\la_n^\text{R}} c^\text{R}_n(x) + 
(1 - e(\tau)) \sqrt{\la_n^\text{M}} c^\text{M}_n(x).
\label{neil_interp}
\een
The eigenvector interpolation function $e(\tau)$ is  taken to be \cite{Bevis:2006mj,Bevis:2010gj},  
\begin{equation}
e(\tau)=\frac{1}{1+\chi[a(\tau)]}\,,
\label{e_Neil}
\end{equation}
where $\chi[a] = a \Omega_{\mathrm{m}}/\Omega_{\mathrm{r}}$ is the ratio between the density fractions of radiation and matter at the given value of the scale factor.

For a given $k$, the source function is defined only at a set of times which are in general not those used by the EB integrator.  The values of $\sqrt{\la_n}c_n(k,\tau)$ at an arbitrary time $\tau$ are found by spline interpolation, with all eigenfunctions set to zero at $\ta = 0$ and for $x > 2000$.

The transition from matter domination to $\Lambda$ domination 
can be treated equivalently, in a manner which we discuss later.

Recently some inconsistencies of this approach have been highlighted \cite{Fenu:2013tea}.
First, the signs of a set of eigenvectors are undetermined, and so a rule must be applied to decide on the relative sign when interpolating between a radiation-era eigenvector and a matter-era one.

As described above, the $n$th radiation and the $n$th matter eigenvectors are matched, and their relative sign 
chosen so  that their scalar product is positive. Abrupt jumps in the shape of the functions are reduced, although the qualitative similarity between the two matched eigenvectors does not hold in all cases. 

However, if one goes beyond the $n$th to $n$th eigenvector sign matching and explores the whole eigenvector scalar product space, usually 
there are  cases where the $n$th eigenvector in radiation has the biggest scalar product ($\approx 1$) with the $m$th eigenvector in matter ($n \neq m$). Even worse, for the higher eigenvectors there is often no clear partner and the matching scheme breaks down.

Even if eigenvectors can be paired off successfully, the set of interpolated source functions are not in general orthonormal, and therefore not eigenvectors.  Finally, the numerically determined eigenvalues can be negative, which means that an arbitrary procedure must be developed to deal with the square root of the eigenvalues in the definition of the source function.

\subsection{Multi-stage eigenvector interpolation}

A second method of generating a set of source functions
\cite{Fenu:2013tea} (see also \cite{Pen:1997ae}), which we call 
multi-stage eigenvector interpolation,  improves on simple eigenvector interpolation by generating 
a set of linear combinations of the pure radiation and pure matter UETCs, 
whose eigenvectors can be more easily matched.
We write the ``transition" UETCs as 
\begin{equation}
{C_i}^\text{RM}(k\tau,k\tau') = f_i {\bC}^{\rm R}(k\tau,k\tau') + (1-f_i) {\bC}^{\rm M}(k\tau,k\tau')\,,
\label{TransitionUETC}
\end{equation}
with $0 \le i \le N_\text{U}$,  $f_0 = 1$, $f_{i+1} < f_i$, and $f_{N_\text{U}} = 0$.
For every transition UETC ${C_i}^\text{RM}$ we will have a set of orthonormal eigenvectors. 
We can have as many {transition UETCs} as we want: in practice we choose $N_\text{U}$ 
so that there is no arbitrariness in the eigenvector matching left: the scalar products between the $i$th and the $(i+1)$th sets of eigenvectors are close to one or close to zero. 
Each set of eigenvectors $c_i^n(k\tau)$ can then be uniquely mapped to its neighbours $i-1$ and $i+1$, with $i=0$ being the pure radiation eigenvectors and $i = N_\text{U}$ the pure matter eigenvectors.

We then divide up the radiation-matter transition era into $N_\text{U}+1$ intervals with a set of $N_\text{U}$ times $\tau_i$, and define a monotonically decreasing interpolating function $f(\tau)$, which will define the linear combination in (\ref{TransitionUETC}) according to 
\ben
f_i = f(\tau_i).
\een
We discuss the interpolating function in Section \ref{ss:IntFun}.

The transition basis functions $c^n(k,\tau)$ are then defined from the set of eigenvectors $c_i^n(k\tau)$ with the help of a set of indicator functions
$J_i(\tau)$ 
\bea
J_0(\tau) & = & \left\{  \ba{cc} 1 &  0 \le \tau \le \tau_{1} \\ 0 & \textrm{otherwise} \ea \right. , \nonumber\\
J_i(\tau) & = & \left\{  \ba{cc} 1 &  \tau_{i} \le \tau \le \tau_{i+1} \\ 0 & \textrm{otherwise} \ea \right. ,  \\  
J_{N_\text{U}}(\tau) & = & \left\{  \ba{cc} 1 &  \tau_{N_\text{U}} \le \tau \le \infty \\ 0 & \textrm{otherwise} \ea \right. .\nonumber
\eea 
The source functions are then
\ben
\sqrt{\la_n}\eVec^n(k,\tau) = \sum_{i=0}^{N_\text{U}} J_i(\tau) \sqrt{\la_{n,i}} \eVec_i^n(x),
\label{f:eVecOld}
\een
We see that the simple eigenvector interpolation is related to multi-stage interpolation with $N_\text{U} = 1$, and would be identical with a step function $e(\tau)$.

This process can also be generalized in the obvious way to take into account the transition from matter domination to $\Lambda$ domination.

\begin{figure}
\centering
\includegraphics[width=0.5\textwidth]{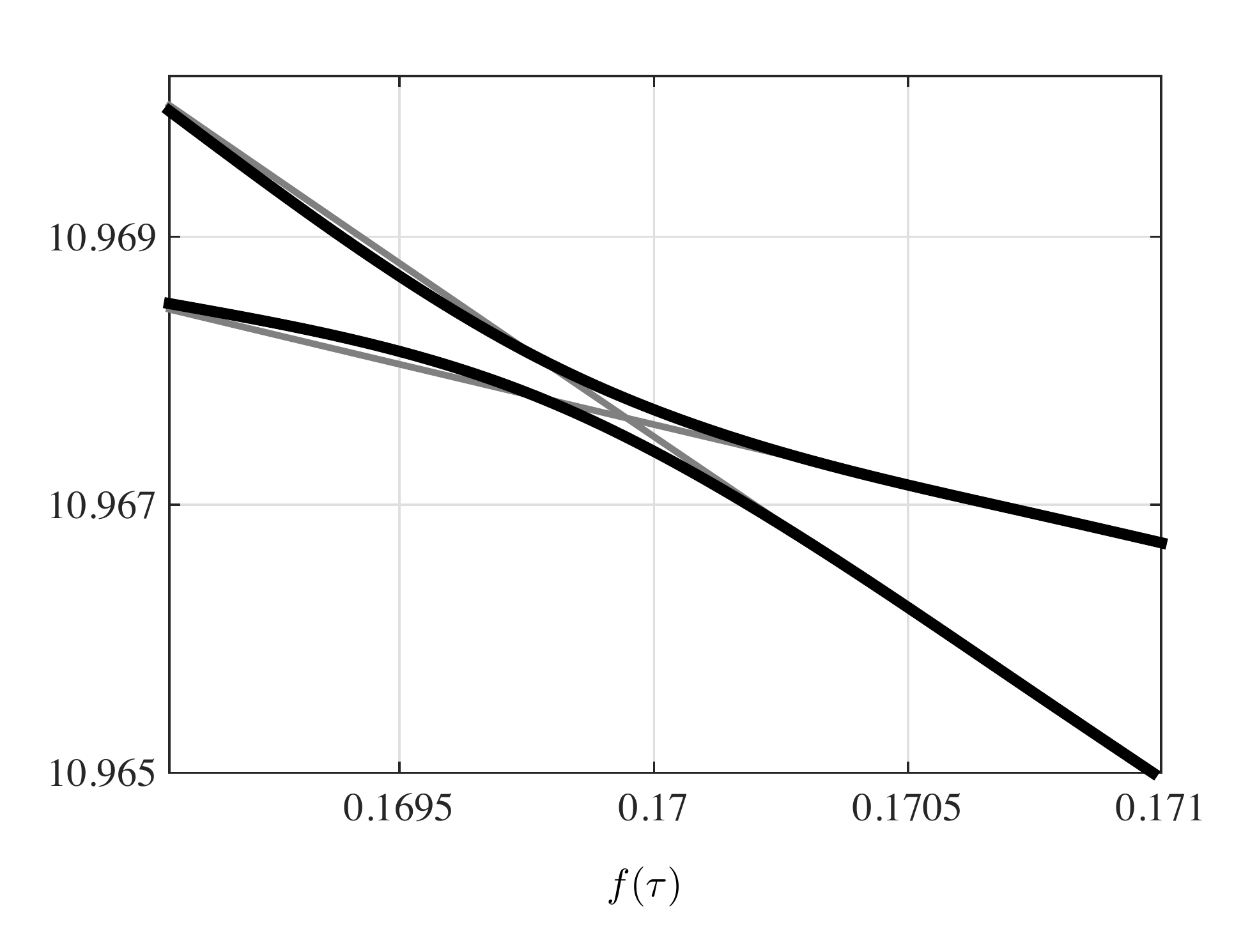}
\caption{
Evolution of the 34$^{\rm th}$ and 35$^{\rm th}$  eigenvalues  of the tensor correlator, as a function of the interpolation parameter $f$ (Eq.\ \ref{TransitionUETC}).
The thin grey lines show eigenvalue crossings, which happen when the time region is divided into few intervals (in this case, 3 intervals). The thick black line, on the other hand, represents the situation when the time region is divided into 18 intervals. It is apparent that  the eigenvalues avoid crossing each other, as they should.
}
\label{AvoidedCrossing}
\end{figure} 

Before discussing the choice of the function $f(\tau)$, we study how the eigenvectors evolve with $f$, the parameter determining the linear combination of the UETCs according to Eq.~(\ref{TransitionUETC}). In particular we can check that a radiation eigenvector evolves into a unique matter eigenvector. We can also plot the value of the corresponding eigenvalue. As each eigenvector is uniquely associated to its eigenvalue, and as the eigenvalues evolve also in a continuous way with $f$, the order of matching eigenvectors can only change if the associated eigenvalues cross along their evolution. However, 
the eigenvalues of a Hermitian matrix which is a continuous function of a parameter $f$ do not in general cross, unless a symmetry appears at a particular value of $f$. 
Hence we can expect that the eigenvectors can be uniquely ordered and matched by their eigenvalues.

In order to illustrate this point, we show explicitly an example in Fig.~\ref{AvoidedCrossing},  where we consider the 34$^{\rm th}$ and 35$^{\rm th}$ eigenvalues of the tensor correlator. Judging only by the scalar product method of the corresponding eigenvectors, the eigenvector corresponding to  the larger eigenvalue on the left (34$^{\rm th}$) has the largest dot-product with the eigenvector of the lowest eigenvalues on the right (35$^{\rm th}$). Following the eigenvalue evolution, they appear to cross.  If  we split this time region into three intervals, and perform a UETC interpolation, we find exactly the same situation  (thin grey lines). However, if we split the time region into more intervals (in our example, 18) to perform the UETC interpolation, 
the crossing of the eigenvalues is avoided. This is apparent in the eigenvalue evolution: the lines repel (thick black lines).

We conclude that we can select an eigenvector $c_i^n$ unambiguously in the sum (\ref{f:eVecOld}), and that corresponding eigenvectors can be found by ordering them by eigenvalue $\la_{n,i}$, and that there is no ambiguity about the construction of the source function $\sqrt{\la_n}c^n_a(k,\ta)$.  However, the source functions are still not orthogonal, a property which is possessed by the eigenfunctions of the true UETC, and numerical eigenvalues can still be negative.  We will see how the third method addresses these two problems in the next section.

\subsection{Fixed-$k$ UETC interpolation}

Going back to the true (non-scaling) UETCs $C_{ab}(k,\tau,\tau')$ we see that it is natural to think of them as symmetric 
functions of $\ta$ and $\ta'$ for a given $k$. 
They contain the full information about the cosmic transitions, and can be discretised and then diagonalised as discussed above. 
This approach also fits very naturally into the scheme used by Einstein-Boltzmann codes, 
which solve the perturbation equations with an outer loop over $k$ and an inner time integration for fixed values of $k$.

\begin{figure}[h]
\centering
\includegraphics[width=0.5\textwidth]{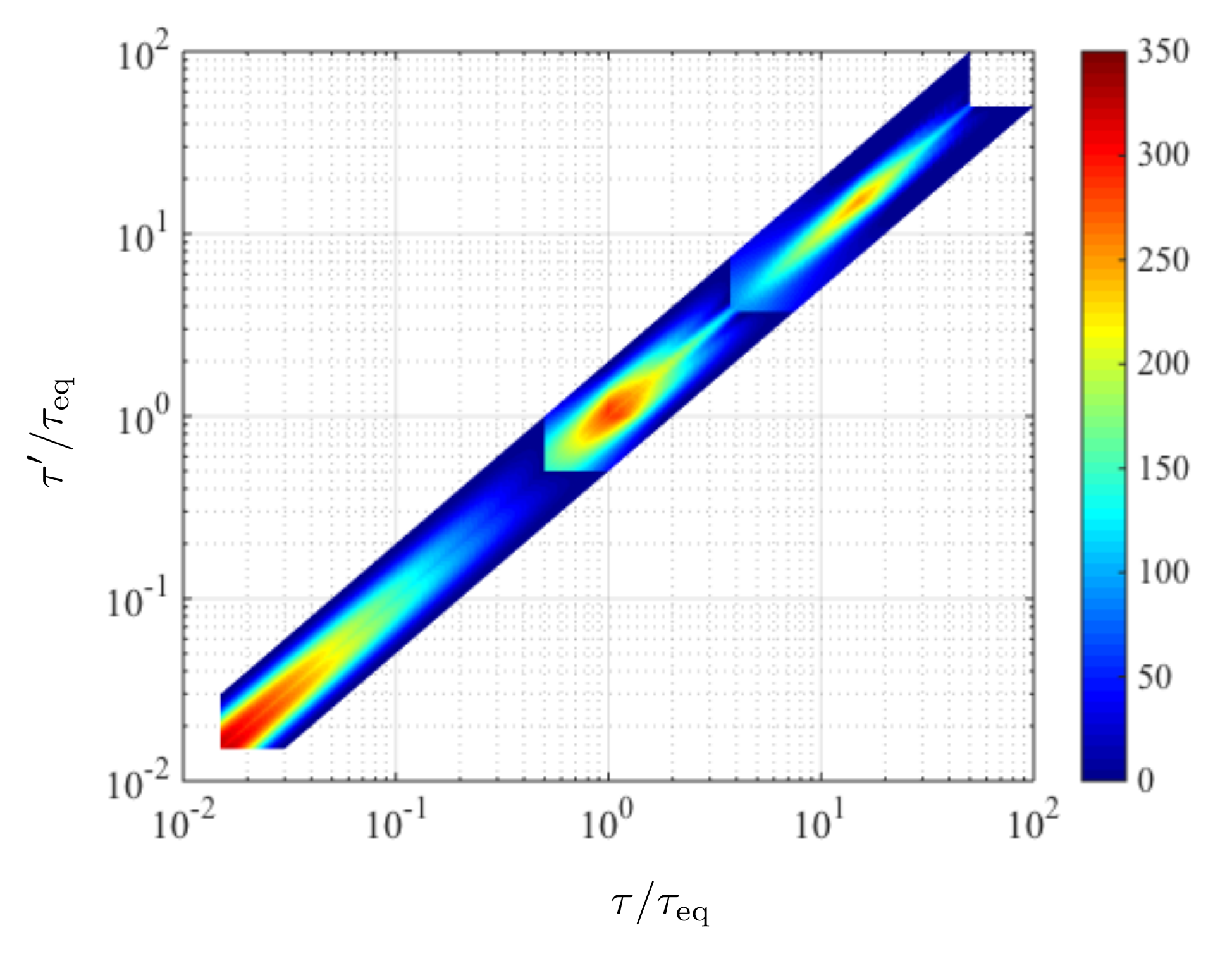}
\caption{The UETC $C_{11}$  plotted for values of $k\tEq$ nearest to $600$, $10$ and $1$, obtained from the radiation-matter transition simulations listed in Table \ref{alpharadmat}. }
\label{f:UETCkPlot}
\end{figure}

In fixed-$k$ UETC interpolation we construct approximations to $C_{ab}(k,\ta,\ta')$ from the scaling matter and radiation sources, 
 at each value of $k$.  The relative mixture of matter and radiation UETCs is determined by $\ta/\tEq$ and $\ta'/\tEq$.
First, we display how a real UETC changes with $k$ in Fig.~\ref{f:UETCkPlot}. The figure shows $C_{11}$ obtained from our seven transition era simulations, plotted against $(\tau/\tEq,\tau'/\tEq)$, for the values of $k\tEq \simeq 600$, $10$ and $1$.

To obtain this graph, we simulated string networks at intermediate stages of the radiation-matter transition, where
the scale factor evolves as 
\begin{equation}
a(\tau)=a_{\rm eq}\left(\left[\left(\sqrt{2} -1\right) \left(\frac{\tau}{\tau_{\mathrm{eq}}}\right) +1 \right]^2 -1 \right)\,.
\label{a_tr}
\end{equation}

These 1k ($1024^3$) numerical simulations had the same Lagrangian parameters as the 4k simulations, with lattice spacing $dx=0.5$, core growth parameter $s=0$, and the same $\tStart$ and $\tDiff$.  
There is limited range between the time when correlator data taking starts at $\tRef = 150$ and the end of the simulation $\tEnd = 300$, 
so each simulation spans only a part of the transition. UETC and ETCs are written at $N_\tau = 50$ logarithmically-spaced intervals between these times.

We performed five independent simulations 
 for seven values of $\tEq$ ($N_{\tEq} = 7$),
so that the simulations covered most of the transition epoch. Table~\ref{alpharadmat} shows the values of $\tEq$ and the time periods that we have simulated, five of which are used in Fig.~\ref{figure_ft_LAH}.
We also give the expansion rate parameter
\begin{equation}
\alpha(\tau) = \frac{d\ln a}{d\ln \tau}\,.
\label{ft_alpha}
\end{equation}

\begin{table}[!ht]
\begin{center}
\renewcommand{\arraystretch}{1.2}
\begin{tabular}{|c||c|c|c|c|c|c|c|}
\hline
$\tEq$ & 600 & 300 & 150 & 80 & 40 & 10 & 3 \\
${\tRef}/{\tEq}$ & 0.25 & 0.5 & 1.0 & 1.875 & 3.75 & 15 & 50 \\
${\tEnd}/{\tEq}$ & 0.5 & 1.0 & 2.0 & 3.75 & 7.5 & 50 & 100 \\
$\alpha(\tRef)$ & 1.05 & 1.09 & 1.17 & 1.28 & 1.44 & 1.76 & 1.91 \\
$\alpha(\tEnd)$ & 1.09 & 1.17 & 1.29 & 1.44 & 1.60 & 1.86 & 1.95 \\
\hline
\end{tabular}
\caption{\label{alpharadmat} Selected parameters for simulations across the radiation-matter transition. The parameters are $\tEq$ in units of $\phi_0^{-1}$, 
the ratio of the reference time $\tRef$ for UETC data-taking  and the simulation end time $\tEnd$ to $\tEq$, and the expansion rate parameters $\al = {d\ln a}/{d\ln \tau}$ at $\tRef$ and $\tEnd$. In the simulations with constant $\alpha$ (see Section \ref{ss:IntFun}), we take the value of $\alpha$ at  $\tau_{\mathrm{ref}}$.
}
\end{center}
\end{table}

In Fig.\ \ref{f:UETCkPlot},  data is taken from the unique UETC which contains a value of $k$ whose product with each of the seven values of $\tEq$ is nearest to the chosen values $600$, $10$ and $1$. For each of these three values of $k\tEq$, we therefore have a $N_{\tEq} \times N_\tau$ array. We plot this array with $\ta' = \tRef$, and also its transpose, 

The general behaviour as a symmetric function peaked near $(\tau/\tEq,\tau'/\tEq) \sim (10/k\tEq,10/k\tEq)$ is clear.  It is also clear that the height of this peak makes a smooth transition from higher values at $k\tEq \gg 1$, where the UETC resembles the UETC in a radiation-dominated universe,   to lower values at $k\tEq \ll 1$, where the UETC resembles the UETC in a matter-dominated universe.

\begin{widetext}
A proposal for the UETCs which models this behaviour across the radiation-matter transition is 
\ben
C_{ab}(k,\ta,\ta') = f\left(\frac{\sqrt{\ta\ta'}}{\tEq}\right)  \barC_{ab}^\text{M}(k\ta,k\ta') + \left(1 - f\left(\frac{\sqrt{\ta\ta'}}{\tEq}\right)\right)  \bC_{ab}^\text{R}(k\ta,k\ta').
\label{f:UETCmodel}
\een
\end{widetext}
This is manifestly symmetric in $\ta,\ta'$.
It approximates the UETC in the entire region $\ta\ta' \sim \tEq^2$ by the linear combination of pure radiation and pure matter era scaling correlators at extreme values of $\ta/\tEq$.  At sufficiently unequal times bracketing $\tEq$ the  true UETC may depart significantly from the model, but this should not matter in practice as the UETC is very small there for any value of $k$. We will see in Sect.\ \ref{sec:compare} that of the UETCs reconstructed from the source functions, the fixed-$k$ UETC interpolation method gives the most accurate results.

\begin{figure*}
\resizebox{0.49\textwidth}{!}{\includegraphics{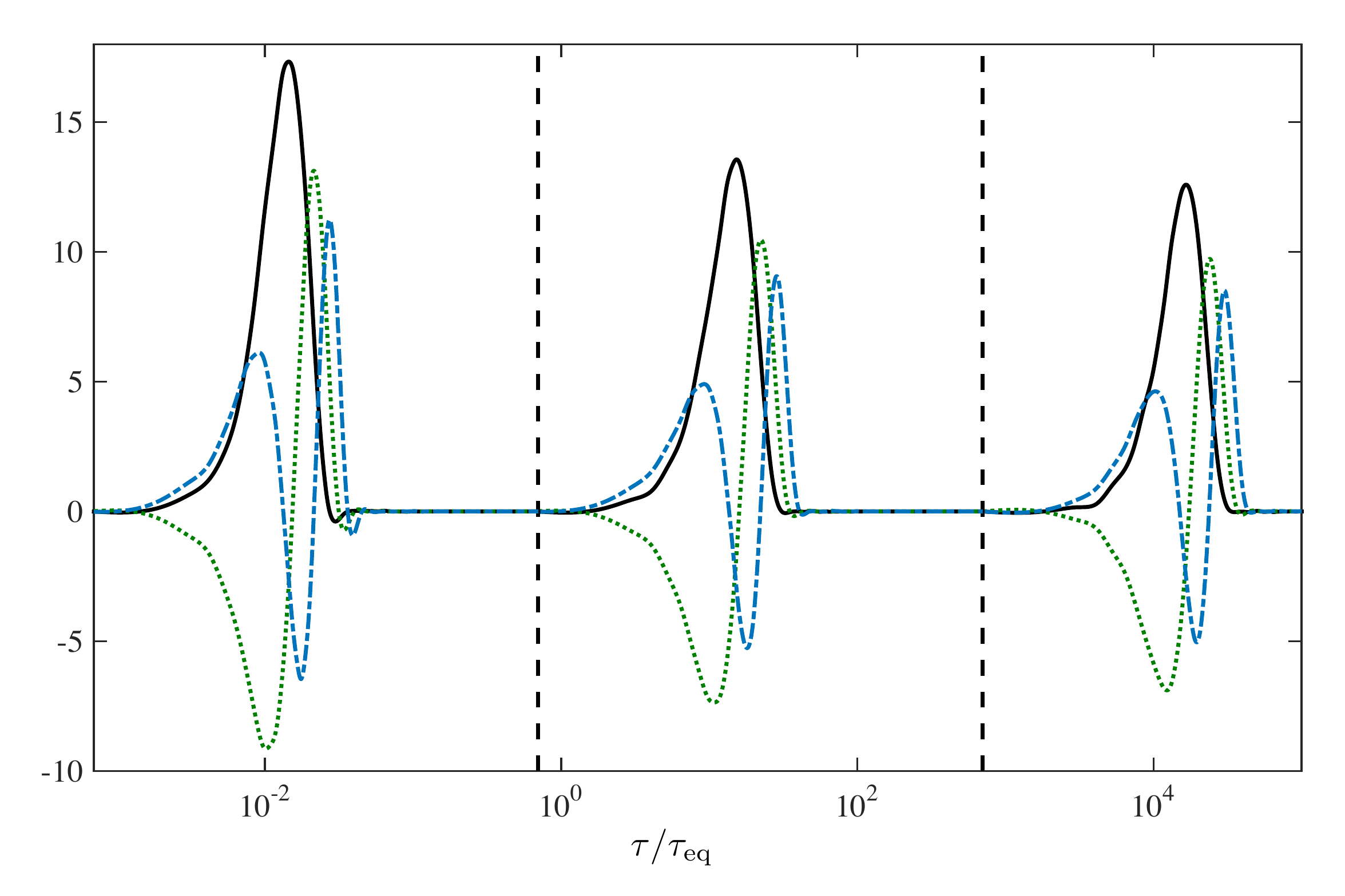}}
\resizebox{0.49\textwidth}{!}{\includegraphics{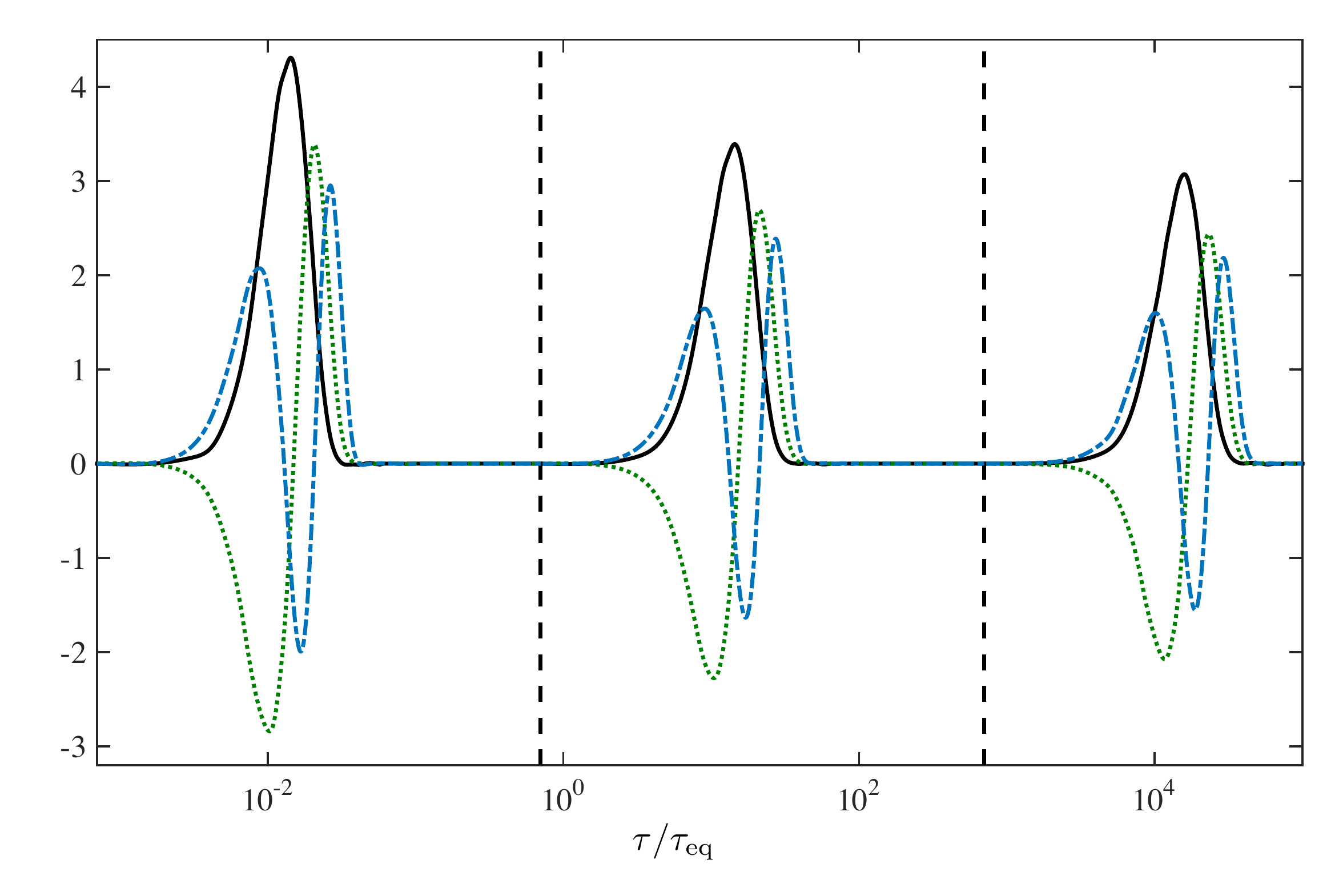}}\\
\resizebox{0.49\textwidth}{!}{\includegraphics{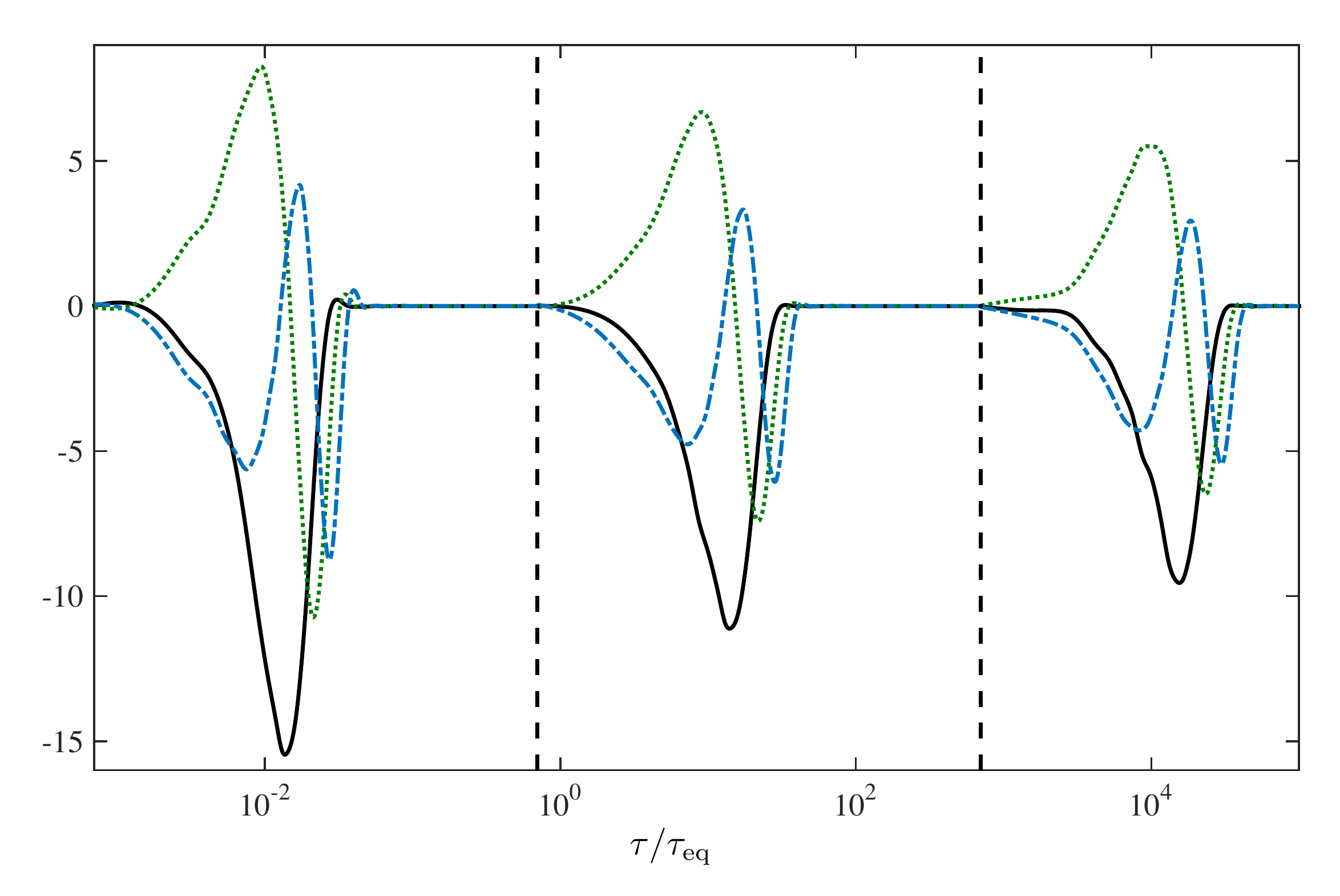}}
\resizebox{0.49\textwidth}{!}{\includegraphics{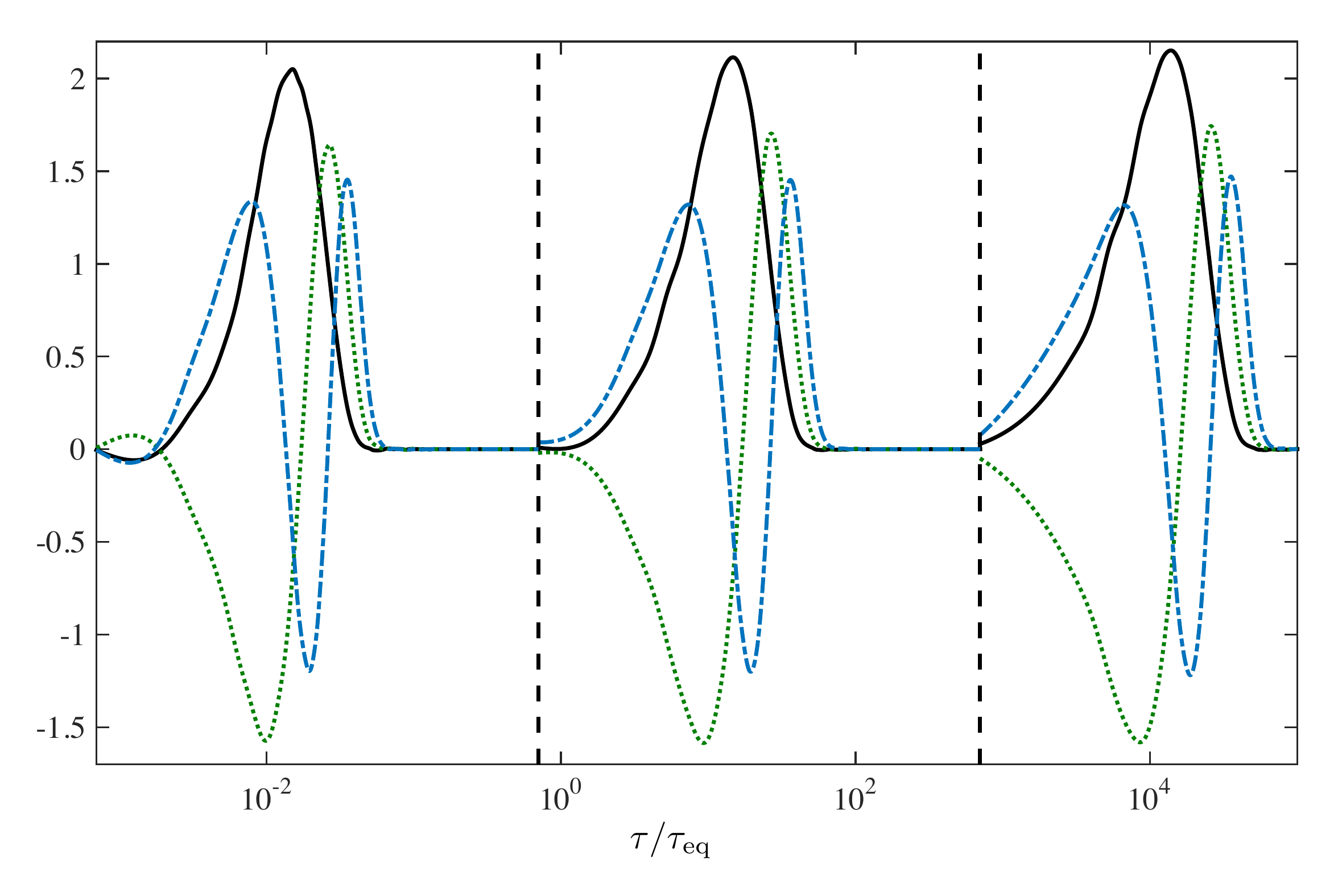}}
\caption{\label{fig_eVec123} First (black solid line), second (green dotted line) and third (blue dash-dotted line) source functions for $k\tau_{eq}=1000,\  1$ and $0.001$ from left to right. The two left figures show scalar $\phi$ (upper pane) and $\psi$ (lower pane) components, whereas the top-right figure is for vector and the bottom-right one for tensors.}
\end{figure*}

We note that the source functions for the EB integrators at a given $k$ are now just the eigenvectors of these model UETCs, multiplied by the square root of the associated eigenvalues, and so they are indeed orthogonal, unlike in the previous two methods.
In Fig.~\ref{fig_eVec123} we show the first three source functions extracted by this method as a function of $\tau/\tEq$, for $k\tEq = 1000$, $1$ and $10^{-3}$. The corresponding UETCs are therefore largest in the radiation, transition, and matter eras respectively.  
It can be seen that the source functions are indeed peaked in different ranges of $\ta$, at around $\ta \sim 10/k$, and that the peak amplitude decreases as $k$ gets smaller, consistent with the matter-era UETCs having a smaller amplitude that the radiation era ones.

\subsection{Interpolating functions $f(\tau)$ and $f_{\Lambda}(\tau)$}

\label{ss:IntFun}

We  adopt  the recipe given in \cite{Fenu:2013tea} to define the function such that it should reproduce the equal-time correlators $E_{ab}(k,\tau) = C_{ab}(k,\tau,\tau)$. First we define
\begin{equation}
f_{ab}(k,\tau)=\frac{E_{ab}^{\rm RM}(k,\tau) - \barE_{ab}^{\rm M}(k\tau)}{\barE_{ab}^{\rm R}(k\tau) - \barE_{ab}^{\rm M}(k\tau)} \ \ \ \ \ \ \ \forall k\,,
\label{ft_ETC}
\end{equation}
where $\barE^{\rm R}(k\tau)$ and $\barE^{\rm M}(k\tau)$ are the scaling ETCs in the radiation and matter eras respectively, and $E^{RM}(k,\tau)$ is the true ETC during the transition.

We will  see that the functions $f_{ab}(k,\tau)$ extracted from our simulations are consistent with being independent of $k$ and thus 
the above definition will reproduce  Eq.~(\ref{TransitionUETC}) when evaluated at equal times.  We will also see that it is a good approximation to take the same function $f(\tau)$ for each of the five ETCs.

\begin{figure}[h!]
\centering
\includegraphics[width=0.5\textwidth]{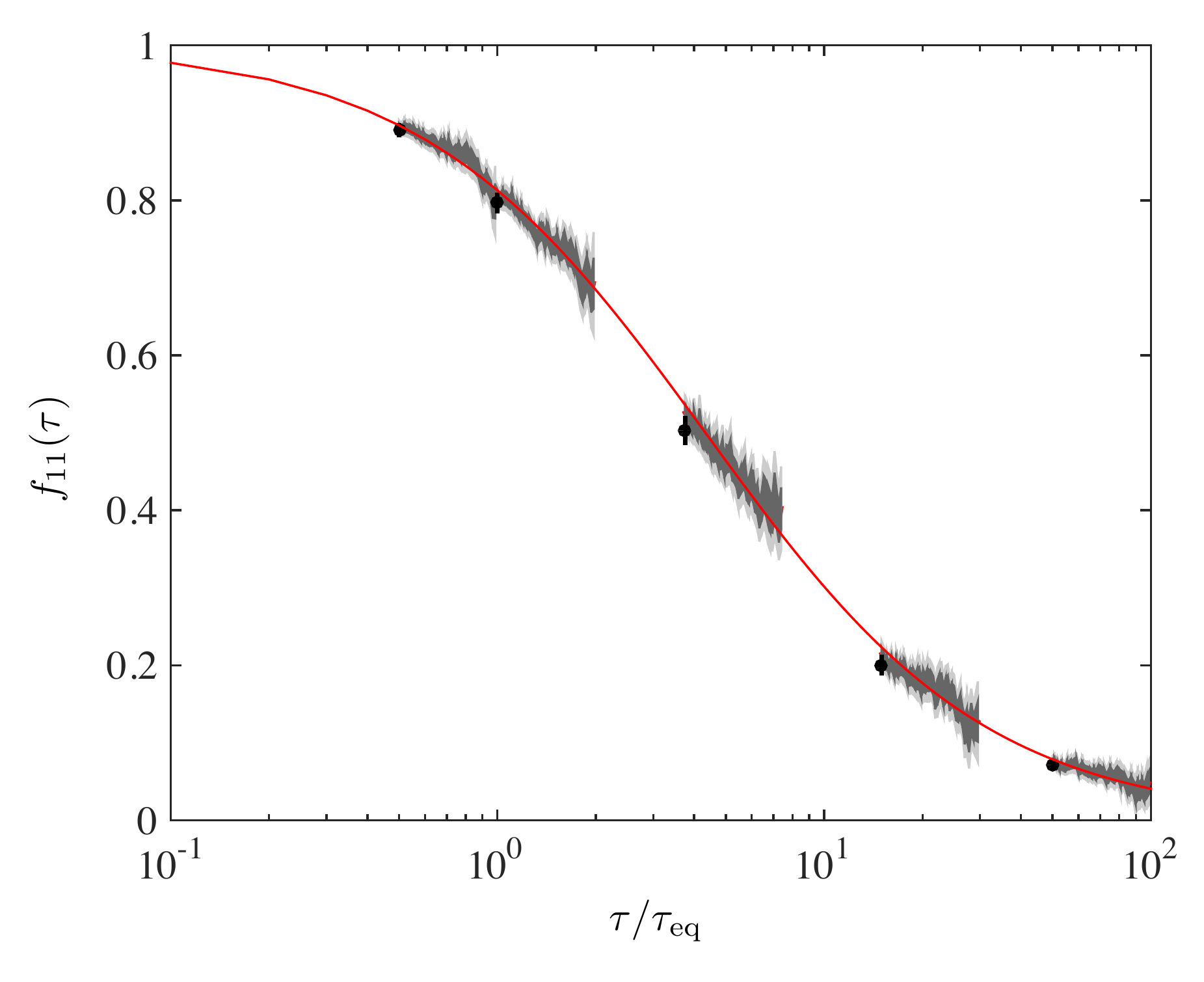}
\includegraphics[width=0.5\textwidth]{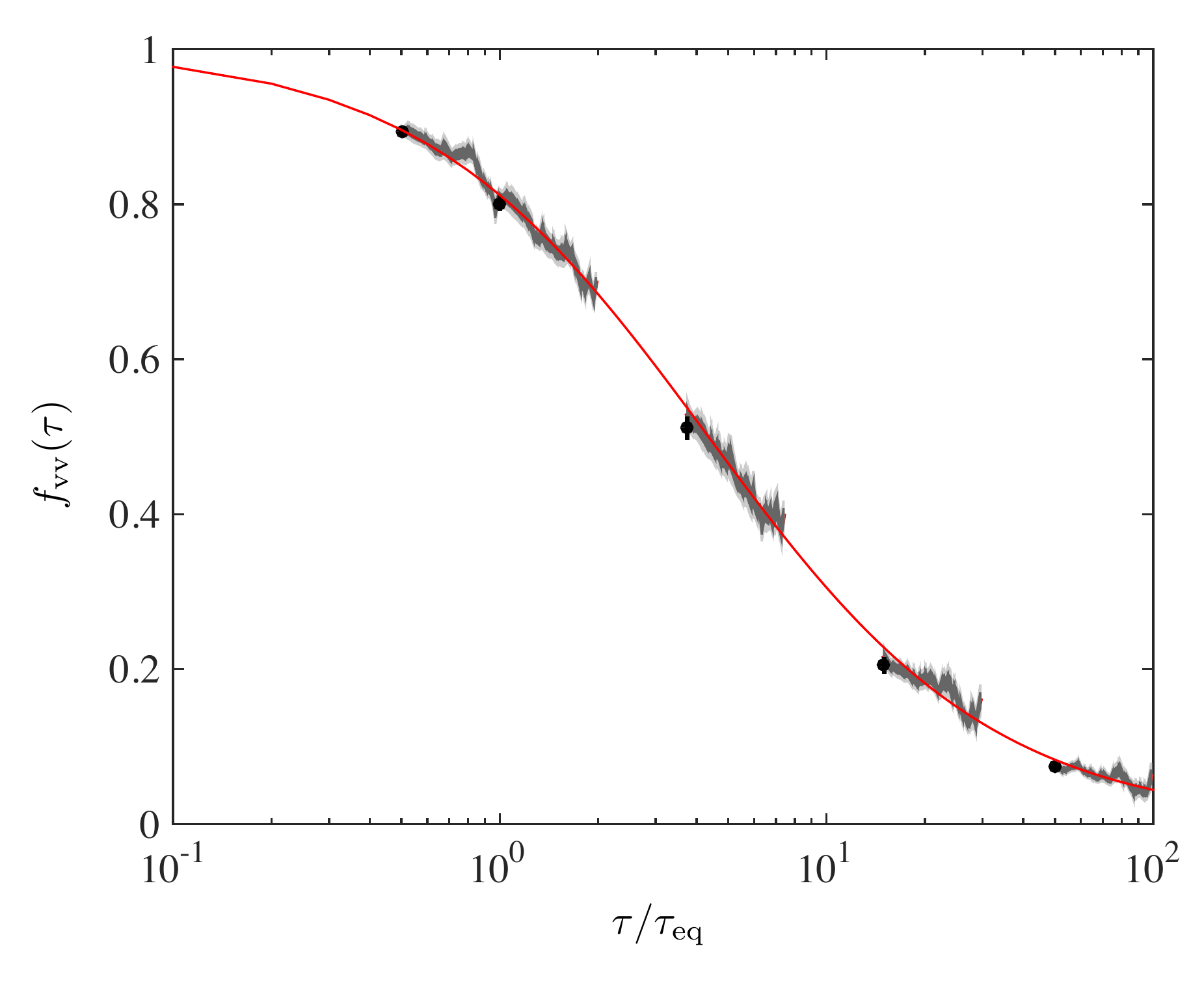}
\caption{UETC interpolation functions derived from simulations performed during the radiation-matter transition (thick grey line). The five patches correspond to 
simulations with $\tEq = 3$, $10$, $40$, $150$ and $300$.
The shaded regions represent the $1\sigma$ and $2\sigma$ deviations from the mean value of the function obtained from Eq.~(\ref{ft_ETC}) calculated from the averaging over $k$. In the upper pane the correlator used is $E_{11}$, while in the lower pane is $E_{\mathrm{vv}}$. The red line, in both cases, corresponds to the function expressed in Eq.~(\ref{ft_LAH}). The black points (with error bars) correspond to constant $\alpha$ simulations, which would mimic the adiabatic cosmology transition. }
\label{figure_ft_LAH}
\end{figure}

We extracted ETCs from 1k simulations with $\tEq = 3$, $10$, $40$, $150$ and $300$, and used Eq.~(\ref{ft_ETC}) to compute the function $f(\tau)$. Fig.~\ref{figure_ft_LAH} shows the results obtained for   correlators  $E_{11}$ and $E_{\mathrm{vv}}$. The five grey shaded regions represent the raw transition functions (\ref{ft_ETC}) obtained during the five transition periods simulated. 
The two grey levels indicate $1\sigma$ and $2\sigma$ deviations from the mean value calculated averaging over a set of wavevectors much less than the inverse string width: $0.12 < |\textbf{k}| < 2$.
We also include in the pictures the best-fit line (solid red line) obtained fitting data using the following functional form:
\begin{equation}
f(\tau)=\left(1+\zeta\frac{\tau}{\tau_{\mathrm{eq}}}\right)^{\eta}\,.
\label{ft_LAH_funcform}
\end{equation}
The narrowness of the shaded regions confirms the initial assumption of the scale independence of the function.

\begin{table*}[t]
\begin{center}
\renewcommand{\arraystretch}{1.2}
\begin{tabular}{|c||c|c|c|c|c||c|}
\hline
 Parameters & $E_{11}$ & $E_{12}$ & $E_{22}$ & $E_{\mathrm{vv}}$ & $E_{\mathrm{tt}}$  & Mean and $\sigma$\\\hline
 $\zeta$ & $0.232\pm0.006$ &  $0.244\pm0.012$ &  $0.246\pm0.010$ & $0.242\pm0.006$ & $0.203\pm0.010$  & $0.235\pm0.004$ \\ \hline
 $\eta$ &  $-1.01\pm0.02$ & $-1.01\pm0.04$ & $-1.03\pm0.03$ & $-0.96\pm0.01$ & $-1.10\pm0.05$ & $-0.984\pm0.008$\\
 \hline
\end{tabular}\\
 \caption{\label{table_fLAH} Mean values together with the standard deviations for the parameters $\zeta$ and $\eta$ of Eq.~(\ref{ft_LAH_funcform}) needed to reproduce the radiation-matter transition. 
 }
\end{center}
\end{table*}

Table~\ref{table_fLAH} shows the mean values and standard deviations for the parameters of Eq.~(\ref{ft_LAH_funcform}); it is clear that the transition applies in a very similar form for all correlators, implying that they evolve in a similar way across the transition. In order to simplify further calculations we consider the following function as the radiation-matter transition UETC interpolation function that applies equally to all correlators of the Abelian-Higgs cosmic string model:
\begin{equation}
f(\tau)=\left(1+0.24\frac{\tau}{\tau_{\mathrm{eq}}}\right)^{-0.99}\,,
\label{ft_LAH}
\end{equation}
We note that the function (\ref{ft_LAH}) is almost the square root of the interpolation function for large-$N$ self-ordering scalar fields obtained in \cite{Fenu:2013tea}
\begin{equation}
f_{N}(\tau)=\left(1+\frac{1}{4}\frac{\tau}{\tau_{\mathrm{eq}}}\right)^{-2}\,.
\end{equation}
The conjecture  \cite{Fenu:2013tea} that the interpolation function is universal is therefore not supported by our findings.

We also compared the transition-era ETCs at time $\tau_n$ with scaling ETCs evaluated with a constant expansion rate parameter $\al(\tau_n)$.
We performed five simulations with constant $\alpha$ chosen to coincide such that the expansion rate fell within the range of expansion rates explored by our simulations across the radiation-matter transition 
(values can be found in Table~\ref{alpharadmat}).  
Defining
\begin{equation}
 f^\al_{ab}(k,\tau)=\frac{\barE_{ab}^{\rm \al}(k\tau) - \barE_{ab}^{\rm M}(k\tau)}{\barE_{ab}^{\rm R}(k\tau) - \barE_{ab}^{\rm M}(k\tau)}, 
\label{ft_ETC_sc}
\end{equation}
we can plot the average values of $f^\al_{ab}$ in Fig.~\ref{figure_ft_LAH}, where the five points (black dot points with corresponding $1\sigma$ bars obtained from $k$-averaging) come from the constant expansion rate simulations.

Interestingly, the best-fit function lies almost on top of the constant expansion rate points $f^\al$.
We therefore conclude that the string network reacts quickly to changes in the expansion rate, and we can treat the ETCs as being adiabatic: 
in other words, the properties of the string network at any given time during the radiation-matter transition corresponds well to the properties of a scaling network at the same instantaneous expansion rate.
In principle we expect the same behavior for other types of defects, therefore,  a good approximation could be found by performing a series of smaller/shorter simulations at intermediate constant expansion rates.

\subsection{Matter-$\La$ interpolation}

We also applied the same procedure to incorporate the effects of the accelerated expansion of our universe, extending our analysis to the matter-$\Lambda$ transition. In a $\Lambda$ dominated universe, one expects the string velocity to decay and the network effectively to freeze with a length scale $\xi_\text{fr}$. 

We computed in \cite{Obradovic:2011mt} the metric perturbations induced by a straight string moving with a velocity $v$, and from the expressions in that article we can see that for $v\rightarrow0$ the scalar potential $\psi$ as well as the vector perturbations vanish,
while the scalar potential $\phi$ and the tensor perturbations remain finite. Based on this we expect that the tensor and the $E_{11}$ UETCs do not vanish, while $E_{12}$, $E_{22}$ and $E_{\mathrm{vv}}$ go to zero.

\begin{figure}[h!]
\centering
\includegraphics[width=0.5\textwidth]{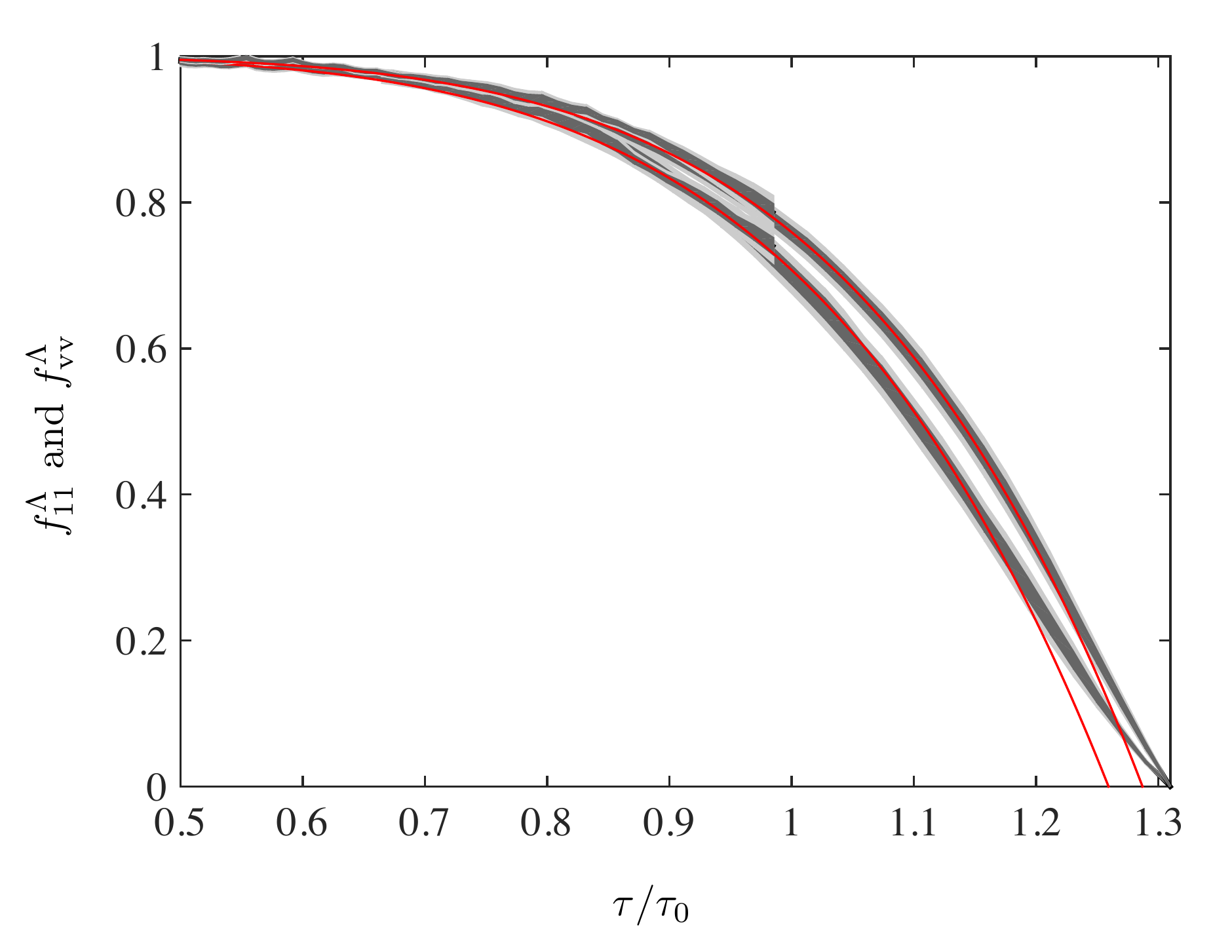}
\caption{Matter-$\Lambda$ UETC interpolation function for different correlators plotted against the conformal time relative to current time ($\tau_0$), from top $E_{\mathrm{vv}}$ and  bottom $E_{11}$. The other two scalar correlators, $E_{12}$ and $E_{22}$, lie roughly between the two red lines. The different shades correspond to $1\sigma$ and $2\sigma$ confidence limits, and the red line corresponds to the best fit.}
\label{f_t_Lambda}
\end{figure}

Therefore, the counterpart of Eq.~(\ref{ft_ETC}) is
\begin{equation}
f_{ab}(k,\tau)=\frac{E_{ab}^{\rm M\Lambda}(k,\tau) - \barE_{ab}^{\Lambda}(k\ta_\text{fr})}{\barE_{ab}^{\rm M}(k\tau) - \barE_{ab}^{\Lambda}(k \ta_\text{fr})} ,
\label{ftl_ETC}
\end{equation}
where $\ta_\text{fr}$ is a time derived from the length scale $\xi_\text{fr}$ of a frozen string network in de Sitter space, and 
$\barE_{12}^{\Lambda}=\barE_{22}^{\Lambda}=\barE_{\mathrm{vv}}^{\Lambda}=0$.
We show the decay of the correlators $E^{\rm M\Lambda}(k,t)$ in Fig.~\ref{f_t_Lambda}, based on simulations evolving in a $\Lambda$CDM background. These simulations covered mainly the late matter dominated era and the beginning of the dark energy domination.  Table~\ref{alphalambda} shows the cosmological parameters at the end of each of the regimes simulated. 
We have been able to go farther towards the $\Lambda$CDM singularity where the conformal time reaches its asymptotic de Sitter value, and where therefore the scale factor diverges as a function of $\tau$ (around $\tau \approx 1.35 \tau_0$ for a value of $\Omega_m=0.315$). 
As our estimate of the de Sitter correlators we  
measure the functions $\barE_{11}^{M\Lambda}$ and $\barE_{\mathrm{tt}}^{M\Lambda}$ at $\tEnd = 1.33$, and take $\ta_\text{fr} = \be^{-1}\xi_\text{fr}$, 
where $\be$ is the slope of the relation between time and network length scale (see Eq. \ref{XiEq}), and we use its value during the matter
era, given in Table \ref{table_beta}.

\begin{table}[h!]
\renewcommand{\arraystretch}{1.2}
\begin{tabular}{|c||c|c|c|c|c}
\hline
$\tau_0 $ & $300$ &$225$ \\
$\tRef/\tau_0$ & $0.5$ &$0.665$ \\
$\tEnd/\tau_0$ & $1$ &$1.33$ \\
$\Omega_m (\tEnd)$ & $0.315$ & $1.29 \cdot 10^{-4}$ \\
$\Omega_r (\tEnd)$ & $9.24\cdot 10^{-5}$ & $2.81\cdot 10^{-9}$ \\\hline
\end{tabular}
\caption{\label{alphalambda} Values of the current conformal time $\tNow$ in simulation time units, the ratio of the reference time $\tRef$ for UETC data-taking  and the simulation end time $\tEnd$ to $\tNow$.  Also given are 
the cosmological parameters $\Omega_m$ and $\Omega_r$ at the end of each simulation across the matter-$\Lambda$ transition} 
\end{table}

The interpolation functions related to each of the different correlators can be fitted by the following set of functions:

\begin{equation}
f^{\Lambda}(\tau)= \left(1+\zeta \left(\frac{\tau}{\tau_{\mathrm{0}}}\right)^\eta\right)\,,
\label{ab}
\end{equation}
where the best fits of the parameters $\zeta$ and $\eta$ for each case are shown in Table~\ref{table_fLam}. 
It can be seen that there is greater variation in the parameters than in the radiation-matter case.
Note that $E_{\rm tt}$ changes little during the transition and so the errors in $f_{\rm tt}$ are very large.  Hence it is a good approximation not to interpolate $E_{\rm tt}$ at all.

\begin{table*}[t]
\begin{center}
\renewcommand{\arraystretch}{1.2}
\begin{tabular}{|c||c|c|c|c|c|}
\hline
 Parameters & $E_{11}$ & $E_{12}$ & $E_{22}$ & $E_{\mathrm{vv}}$ & $E_{\mathrm{tt}}$  \\\hline
 $\zeta$ & -0.302$\pm$0.003 &  -0.276$\pm$0.003 &  -0.292$\pm$0.003 & -0.241$\pm$0.001 & -- \\ \hline
 $\eta$ &  5.2$\pm$0.1 & 5.4$\pm$0.1 & 5.3$\pm$0.1 & 5.63$\pm$0.03 & --\\
 \hline
\end{tabular}
 \caption{\label{table_fLam} Best fit values for parameters $\zeta$ and $\eta$ in Eq.~(\ref{ab}) corresponding to the Matter-$\Lambda$ transition function.}
\end{center}
\end{table*}

We can anticipate that the effect of taking into account a $\Lambda$CDM background cosmology will slightly decrease the amplitude of the late time correlators. Consequently, this decay will affect the power spectra at lower multipoles, decreasing the contribution at scales that entered late the horizon.

\subsection{Comparison of interpolation methods\label{sec:compare}}

In this section we compare simple and multistage eigenvector interpolation, as used in \cite{Bevis:2006mj} and \cite{Fenu:2013tea}, with the fixed-$k$ UETC interpolation introduced in this paper.
We perform the comparison by reconstructing the UETC from the 
interpolated source functions 
\ben
C^{\mathrm{rc}}(k,\tau,\tau') = \sum_n \la_n c_n(k,\tau) c_n^* (k,\tau').
\label{UETC_rc}
\een
This is then compared with a measured transition correlator. 
We choose an intermediate stage of the radiation-matter transition, $1 < \tau/\tau_{eq} < 1.5$ and restrict the analysis to scales around the peak of the correlator ($8.3<k\tau<30$), where the most important contribution is encapsulated. Note that though the eigenvectors of the time-interpolated UETCs do not strictly form an orthonormal set, their product forms an effective UETC, see Eq.~ (\ref{UETC_rc}). Time evolving eigenvectors for the eigenvector interpolation method, in turn, are calculated using (\ref{neil_interp}) and (\ref{e_Neil}).

\begin{figure}
\resizebox{0.49\textwidth}{!}{\includegraphics{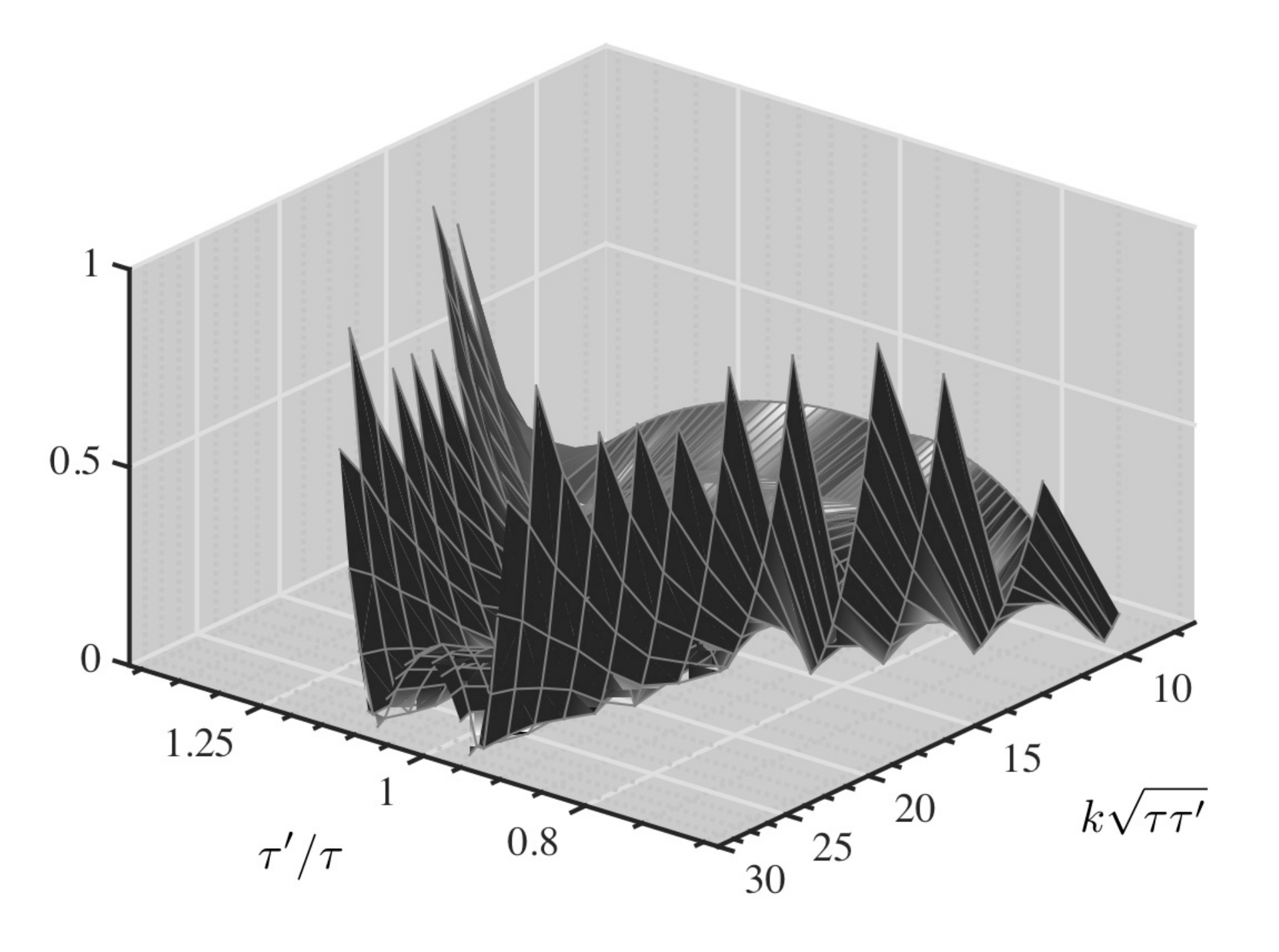}}
\resizebox{0.49\textwidth}{!}{\includegraphics{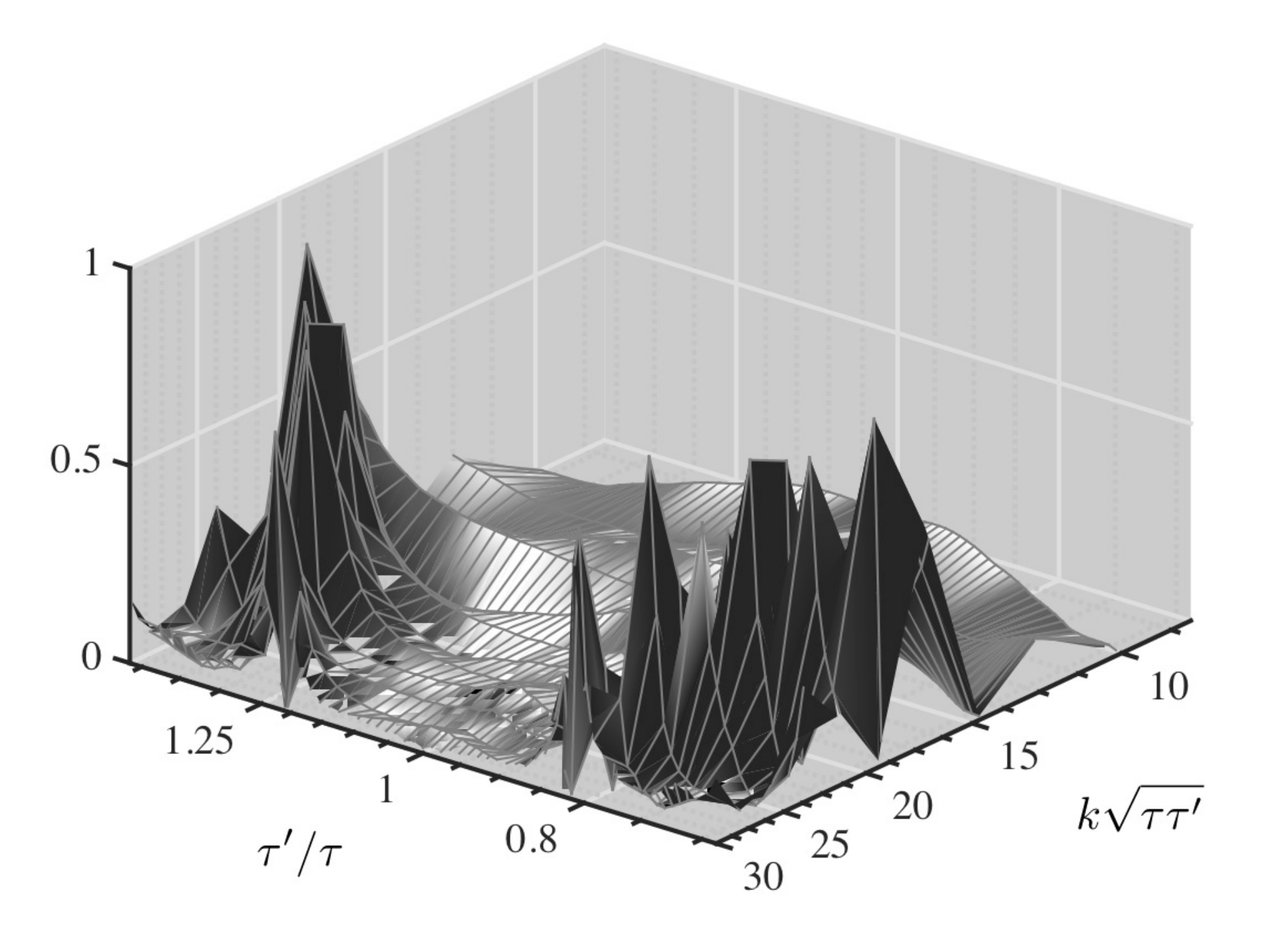}}
\resizebox{0.49\textwidth}{!}{\includegraphics{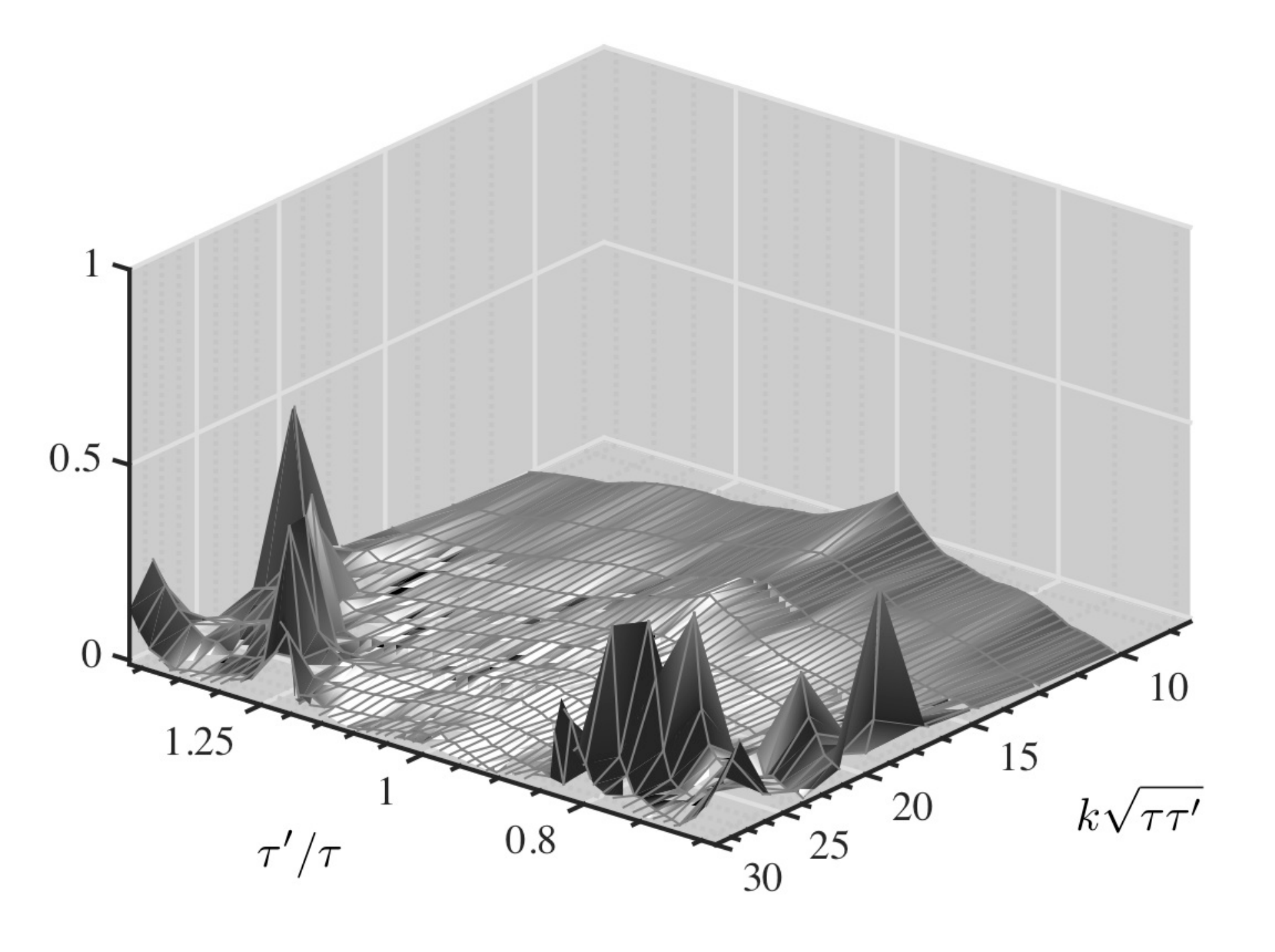}}\\
\caption{\label{C11_reldiff} Relative difference of the reconstructed scalar $C_{11}$ UETC with respect to the measured UETC. The UETCs are reconstructed using three different methods: simple eigenvector interpolation (first pane), multi-stage eigenvector interpolation (second pane)  and fixed-$k$ UETC interpolation (third pane). In the simple eigenvector interpolation method the interpolating function is (\ref{e_Neil}), whereas for the multi-stage eigenvector interpolation at the correlator level we use (\ref{ft_LAH}). Only the region around the peak of the correlator, $1/1.5 < r < 1.5$ and $8.3<k\tau<30$, is shown. Note that in order to represent the relative difference as neatly as possible, we remove values greater than 1 from the two top pictures.}
\end{figure}

We show in Fig.~\ref{C11_reldiff} the relative difference of the reconstructed UETC using 128 eigenvectors for the scalar $C_{11}$ function for the three proposed methods. There can be seen that the resemblance of the fixed-$k$ interpolation to the real case is the highest and is clearly better than the multi-stage eigenvector interpolation method, which is in turn better than simple eigenvector interpolation. The values of the relative differences at $z=10$, near the peak of the UETCs, are approximately  $0.03$, $0.09$, and $0.2$ respectively.

\section{Discussion and Conclusions}
\label{sec:Disc}

In this work we compute the unequal time correlators (UETCs) from numerical simulations of Abelian Higgs strings, and 
describe and implement a new method of deriving source functions from them. Those source functions can be used by
source-enabled versions of Einstein-Boltzmann integrators to obtain CMB and matter perturbations.
 
Our  numerical simulations have been improved considerably from our previous works due to improvements in both the hardware resources  and the software used. 
The Abelian Higgs code uses the recently-released LATfield2 \cite{LatField2d}, allowing the efficient numerical integration of the field equations in parallel, and portable parallel fast-fourier transforms on large grids.

Our new production runs took place on lattices of $4096^3$ sites, distributed over $34816$ CPUs, a great improvement over our previous lattice sizes of $1024^3$ \cite{Bevis:2010gj}. Bigger lattices mean larger dynamical ranges: our simulations cover a larger portion of the evolution of the universe, both in space and time.  As the space simulated is 64 times larger, we can obtain more accurate statistics. A factor of 4 longer evolution time allows for a more accurate study of the scaling of the network, and we can explore regions of the UETCs that could only be reached by extrapolation in previous work. Not only that, one of our key approximations in earlier works, string core growth, can be dropped, and thus the {\it true} equations of motion have been solved, for the first time in the matter era.  We confirm the extrapolations from our older simulations.

In summary, our new UETC measurements span a much larger time ratio  than in previous Abelian Higgs string simulations when using string core growth, and solve the true equations of motion  over a long enough period to achieve scaling and measure the UETCs. 
We have combined two complementary sets of simulations, one with string core growth, and one with the true equations of motion, to obtain our final correlation functions.
 
Close to the peak of the UETCs, at near-equal times, we use the simulations with the true equations of motion. Outside this region we use the simulations with string core growth. In order to merge the UETCs, their normalization at equal times was matched, but no other adjustment was necessary.  The new UETCs are consistent with our previous measurements near the peak of the correlators, reach the horizon scale for the first time, and confirm the power-law behaviour of the correlators at large wave-numbers.  
The normalization is slightly higher in the high wave-number tails, due to a small increase in the string density.

Numerical simulations of Abelian Higgs strings across cosmological transitions have been performed for the first time.  The radiation-matter transition is particularly important for the accurate computation of CMB perturbations at around a degree scale.  We also performed simulations across the matter-$\Lambda$ transition, important for large angular scales.
We have introduced and investigated a new method for calculating the source functions for Einstein-Boltzmann integrators, which better accounts for cosmological transitions. The method is more accurate than two previous methods  \cite{Bevis:2006mj,Fenu:2013tea}, and is also consistent with the underlying idea of decomposing the UETC into its component eigenvectors. It is also easier to implement.

Armed with the new simulations and an improved procedure to overcome the difficulties of the cosmological transitions, we can compute new and more accurate predictions for the temperature and polarization anisotropies in the CMB due to cosmic strings, which we will report in a future publication.

After our article appeared in the arXiv, further work has been done using the Unconnected Segment Model to approximate the cosmic string network \cite{Charnock:2016nzm}. The free parameters of the model can be tuned to mimic the TT power spectrum of an Abelian Higgs network, and the resulting BB spectrum also agrees quite well. There are differences in detail: for example, the tensor contributions appear low, and it would be interesting to investigate further.


\begin{acknowledgments}
We thank Neil Bevis and Ruth Durrer for helpful discussions. This work has been supported by two grants from the Swiss National Supercomputing Centre (CSCS) under project IDs s319 and s546. In addition  this work has been possible thanks to the computing infrastructure of the i2Basque academic network, the COSMOS Consortium supercomputer (within the DiRAC Facility jointly funded by STFC and the Large Facilities Capital Fund of BIS), and the Andromeda/Baobab cluster of the University of Geneva.
JL and JU acknowledge support from Eusko Jaurlaritza (IT-559-10) and  Ministerio de Econom\'\i a y Competitividad (FPA2012-34456) and  Consolider-Ingenio  Programmes CPAN (CSD2007-00042) and EPI (CSD2010-00064). DD and MK acknowledge financial support from the Swiss NSF. MH acknowledges support from the Science and Technology Facilities Council (grant number ST/L000504/1).
\end{acknowledgments}

\appendix

\section{Equal time correlators}

In this appendix we show the ETCs in the pure radiation and pure matter era simulations, compared to those we found in Ref.\ 
\cite{Bevis:2010gj}.  We recall that the simulations reported in this work have four times the spatial volume and run for four times as long as the previous ones, and solve the true field equations ($s=1$ in Eqs.\ (\ref{AHMod1}) and  (\ref{AHMod2})) rather than those with the core growth approximation ($s=0$).  There are now 7 realisations in each case, rather than 3. We also use a different method for extracting the scaling form of the ETCs, detailed in Section \ref{sec:UETC}.

\begin{figure}[t]
\includegraphics[width=0.48\textwidth]{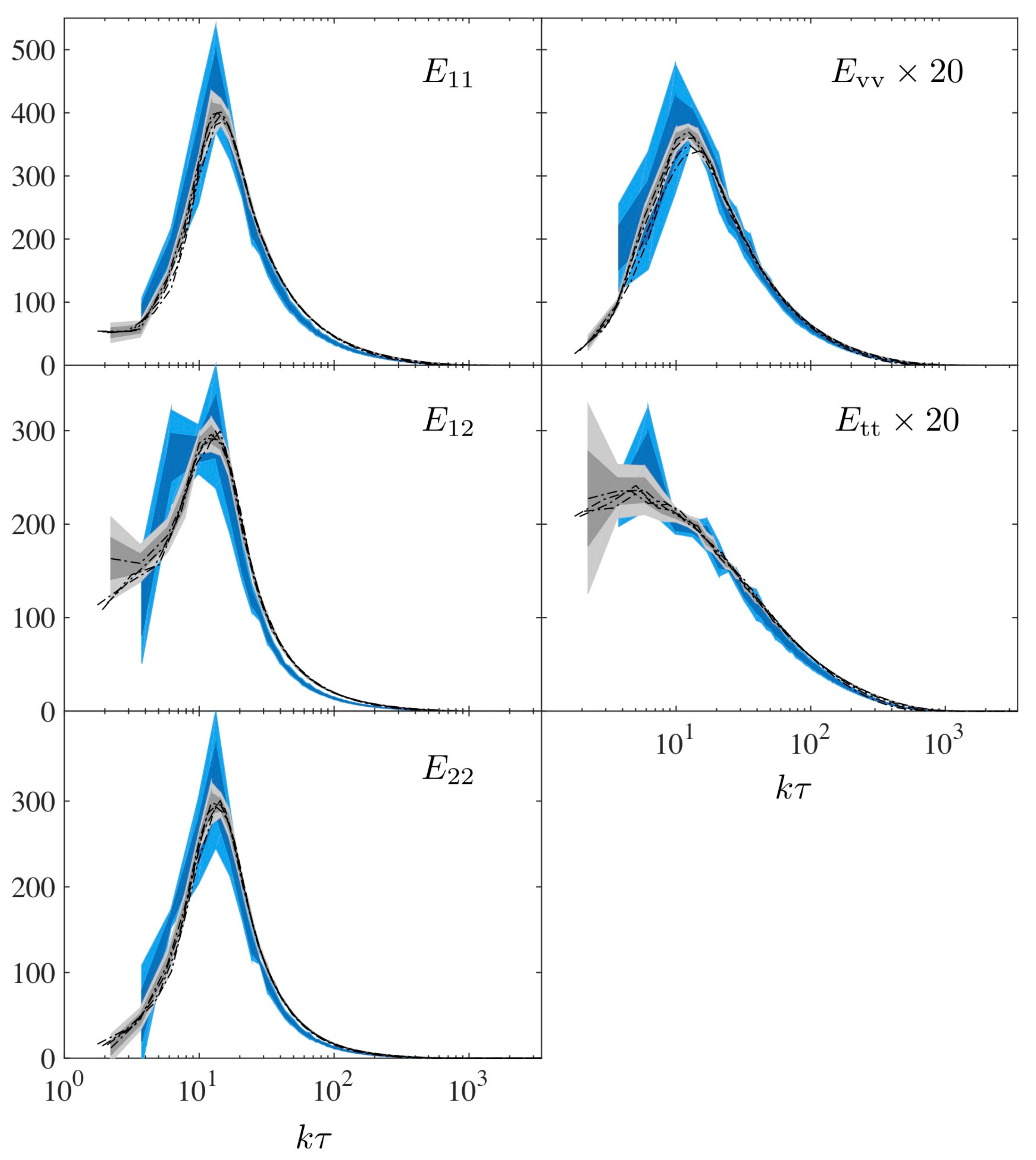}
\caption{\label{f:ETCcomRad} Equal time correlators in the radiation era for the simulations described in this work (grey), compared with those obtained in the smaller simulations of Ref.~\cite{Bevis:2010gj} (blue). The times shown are $\ta = 450$, $500$, $550$ and $600$ (in units of $\phi_0^{-1}$), with  the grey shaded regions showing 1$\si$ and 2$\si$ fluctuations over 7 realisations at $\ta = 600$. The blue shaded regions show the fluctuations for 3 realisations at $\ta = 300$.}
\end{figure}

\begin{figure}[t]
\includegraphics[width=0.48\textwidth]{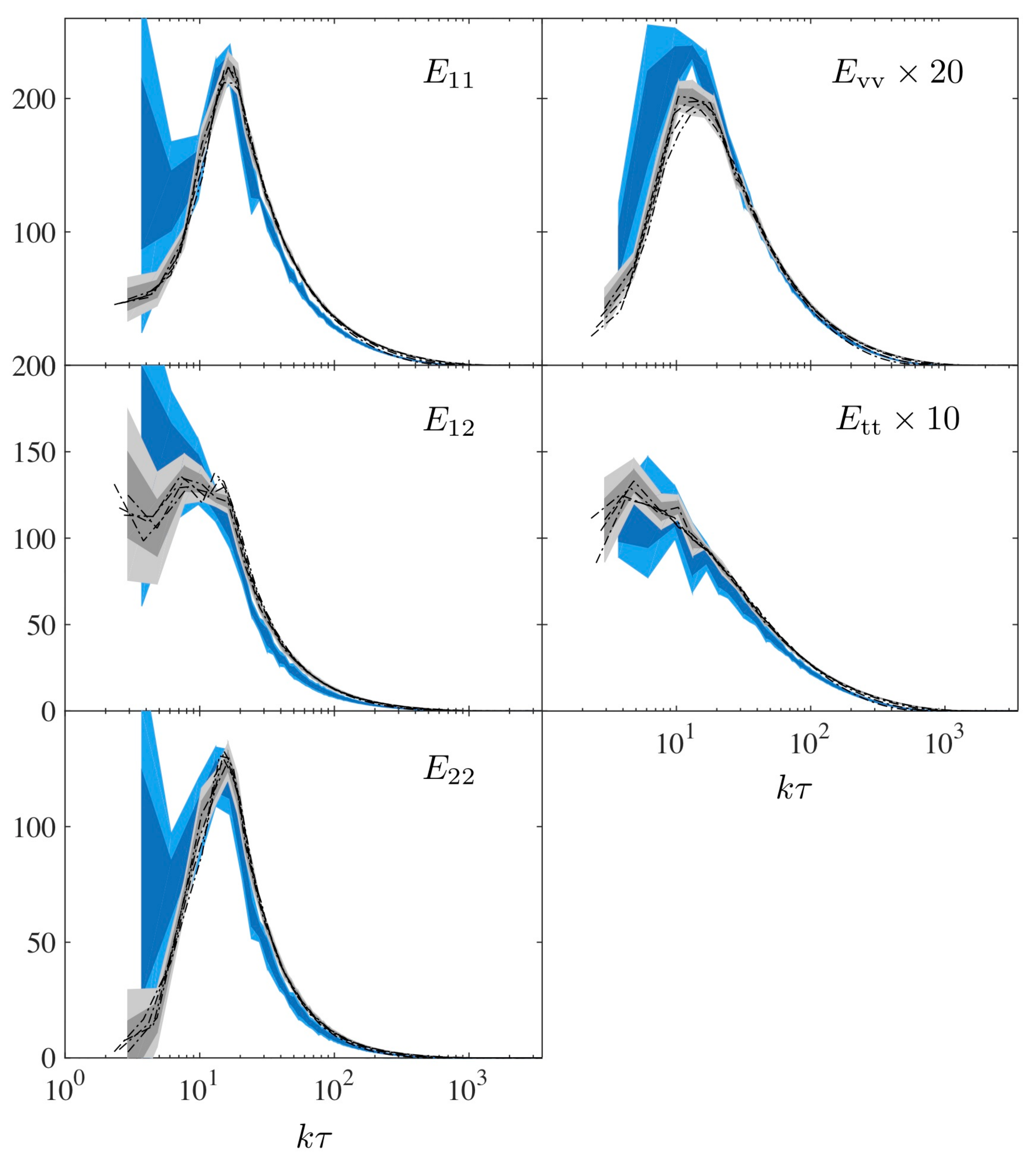}
\caption{\label{f:ETCcomMat} Equal time correlators in the matter era for the simulations described in this work (grey), compared with those obtained in the smaller simulations of Ref.~\cite{Bevis:2010gj} (blue). The times shown are $\ta = 600$, $666$, $733$ and $800$ (in units of $\phi_0^{-1}$), with  the grey shaded regions showing 1$\si$ and 2$\si$ fluctuations over 7 realisations at $\ta = 800$. The blue shaded regions show the fluctuations for 3 realisations at $\ta = 300$.}
\end{figure}

In Figs.~\ref{f:ETCcomRad} and \ref{f:ETCcomMat}, it can be seen that the shapes of the correlation functions are very similar, with small differences in normalisation at $k\tau$ beyond the peak, although the peak positions are consistent.  This could be evidence that for even greater accuracy we should distinguish between the string correlation length (which sets the peak position) and the string separation (which sets the amplitude via the string density).
The largest difference is for the radiation era vector ETC, which is lower by about 20\% at the peak.

In Ref.\ \cite{Bevis:2010gj} we argued, from $s=1$ radiation era simulations and $s=0.5$ matter era simulations, that the core growth approximation was not a significant source of error. Now that we are able to perform $s=1$ simulations in both matter and radiation eras, the consistency of the two sets ETCs demonstrates that $1024^3$ simulations at $s=0$ with lattice spacing $\Delta x = 0.5\phi_0^{-1}$ give a reasonable approximation to the correlation functions, to within about 20\% at the peak.  The smaller simulations do not however give an accurate measurement of the super-horizon ETCs, whose mean values differ by up to O(1) in the case of $E_{11}$.

We also recall that the quality of the ETCs provides a scale-dependent test of the scaling hypothesis for the string network.  If the network is scaling properly, the ETCs should be a function of $k\tau$ only, and the ETCs from different times in the simulation should collapse onto a single line when plotted against this variable. It is clear that the scaling around the peak over the time intervals from which data is taken (Table \ref{table_tref}) is very good. Scaling is violated for wavenumbers near the inverse string width, and we do not use these scales for the UETC construction. This is not an important  source of error for CMB calculations, as the ETCs are extremely small at high $k\tau$.

\begin{figure*}[t]
\resizebox{0.32\textwidth}{!}{\includegraphics{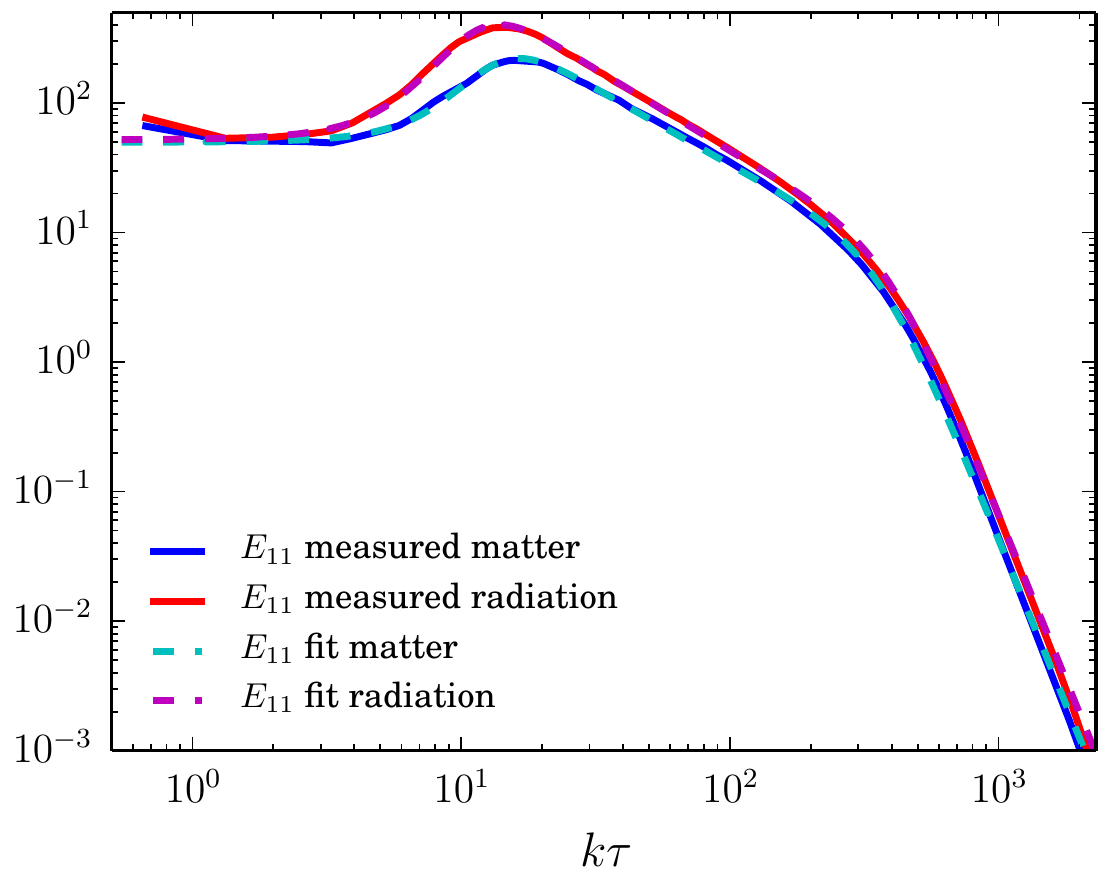}}
\resizebox{0.32\textwidth}{!}{\includegraphics{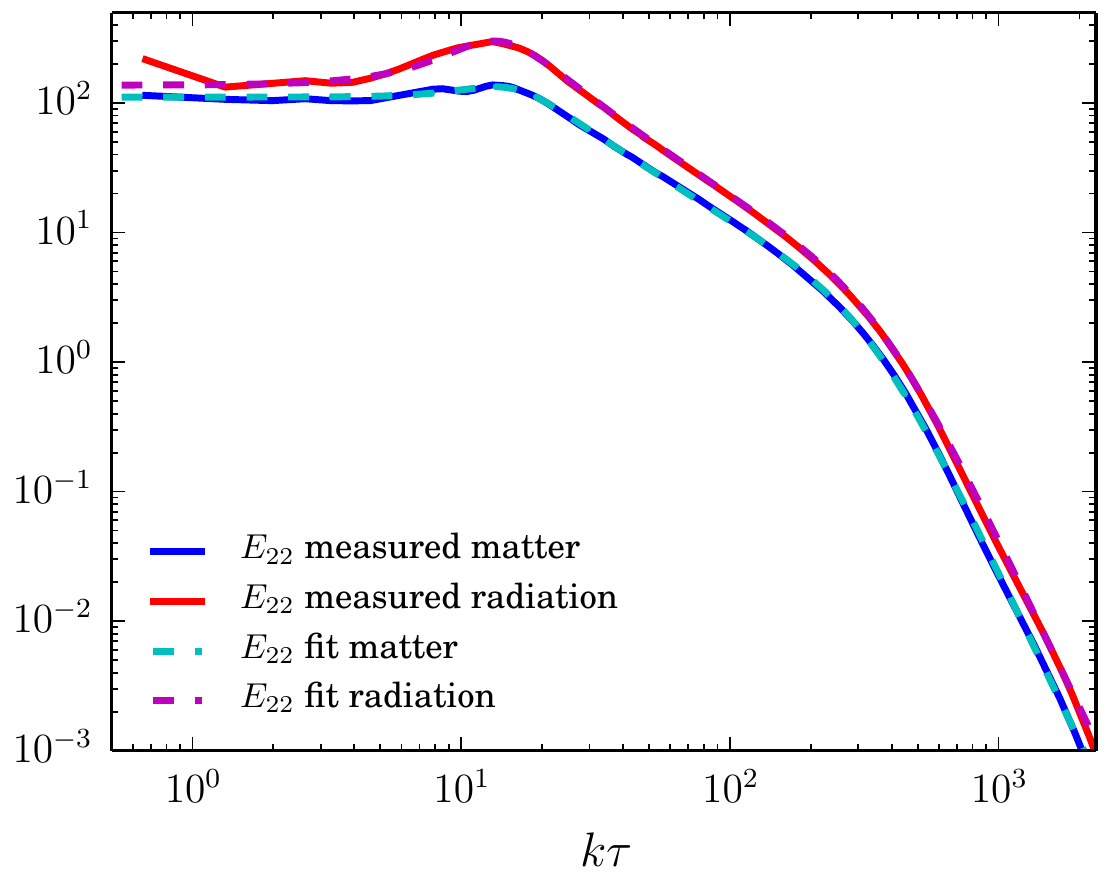}}
\resizebox{0.32\textwidth}{!}{\includegraphics{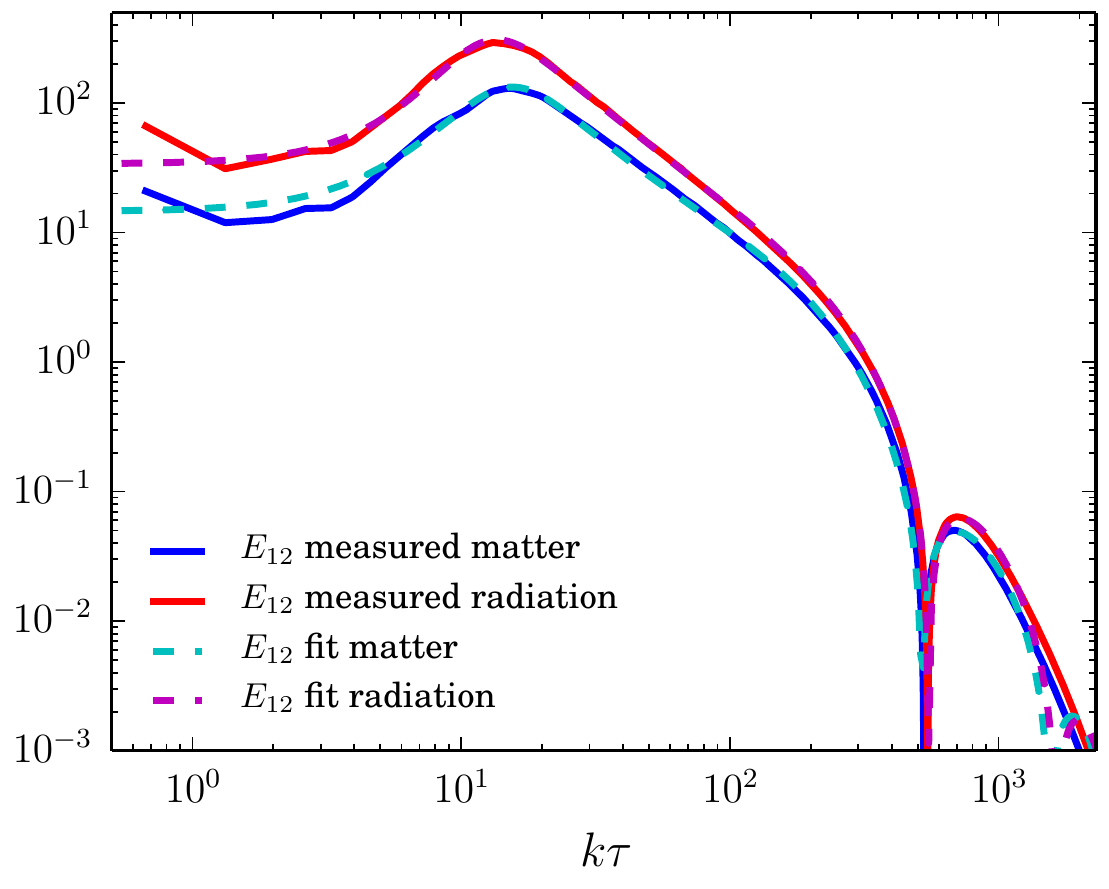}}
\resizebox{0.32\textwidth}{!}{\includegraphics{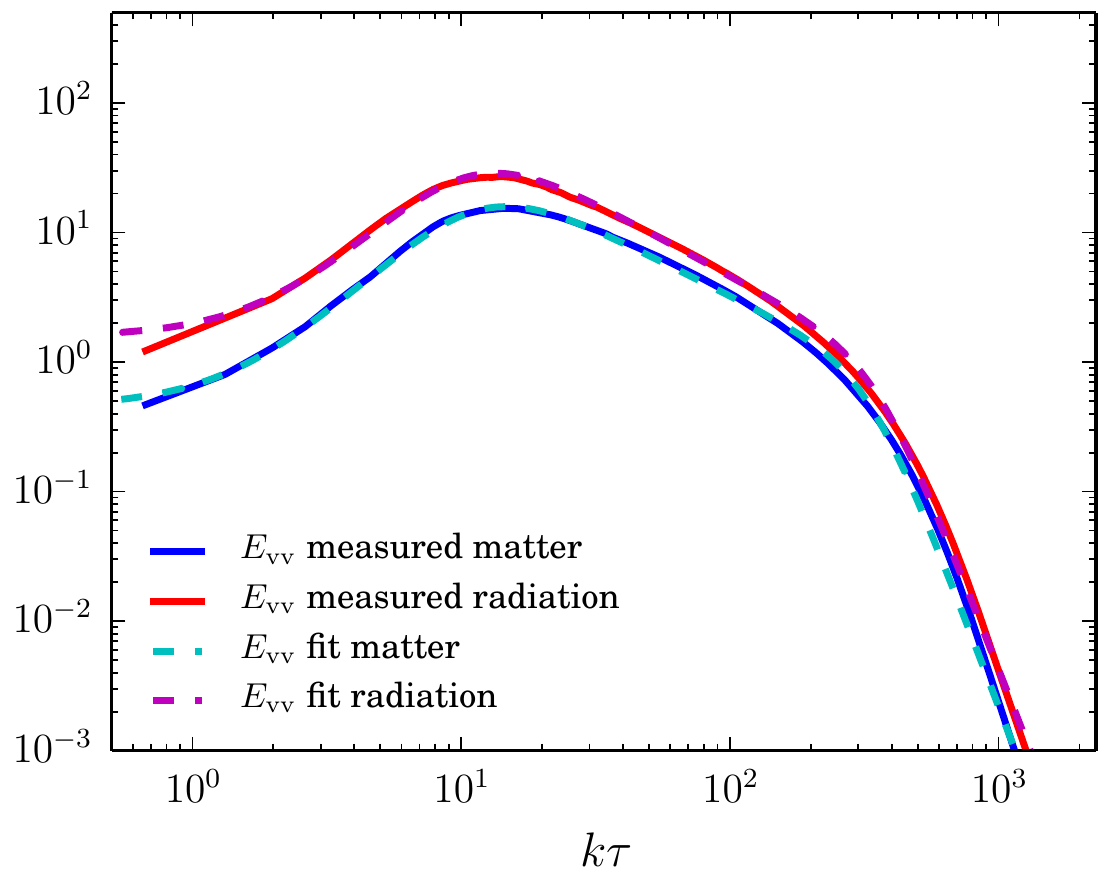}}
\resizebox{0.32\textwidth}{!}{\includegraphics{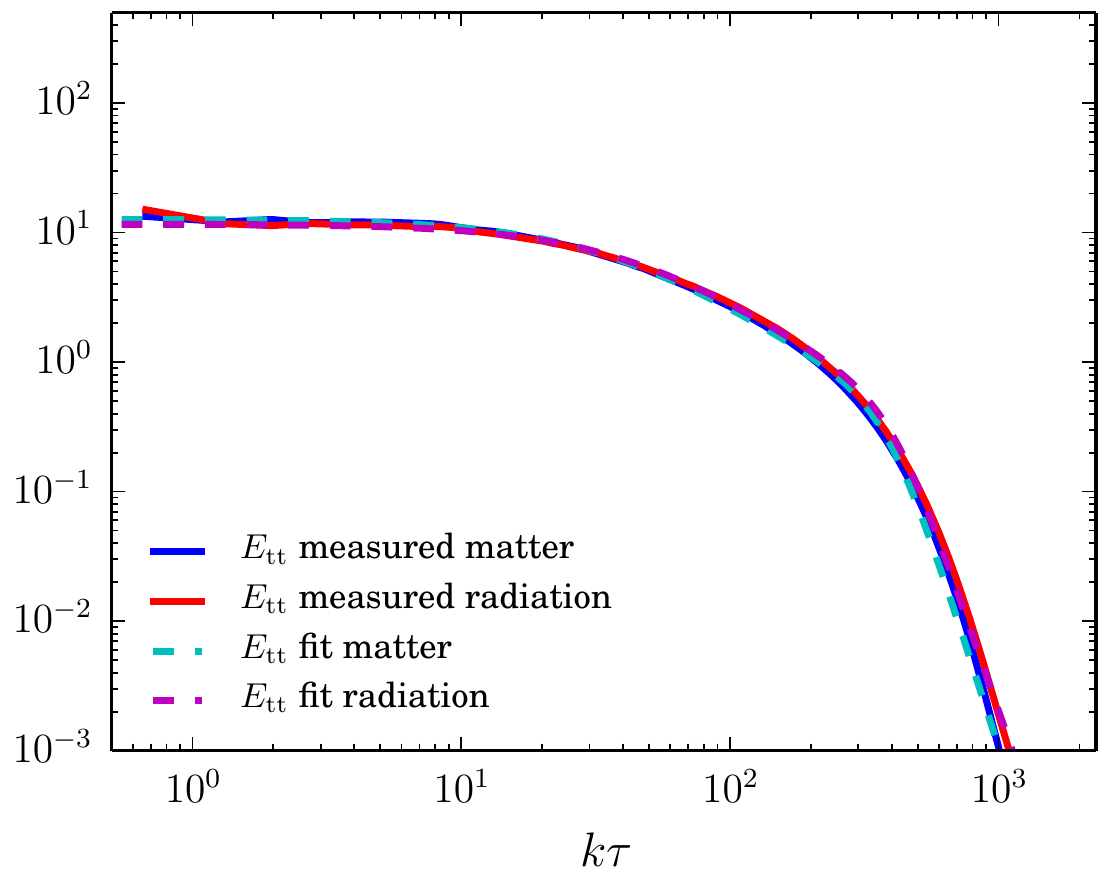}}
\caption{\label{fig:etc_fit} Fits of the ETCs $E(k\tau)$ of the source functions at the reference time to the function given in Eq.\ (\ref{e:ETCfit}). The solid curves show the measured ETCs
and the dashed curves the fits, for the numerical values given in Table~\ref{table_etcfit}.}
\end{figure*}

\begin{table*}[tb]
\begin{center}
\begin{tabular}{|c||c|c|c|c|c|c|c|c|c|c|}
\hline
Parameters & $E_{11}$ mat & $E_{11}$ rad & $E_{22}$ mat  &  $E_{22}$ rad  & $E_{12}$ mat  &  $E_{12}$ rad  & $E_{\rm vv}$ mat  &  $E_{\rm vv}$ rad  & $E_{\rm tt}$ mat  &  $E_{\rm tt}$ rad  \\\hline
 $a_0$ & $49.7$                    & $51.9$      & $111$               &  $137$             &  $14.6$            &  $33.8$            & $0.45$            & $1.6$               & $12.7$               & $11.6$    \\
 $a_2$ & $0.290$                  & $0.77$      & $0.046$            &  $0.96$            &  $0.64$            &  $1.3$            & $0.21$              & $0.42$               & \textemdash        & \textemdash   \\
 $a_4$ & $0.0076$                & $0.030$  & $0.0024$           & $0.0077$         & $0.0017$         &  $0.013$        & \textemdash     & \textemdash  & \textemdash         & \textemdash      \\
 $z_p$ & $14.2$                    & $11.9$      & $15.5$               & $13.8$            &  $15.4$             &  $12.8$          & $12.0$            & $11.4$               & $37.5$               & $44.1$     \\
 $d$    & $1.14$                     & $1.23$      & $1.28$              & $1.34$            &  $1.18$             &   $1.51$          & $1.05^a$   & $1.13^a$        & $1.39^b$            & $1.40^b$  \\
 $w$    & $362$                     & $394$       & $340$               &  $363$            &   $294$             &   $389$           & $332$                & $354$               & $395$               & $416$    \\
 $\delta$& $4.0$                   & $3.9$        & $3.1$                & $3.0$              &   $2.7$             &   $2.7$             & $4.4$                & $4.2$               & $5.2$               & $4.9$    \\
 $\omega$& \textemdash      & \textemdash  & \textemdash & \textemdash  &   $0.0030$           & $0.0029$      & \textemdash   & \textemdash  & \textemdash  & \textemdash  \\
 \hline
\end{tabular}
 \caption{\label{table_etcfit} The numerical values of the fits of the ETC shown in Fig.\ \ref{fig:etc_fit} to Eq.\ (\ref{e:ETCfit}). \\${}^a$: Fit to $E(x) = (a_0 + a_2 x^2)/\left[(1 + (x/z_\text{p})^{2+d})(1 + (x/w)^\delta)\right] $ \\${}^b$: Fit to $E(x) = a_0/\left[(1 + (x/z_\text{p})^{d})(1 + (x/w)^\delta) \right]$}
\end{center}
\end{table*}

\section{Global modelling of the UETCs\label{app:fit}}

In this appendix we provide global fits that describe the merged UETCs obtained with our simulations. We 
first fit only the equal-time part and then extend the fits to take into account the temporal decoherence
properties as well. A fit for the full UETCs can be obtained by simply combining the two fitting functions.
These fits are not optimised to give the best possible interpolation, but rather to minimise the number of
free parameters while still allowing for a reasonable representation.

In order to perform the fits, the  $1\sigma$ fluctuations were found with 10 bootstrap samples from the 7 $s=0$ and the 7 $s=1$ realisations. 
The statistical error bars are the smallest at high $k$ as there are many more modes there than on large scales. To avoid having those scales (which probe the string structure rather than the scaling UETCs) dominate the fits, and to have more uniform error bars, we scale the `raw' standard deviations from the bootstrap by $\sqrt{k}$ and re-estimate the overall error amplitude simultaneously with the fitting parameters.

\subsection{Fits of the ETCs}

Looking at the equal-time line of Fig.\ \ref{fig:UETCs} we see that we generally expect a plateau on super-horizon
scales, $k\tau \lesssim 1$, possibly a peak around the horizon scale, and then a decay for $k\tau \gg 1$. From
Fig.\ \ref{ETC_PL} we gather that the high-$k\tau$ decay consists of a power-law and then faster decay once we
hit the string scale, that can also be modelled with a power law. Based on these considerations,
we propose the following fitting functions for the final ETCs $E(x) = C(z,r=1)$ of Eq.\ (\ref{eq:finalC}) (with $x\equiv k\tau$, $x'\equiv k\tau'$, $z\equiv k \sqrt{\tau \tau'}$ and $r=\tau/\tau'$):
\begin{equation}
E(x) = \frac{a_0 + a_2 x^2 + a_4 x^4}{\left( 1+ (x/z_p)^{4+d} \right) \left(1+ (x/w)^\delta \right) } \cos(\omega x) \, .
\label{e:ETCfit}
\end{equation}
Here $w \gg z_p$ so that $d$ gives the cosmologically relevant decay law while $\delta$ is relevant inside the string
core. Not all ETCs require all parameters. The oscillation frequency $\omega$ only applies to $E_{12}$, it is zero
for all other ETCs. The vector ETCs are well modelled only with the constant and quadratic term in the numerator
and we set $a_4 = 0$, correspondingly we use $2+d$ for the decay exponent in the denominator. The tensor
ETCs are well described even by just a constant numerator, and for this reason we only use $d$ as the
exponent in the denominator. 

The ETCs that we fit were obtained at the reference times $\tau_\mathrm{ref}$ listed in Table~\ref{table_tref} for
$s=1$. The large-scale behaviour is independent of this choice thanks to scaling, but the string width is not.
For cosmological applications, the string width should be much smaller than the horizon scale, which means
that one should set $w \rightarrow \infty$ in this case.

\begin{figure*}[t]
\resizebox{0.32\textwidth}{!}{\includegraphics{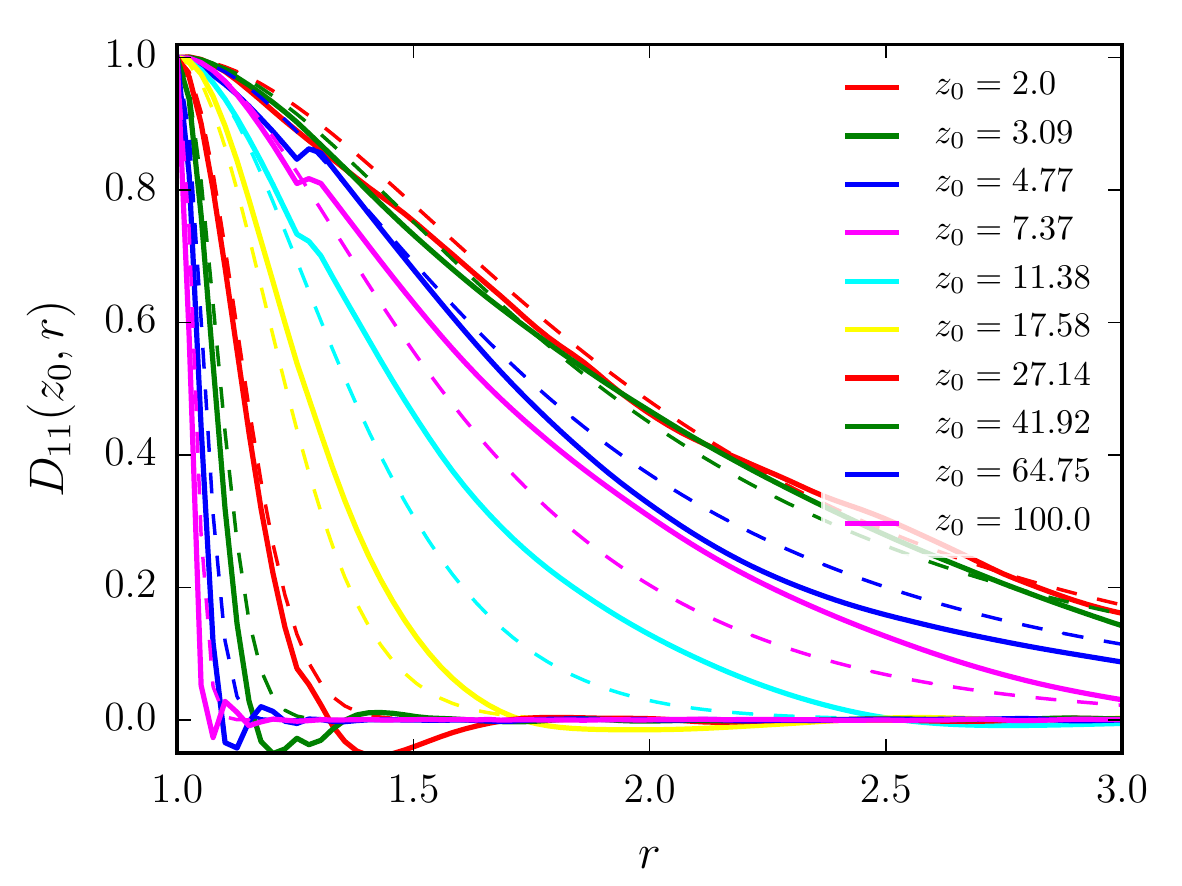}}
\resizebox{0.32\textwidth}{!}{\includegraphics{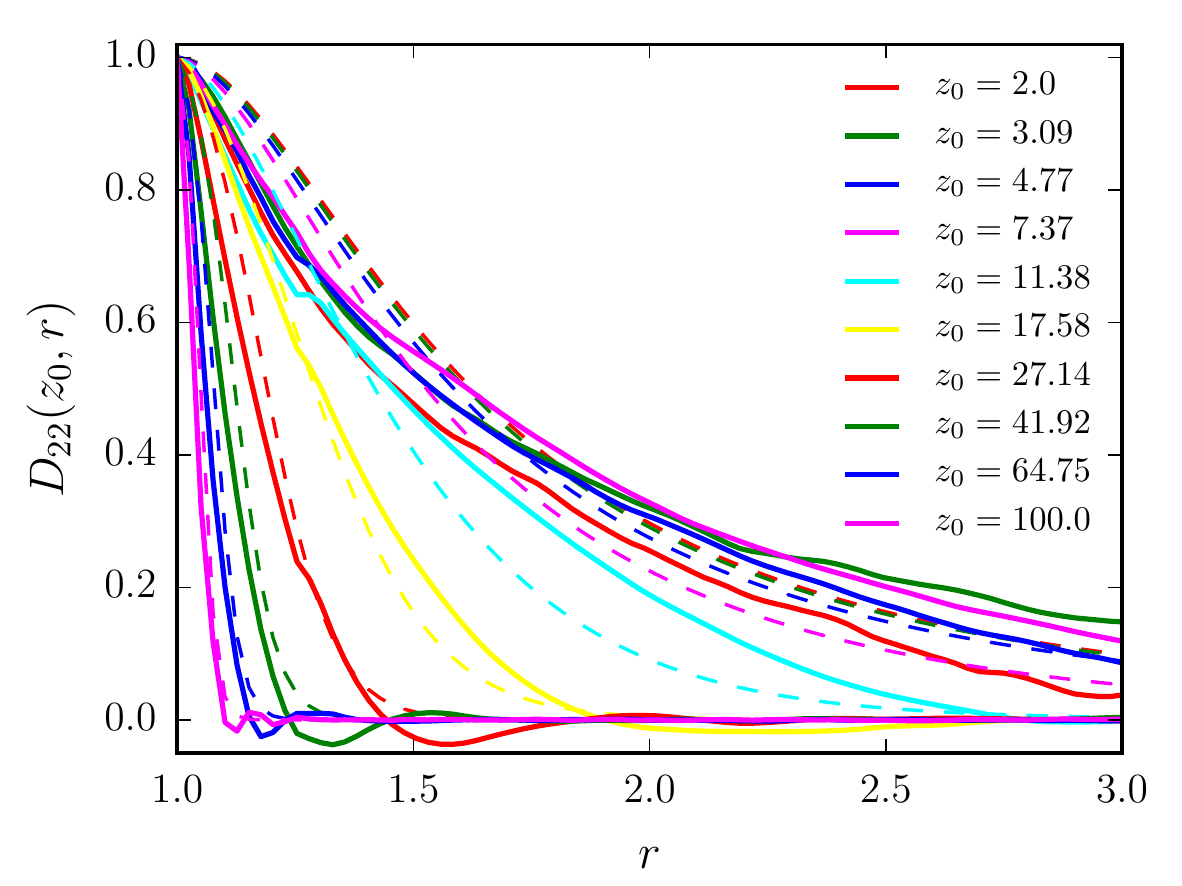}}
\resizebox{0.32\textwidth}{!}{\includegraphics{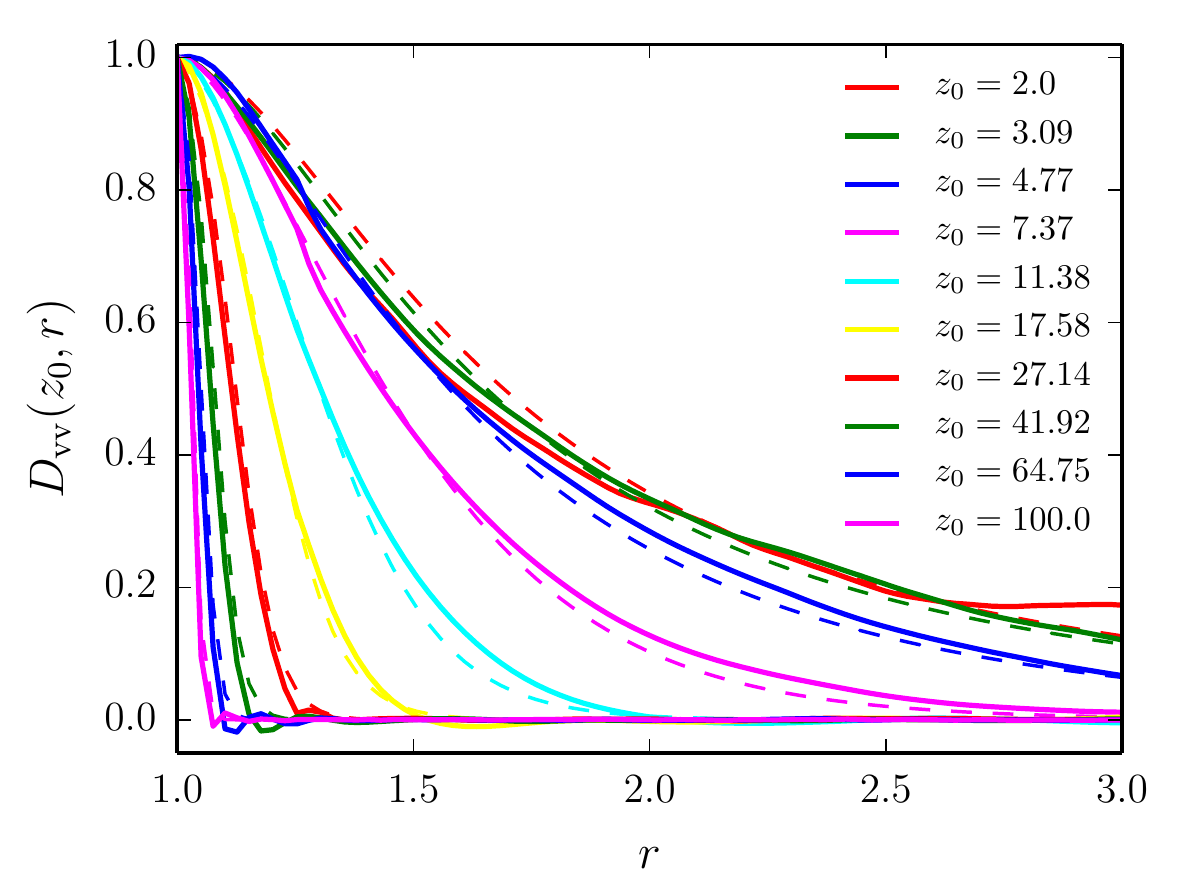}}
\resizebox{0.58\textwidth}{!}{\includegraphics{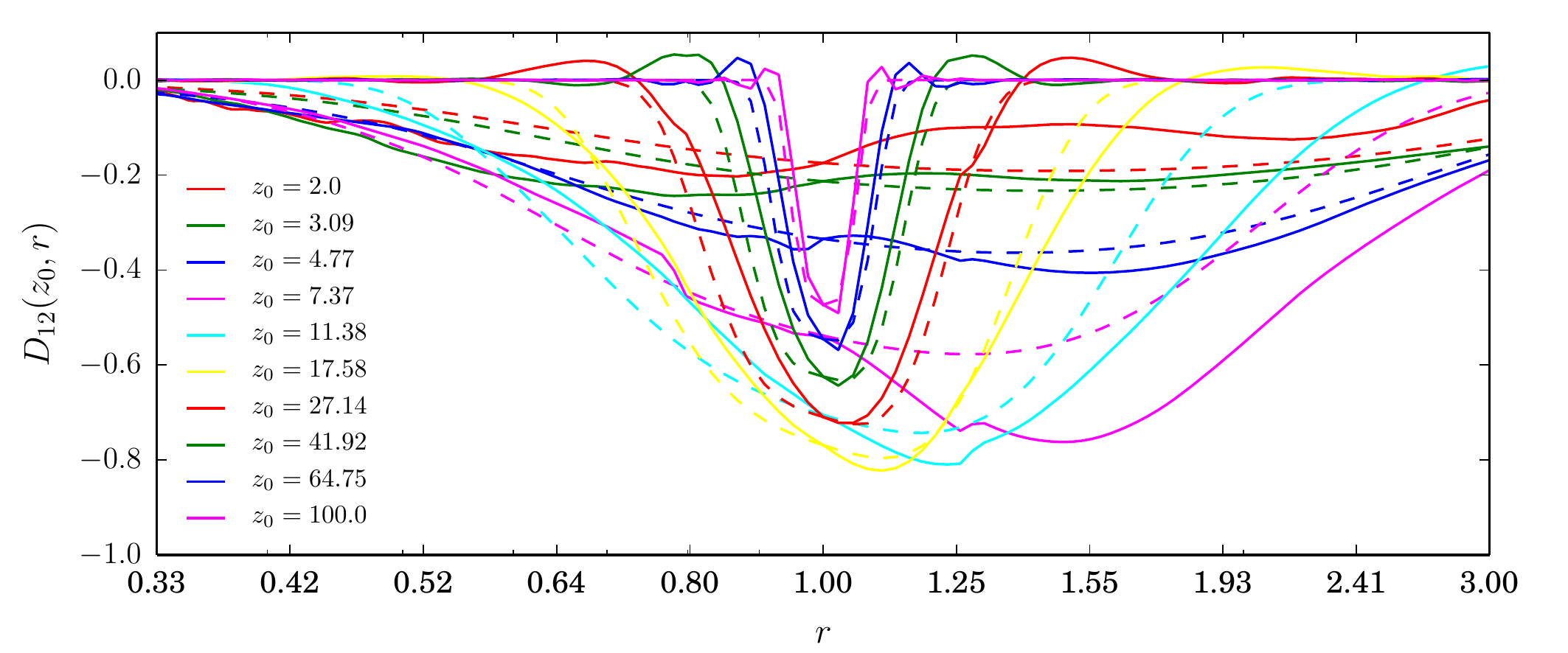}}
\resizebox{0.32\textwidth}{!}{\includegraphics{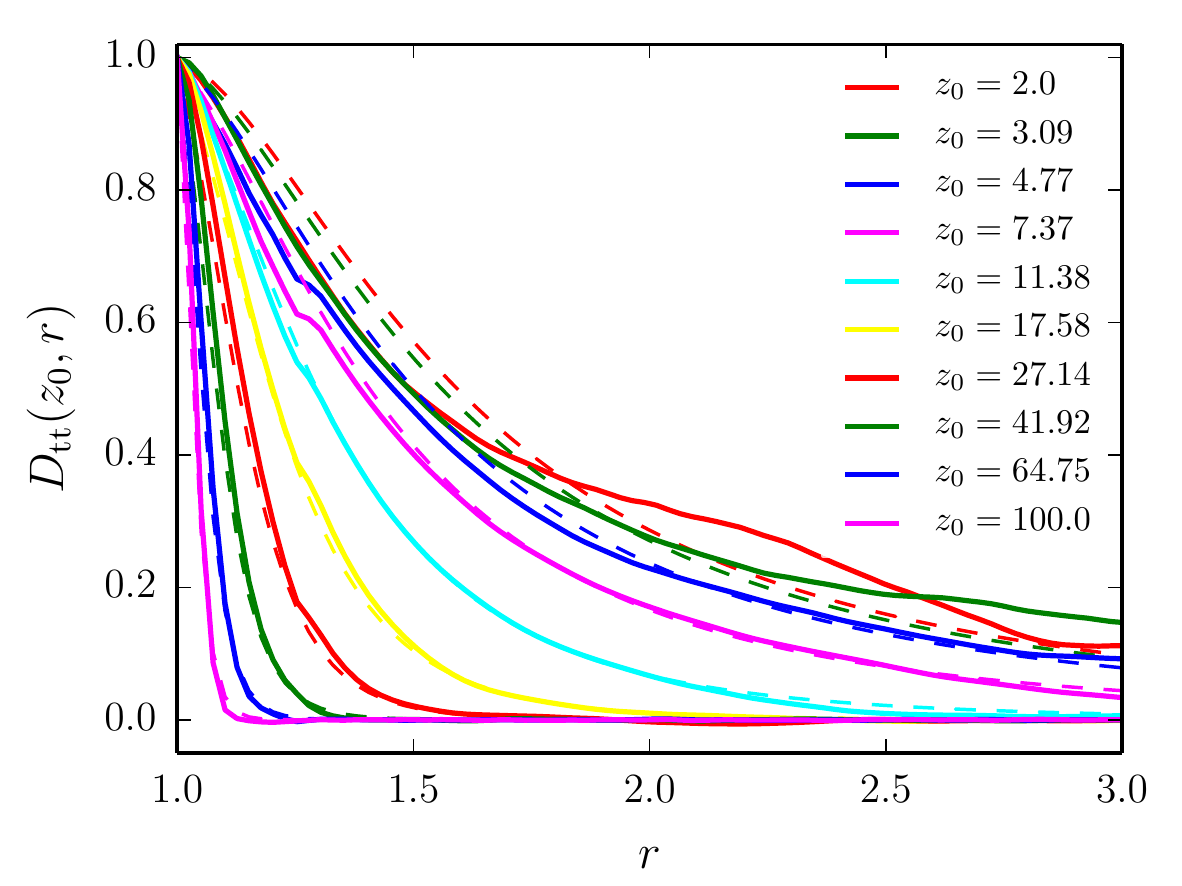}}
\caption{\label{fig:corr_fit} Fits of the correlation functions $D(z,r)$ of the matter era source functions as a function of $r$ for a range of values of $z$, to the functions (\ref{e:Dfit1},\ref{e:Dfit2}). The solid curves show the measured correlation functions
and the dashed curves the fits, for the numerical values given in Table~\ref{table_corrfit}.}
\end{figure*}

\begin{table*}[tb]
\begin{center}
\begin{tabular}{|c||c|c|c|c|c|c|c|c|c|c|}
\hline
Parameters & $D_{11}$ mat & $D_{11}$ rad & $D_{22}$ mat  &  $D_{22}$ rad  & $D_{12}$ mat  &  $D_{12}$ rad  & $D_{\rm vv}$ mat  &  $D_{\rm vv}$ rad  & $D_{\rm tt}$ mat  &  $D_{\rm tt}$ rad  \\\hline
 $\alpha$     & $3.3$           & $2.5$               & $2.8$               &  $2.3$             &  $4.0$            &  $2.4$        & $2.6$            & $2.8$               & $2.8$               & $3.6$    \\
 $r_p$          & \textemdash & \textemdash   & \textemdash     & \textemdash   &   $1.5$           & $1.9$            & \textemdash   & \textemdash    & \textemdash  & \textemdash  \\
 $A$             & $6.5$          & $2.5$ & $0.30$         &  $0.20$          &  $26$            &  $16$             & $0.37$            & $0.68$             & $0.27$          & $0.27$   \\
 $\sigma$    & $2.0$           & $1.6$             & $1.4$               & $1.3$              & $3.0$         &  $0.18$             & $0.98$           & $1.3$                & $0.65$          &  $0.70$      \\
 $\beta$      &  $2.0$           & $2.0$             & $1.4$               & $1.4$             &  $3.6$             &  $3.0$            & $1.4$              & $1.7$               & $0.85$           & $0.95$     \\
 \hline
\end{tabular}
 \caption{\label{table_corrfit} The numerical values of the fits of the correlation functions shown in Fig.~\ref{fig:corr_fit} to the functions 
 (\ref{e:Dfit1},\ref{e:Dfit2}). All the correlation functions except $D_{12}$ are
 symmetric, which implies that the peak shift is absent, $r_p=1$.}
\end{center}
\end{table*}

\subsection{Fits of the decoherence functions}

We define the decoherence functions $D$ as
\begin{equation}
D(x,x') = C(x,x')/\sqrt{E(x) E(x')} \, .
\end{equation}
Together with the ETCs they parameterize the full UETCs. The functions $D$ describe
how the UETCs decohere away from the equal-time line, and are defined such that $D(x,x)=1$. 
We also find that on super-horizon scales ($z\ll 1$) their width is constant in $r$, while on sub-horizon scales their width instead is constant in $|x-x'|$ (i.e.\ inversely proportional to $z$).

The correlation functions are symmetric under $x \leftrightarrow x'$, which corresponds to $r \leftrightarrow 1/r$, combined with complex conjugation. The UETCs are in fact real, so the conjugate need not be taken. The exception is $D_{12}$ which is the cross-decoherence function of the two scalar sources, and does not need to be symmetric. To allow for this, we introduce a peak location parameter $r_p$, which is equal to unity for all other decoherence functions. The cross-decoherence function is also different from the others as we define it through $D_{12}(x,x') = C_{12}(x,x')/\sqrt{E_{11}(x) E_{22}(x')}$. The equal-time line $D_{12}(x,x)$ quantifies then the correlation between the sources for $\phi$ and $\psi$ at equal times, which is not unity in general. However, it can be straightforwardly computed from the ETC fits given above.

The fitting formula that we use for the correlation functions is then
\begin{equation}
D(z,r) = (2+A)\left[ r^\alpha + r^{-\alpha} + A \exp\left\{ (0.5 \Delta/\sigma)^\beta \right\} \right]^{-1}
\label{e:Dfit1}
\end{equation}
where we set $\Delta = z |\log r|$ to obtain the correct scaling of the decay width for $z \gg 1$.
The correlation functions also show small residual oscillations, however they are small and not
well measured in our simulations as $D$ decays quite rapidly away from the equal time line, and
we decided to neglect these oscillations in the fits given here.

For $D_{12}$ we took
\begin{eqnarray}
\label{e:Dfit2}
D_{12}(z,r) &=& \left(r_p^\alpha + r_p^{-\alpha} +A \right) D_{12}(z,1) \times \\
&&\left[ \left(\frac{r}{r_p}\right)^\alpha + \left(\frac{r}{r_p}\right)^{-\alpha} + A \exp\left\{ (0.5 \Delta/\sigma)^\beta \right\} \right]^{-1} \nonumber
\end{eqnarray}

We show the correlation functions and the fits in Fig.~\ref{fig:corr_fit} for the matter dominated evolution of the universe. The correlation function during radiation domination look qualitatively similar. The fitting values are given in Table~\ref{table_corrfit} for both epochs.

As mentioned earlier,
in order to reconstruct the full UETCs for cosmological purposes, one combines the ETCs and the correlation functions by
setting
\begin{equation}
C(x,x') = D(x,x') \sqrt{E(x) E(x')} \, .
\end{equation}
In this situation one should also set $w=\infty$ in the ETC parameterisation.


\vspace{3mm}

\bibliography{CosmicStrings.bib}


\section*{Erratum}
In this paper we presented new results for the energy-momentum Unequal Time Correlators (UETCs) of Abelian Higgs cosmic strings. The new modelling includes contributions from simulations of strings with constant physical width ($s=1$), improving on previous simulations which used constant comoving width ($s=0$). The dramatically increased volume also allowed the UETCs to be measured over greater conformal time ratios and a wider range of wavenumber.

In the paper, we neglected to take into account the fact that covariant conservation of the energy-momentum (EM) tensor constrains the superhorizon behaviour of the vector correlators. The constraints come from the fact that EM conservation links the two vector parts of the strings' EM tensor through:
\begin{equation}
\dot T_{0i} + H T_{0i} + ik_j T_{ij} = 0.
\end{equation}
There is no constraint on the spatial parts $T_{ij}$, so as the strings are created by a random causal process (the phase transition), the power spectral density is just white noise, or $\vev{|T_{ij}|^2} \propto k^0$. Energy momentum conservation means that the spectral density of $T_{0i}$ must be consistent with the white noise behaviour of $T_{ij}$:
hence $\vev{|T_{0i}|^2} \propto k^2$.

Our fitting in Appendix B of the paper did not consider this constraint and thus vector equal time correlator (ETC) does not follow the behaviour dictated by it at low-$k\tau$. This erratum aims to highlight that the decay of the vector UETCs at superhorizon scales must be corrected to be $\sim z^2$ where $z=k\sqrt{\tau\tau'}$. 

We apply this correction to the vector correlator obtaining the values corresponding to the $\bk$ bins in the range $0.17<z_{\xi}\exp(-|\ln(r_{\xi})|)<1.29$ by extrapolating the value of the correlator at the upper end of the range. 
Using the notation of the published paper,
\begin{equation}
C_{\rm vv}^{({\rm tot},\xi)} (z_{\xi},r_{\xi}) = C_{\rm vv}^{({\rm tot},\xi)} (z_{\xi}^*,r_{\xi}^*) \left(\frac{z_{\xi}^{2}}{z_{\xi}^{*2}} \right)\, ,
\end{equation}
where $z^*_{\xi}\exp(-|\ln(r^*_{\xi})|)=1.29$.
 
Only the vector perturbations are affected by this constraint, hence the assumptions and extrapolations applied to other perturbation correlators are not affected. 

The correction only affects the CMB vector power spectrum at $\ell\sim30$, thus Planck constraints will not be affected. In order to support this statement we have tested some cases, and checked that there is no difference to the constraints on the string tension $G\mu$ presented in \cite{Lizarraga:2016onn}.
 
 Here we give the fit results presented in Appendix B taking into account the constraint discussed above, {\em i.e.}\ with low $x$ behaviour of the ETC $E(x)$ going as $x^2$. The fitting formula (B1) now becomes
\begin{equation}
E(x) = \frac{a_2 x^2}{\left( 1+ (x/z_p)^{2+d} \right) \left(1+ (x/w)^\delta \right) } \, .
\label{e:ETCfiterr}
\end{equation}
The numerical values reported in Table X change slightly, but overall remain consistent with the previous fit. We give them in Table \ref{table_etcfiterr}. We also show in Fig.\ {\ref{fig:etc_fiterr} the (updated)  relevant panel of  Fig.\ 17 of  the published paper.

\begin{table}[!htb]
\begin{center}
\begin{tabular}{|c||c|c|}
\hline
Parameters & $E_{\rm vv}$ mat  &  $E_{\rm vv}$ rad    \\\hline
 $a_2$        &  $0.22$              & $0.49$                  \\
 $z_p$         & $12.2$            & $10.8$                    \\
 $d$            & $1.14$            & $1.16$      \\
 $w$            & $384$                & $387$                \\
 $\delta$      & $4.7$                & $4.4$                 \\
 \hline
\end{tabular}
 \caption{\label{table_etcfiterr} The numerical values of the fits of the vector ETC shown in Fig.\ \ref{fig:etc_fiterr} to Eq.\ (\ref{e:ETCfiterr}).}
\end{center}
\end{table}

\begin{figure}[!htb]
\resizebox{0.32\textwidth}{!}{\includegraphics{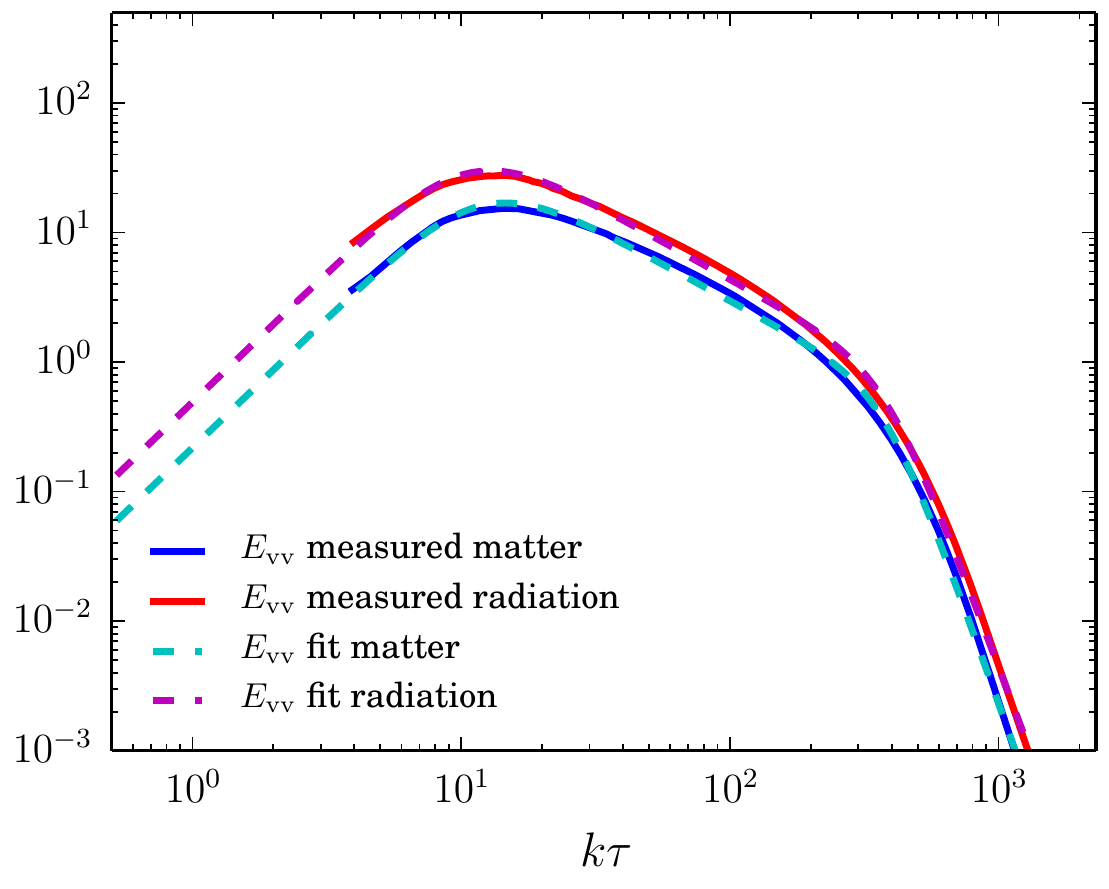}}
\caption{\label{fig:etc_fiterr} Fits of the ETC $E_{\rm vv}(k\tau)$ of the vector source function at the reference time to the function given in Eq.\ (\ref{e:ETCfiterr}). The solid curves show the measured ETCs
and the dashed curves the fits, for the numerical values given in Table~\ref{table_etcfiterr}.}
\end{figure}


\end{document}